\documentclass[11pt,a4paper]{article}
\usepackage{geometry}
\usepackage[utf8]{inputenc}
\usepackage[T1]{fontenc}
\usepackage[english]{babel}
\usepackage{graphicx}
\usepackage{caption}
\usepackage{amssymb,amsmath,amsfonts}
\usepackage{mathrsfs}
\usepackage{color}
\usepackage[normalem]{ulem}
\usepackage{enumitem}
\usepackage{hyperref}
\usepackage{mathptmx}
\usepackage{bm}

\title{Arrival-time distributions as a probe of the preferred foliation\\
in relativistic Bohmian mechanics}
\author{Arnaud Amblard \and Aurélien Drezet}
\date{}

\begin{document}

\maketitle

\begin{center}
\textit{Institut Néel, UPR 2940, CNRS-Université Grenoble Alpes,\\
25 avenue des Martyrs, 38000 Grenoble, France}
\end{center}

\begin{abstract}
    Relativistic extensions of de Broglie-Bohm theory postulate a preferred foliation of space-time, an additional structure essential for defining simultaneous configurations on Minkowski space-time, but conventionally believed to be empirically undetectable at quantum equilibrium. In this paper, we outline an experimental protocol for empirically detecting the preferred foliation, which is assumed to be flat for simplicity. Building on the arrival-time distributions for spin-1/2 particles predicted by Das and D\"urr [S.\ Das and D.\ D\"urr, \href{https://www.nature.com/articles/s41598-018-38261-4}{Sci.\ Rep.\ \textbf{9\,}: 2242} (2019)], we show that in an EPRB-type experiment with spacelike-separated spin and arrival-time measurements, the observed arrival-time statistics will depend crucially on the temporal order of these measurements relative to the preferred foliation of space-time. This dependence offers a potential experimental signature of the preferred foliation postulated by relativistic Bohmian models. Moreover, it implies the possibility of superluminal signaling.
\end{abstract}

 \section{Introduction}\label{sec:introduction}

The widely accepted ``peaceful coexistence'' between quantum mechanics (QM) and special relativity rests on a crucial mathematical pillar: no-signaling theorems \cite{eberhard1978bell,ghirardi1980general}. These theorems guarantee that spacelike-separated quantum measurements performed on entangled particles cannot be used for superluminal communication. However, they are not unconditional. Their proofs rely--explicitly or implicitly---on the assumption that every conceivable experiment yields statistics described by a Positive Operator-Valued Measure (POVM) \cite{cavendish2026} \cite[Sec.\ 5.5.9 and\ 7.6.2]{tumulka2022foundations}. If a quantum experiment were to produce statistics outside the scope of the POVM framework, the central premise of these theorems would break down, and the door to superluminal signaling would, in principle, be opened.

Arrival-time measurements present a promising setting for investigating this possibility. This is because, unlike position or momentum, time is not represented by a Hermitian operator in standard QM, and despite decades of research, no satisfactory POVM for describing arrival-time experiments has been established. Existing proposals---such as the Aharonov–Bohm–Kijowski distribution \cite{aharonov1961time,kijowski1974time}---are essentially limited to freely moving particles and lack compelling extensions to general contexts \cite{mielnik2011timeoperatorchallengepersists,das2021times}. More recent POVM proposals, such as those based on an absorbing boundary condition \cite{tumulka2022distribution}, have been shown to be empirically inadequate \cite{cavendish2025absorbing}.  This widely recognized theoretical gap, the so-called ``quantum arrival-time problem'' \cite{muga2000arrival}---the absence of a widely accepted, empirically promising POVM for arrival times---makes arrival-time experiments a privileged arena for exploring potential violations of no-signaling constraints.
\medskip

In 2019, Das and D\"urr (DD) \cite{das2019arrival} introduced spin-dependent arrival-time distributions calculated within the framework of de Broglie–Bohm theory (dBB), also known as Bohmian mechanics (BM). These distributions, which have recently been proven to be non‑POVM by Goldstein, Tumulka, and Zangh\`i (GTZ) \cite{Goldstein_2024,Goldstein_2024_SR}, offer a concrete example of statistics that that cannot be described (or even approximated) by any POVM, thus violating the key assumption of the no-signaling theorems. In fact, their spin‑dependent arrival‑time distributions for a spin‑$\tfrac12$ particle moving in a waveguide exhibit \textit{drastic} deviations from POVM statistics, thereby providing a concrete, experimentally feasible opportunity to investigate superluminal‑signaling and foliation‑detection protocols.

The insight that DD's predictions could, in principle, enable superluminal signaling was, to our knowledge, first recognized by Sandro Donadi and Siddhant Das (private communication with Siddhant Das). Tim Maudlin subsequently popularized this idea in several public discussions~\cite{video1Maudlin,video2Maudlin,video3Maudlin}, though without specifying a concrete experimental protocol. These remarks directly inspired GTZ to outline the first schematic superluminal signaling protocol based on DD's arrival‑time statistics~\cite{Goldstein_2024}. A similar schematic proposal was also sketched by one of the present authors (Aurélien Drezet) in~\cite{drezet2024arrivaltimebohmianmechanics}. Notably, both Ref.\ \cite{drezet2024arrivaltimebohmianmechanics} and Ref.\ \cite{Goldstein_2024} express strong skepticism about the actual feasibility of such signaling. Indeed, Drezet has previously argued that empirical confirmation of DD's predictions ``would imply an entirely new physics'' beyond the standard POVM‑based measurement framework and would threaten the peaceful coexistence between QM and special relativity \cite{drezet2024arrivaltimebohmianmechanics}.

In the present work, we adopt a different perspective: we explore the consequences that would follow if these arrival-time distributions---as drastic as they are---were even \textit{approximately} correct. Specifically, we extend DD's single-particle setup into a two-particle EPRB-like setup in Minkowski space-time. From this extension, we derive an explicit protocol for superluminal signaling. More importantly---and this constitutes the central novel contribution of this paper---we demonstrate that, if DD's predictions are approximately correct, they provide a method for the direct experimental detection of the \textit{preferred foliation}, a defining feature of all relativistic Bohmian models. As a concrete framework for our analysis, we adopt a Hypersurface Bohm–Dirac (HBD) model \cite{duerr_hypersurface_1999} equipped with a flat foliation of Minkowski space-time into spacelike hyperplanes; this model is physically equivalent to the pilot-wave theory originally proposed by Bohm and Hiley for the Dirac equation \cite[Chap.\ 12]{bohm2006undivided}. For simplicity, however, our explicit calculations are carried out in the formalism of the Bohm-Pauli theory, which is a pilot-wave theory for spin-$\tfrac12$ particles moving in Galilean space-time that emerges as the non-relativistic limit of a HBD model \cite[Sec.\ 10.4]{bohm2006undivided}.

While the potential connection between DD's predictions and superluminal signaling has been noted before, the possibility of using these predictions to detect the preferred foliation has received less attention. To our knowledge, Maudlin was the first to publicly mention this possibility \cite{video1Maudlin,video2Maudlin,video3Maudlin}, which was also mentioned in \cite{manero2025disputableassumptionempiricalequivalence}, though without providing a concrete experimental protocol. The original contribution of this paper is to extend Maudlin's insight into a precise, operational procedure for empirically mapping the preferred foliation.

\medskip

We are acutely aware that this proposal rests on the empirical validity of non-POVM predictions, a matter of intense debate within the Bohmian community \cite{Goldstein_2024,das2023comment,beck2025povm}. Three points must be emphasized here. First, the absence of any universally accepted POVM for arrival-time or time-of-flight experiments---despite more than half a century of intense investigation---provides genuine motivation for exploring the physical consequences of non-POVM arrival-time predictions. Second, and crucially, our protocols do not require exact confirmation of DD's predictions. Their ``exotic'' distributions are so radically non-POVM that our proposals remain viable even under substantial deviations from their specific functional form. This robustness will become clear in Section \ref{sec:Alice_measures_first}, where we recall and contrast the qualitative features of two spin-dependent distributions relevant to our considerations. Third, as we shall emphasize in Section\ \ref{sec:from_bohmian_trajectories_to_arrival_time_statistics}, these predictions are based on a predictive method that has a well-established empirical track record spanning a wide range of quantum phenomena.

\medskip

The paper is organized as follows. Section~\ref{sec:from_bohmian_trajectories_to_arrival_time_statistics} introduces the \textit{de Broglie predictive method}---the trajectory-based method for extracting empirical predictions in BM---and summarizes DD's striking spin-dependent arrival-time distributions. Section~\ref{sec:the_preferred_foliation} explains why any relativistic extension of BM requires a \textit{preferred foliation} of space-time and provides a brief introduction to a HBD model equipped with a flat foliation. Section \ref{sec:Setup} describes our proposed variant of an EPRB-Bell-type experiment, which is augmented with DD's arrival-time setup on one side. Section~\ref{sec:A_General_Bohmian_Formalism} presents a general Bohmian formalism for non-relativistic spin-$\frac{1}{2}$ 2-particle systems. Section~\ref{sec:Bohmian_dynamics_of_entangled_singlet_state} applies this formalism to the entangled singlet state in the proposed experimental setup. Section~\ref{sec:definition_of_the_temporal_order} defines precisely the notion of temporal order relative to the preferred foliation that our derivations require. Sections \ref{sec:Alice_measures_first} and \ref{sec:Bob_measures_first} present a simplified derivation of the foliation‑dependent guiding fields and show how they lead to the arrival‑time distributions predicted by DD. These predictions are summarized and analyzed in Section~\ref{sec:summary_and_analysis_of_the_results}. Based on these results, Section~\ref{sec:Foliation_Detection} presents a protocol for empirically detecting the preferred foliation and Section~\ref{sec:superluminal_signaling} presents a protocol for superluminal signaling. We conclude  this article with a general  summary and analysis of our proposal and its connection with earlier works.

\section{From Bohmian Trajectories to Arrival-Time Statistics}\label{sec:from_bohmian_trajectories_to_arrival_time_statistics}

Arrival-time experiments probe a foundational question in QM: When does a particle, prepared in a known initial state, trigger a detector placed at a specified location? In a typical experimental setup, an initially well-localized particle---described by an initial wave function $\Psi_0$---is released from a source and propagates toward a detection surface $\mathcal{D}$, leading to a detection event at some time \(\tau\). By repeating the experiment under identical conditions, one accumulates a statistical distribution for $\tau$. Any satisfactory quantum theory should predict this distribution: a probability density $\Pi^{\Psi_0}(\tau)$ such that $\Pi^{\Psi_0}(\tau)d\tau$ gives the probability that the particle is detected in $\mathcal{D}$ between times $\tau$ and $\tau+d\tau$.

Within the standard formalism of QM, however, predicting such a distribution remains an open problem, the so-called ``quantum arrival-time problem'' \cite{muga2000arrival}. The main roadblock is that unlike position or momentum, time is not represented by a Hermitian operator; consequently there is no Born rule and associated projection postulate for arrival-time experiments. Consequently, multiple competing proposals have been advanced to define and compute $\Pi^{\Psi_0}(\tau)$, but no consensus has emerged \cite{muga2000arrival}.

In this landscape, the dBB theory, given its explicit particle ontology, offers a distinctive conceptual advantage over most existing quantum theories. In the words of D\"urr and Teufel, this theory is ``tailor-made to answer [the arrival-time question], since the notion of where and when the particle crosses a surface is a natural one when trajectories exist'' \cite[pp.\ 345]{durr2009bohmian}. By providing explicit, deterministic trajectories, the Bohmian approach yields arrival-time predictions with conceptual clarity and mathematical precision. This trajectory-based method is applicable to a wide range of physical scenarios---an advantage that stands in sharp contrast to the severely restricted operator-based proposals \cite{aharonov1961time,kijowski1974time}. Pioneered by Leavens \cite{leavens1993arrival}, the Bohmian-trajectory based treatment of arrival-time experiments has since been consolidated and extended, becoming a central tool in the field (see  [56, 61, 110–124] of \cite{das2023detlefdurrarrivaltimedistributions}). We now outline how this method works.

\subsection{The de Broglie-Bohm Theory for a Non-Relativistic Spin-{1/2} Particle}

The de Broglie-Bohm (dBB) theory\footnote{For more detailed presentations of this theory see \cite{bohm1952hiddenII,bricmont2016making,bell_speakable_1987,sep-qm-bohm,tumulka2022foundations,durr2020understanding}.}, also known as Bohmian mechanics (BM), is a heterodox quantum theory that posits the existence of point-like particles following well-defined trajectories at all times, including outside measurement contexts. In this theory, a particle follows a deterministic trajectory $\mathbf{X}(t)$ determined by its initial position \(\mathbf{X}(0)=\mathbf{X}_0\) and the initial quantum state \(\Psi(\textbf{x},0)=\Psi_0(\textbf{x})\). The quantum state of a non-relativistic spin-$\frac{1}{2}$ particle is a two-component spinor: $ \Psi \in L^2(\mathbb{R}^3)\otimes\mathbb{C}^2$, governed (in the absence of external magnetic fields) by the Schr\"odinger-Pauli equation:
\begin{equation}\label{eq:single_particle_pauli_eq}
    i\hbar\, \frac{\partial \Psi}{\partial t}(\mathbf{x}, t) = \left[ -\,\frac{\hbar^2}{2m} \left(\boldsymbol{\sigma}\cdot\nabla\right)^2 + V(\mathbf{x}, t) \right] \Psi(\mathbf{x}, t),
\end{equation}
where $\boldsymbol{\sigma}=(\sigma_x,\sigma_y,\sigma_z)$ denotes the vector of Pauli matrices. For a spin-$\frac{1}{2}$ particle, the probability density is given by the usual Born rule: $\rho(\mathbf{x},t) = \Psi^\dagger(\mathbf{x},t)\Psi(\mathbf{x},t)$.

As in the spinless case, a Bohmian particle is guided by a velocity defined as the ratio of a probability current to the probability density:
\begin{equation}\label{eq:single_spin_guiding_equation}
    \dot{\mathbf{X}}(t)=\boldsymbol{v}_{\text{dBB}}\left(\mathbf{X}(t), t\right),\quad \boldsymbol{v}_{\text{dBB}}\left(\mathbf{x}, t\right)= \frac{\mathbf{J}_{\text{Pauli}}(\mathbf{x},t)}{\rho(\mathbf{x}, t)}.
\end{equation}
This guiding equation ensures the equivariance of the Bohmian dynamics. The Pauli current $\mathbf{J}_{\text{Pauli}}$, together with the probability density $\rho(\mathbf{x}, t) = \Psi^\dagger \Psi$, solve the quantum continuity equation:
\begin{equation}\label{eq:single_Pauli_continuity_equation}
    \frac{\partial\rho}{\partial t} (\mathbf{x},t)+\nabla\cdot\mathbf{J}_{\text{Pauli}}(\mathbf{x},t)=0
\end{equation}
which follows from the single-particle Pauli equation~\eqref{eq:single_particle_pauli_eq}.

Some authors \cite{norsen2014pilot} propose to define the guiding equation~\eqref{eq:single_spin_guiding_equation} by relying on the simplest current that satisfies this continuity equation:
\begin{equation}\label{eq:convective_only}
    \mathbf{J}_{\text{Pauli}}(\mathbf{x}, t) = \frac{\hbar}{m}\operatorname{Im}\!\left(\Psi^\dagger\nabla\Psi\right)\!.
\end{equation}
However, this expression is not unique. The continuity equation~\eqref{eq:single_Pauli_continuity_equation} admits a gauge freedom: given any scalar function $a(\mathbf{x},t)$ and vector field $\mathbf{b}(\mathbf{x},t)$ satisfying $\partial_t a + \nabla\cdot\mathbf{b}=0$, the transformations $\rho \to \rho + a$ and $\mathbf{J} \to \mathbf{J} + \mathbf{b}$ preserve the conservation law~\eqref{eq:single_Pauli_continuity_equation} \cite{holland2003uniqueness}. Such transformations preserve probability conservation but can significantly alter the resulting Bohmian trajectories. The expression for the probability current density $\mathbf{J}_{\text{Pauli}}$ is not uniquely determined by the continuity equation~\eqref{eq:single_Pauli_continuity_equation} alone.

The resolution lies in an additional physical constraint: the Pauli current must arise as the non-relativistic limit of the Dirac current, thereby inheriting the correct relativistic kinematics. Following the analyses of Holland \cite{holland2003uniqueness} and Hodge \textit{et al.} \cite{hodge2014electron}, and adopting the expression used by Das and D\"urr \cite{das2019arrival}, we take:
\begin{equation}\label{eq:pauli_current_general}
    \mathbf{J}_{\text{Pauli}}(\mathbf{x}, t) = \underbrace{\frac{\hbar}{m} \operatorname{Im}\left[ \Psi^\dagger \nabla \Psi \right]}_{\text{\textit{convective flux}}} + \underbrace{\frac{\hbar}{2m} \nabla \times \left( \Psi^\dagger \boldsymbol{\sigma} \Psi \right)}_{\text{\textit{spin flux}}}.
\end{equation}
This current is comprised of two physically distinct contributions. The first term, the \textit{convective flux}, resembles the well-known current density of a spinless particle and captures the motion resulting from the phase gradient of the wave function. The second term, the \textit{spin (magnetization or Gordon) flux}, arises from the spin degrees of freedom.\footnote{ A word of caution is in order here, since in BM the particles possess no spin: their only beables are their positions, and the outcome $\pm1$ of a Stern-Gerlach ``measurement'' is fixed contextually by the particle's position within the incoming wave packet rather than revealing a pre-existing value \cite{norsen2014pilot}. What the spin flux involves is not such a value but the spinorial structure of $\Psi$ itself---in particular its local polarization $\Psi^\dagger\boldsymbol{\sigma}\Psi$. Since $\Psi$ is part of the Bohmian ontology, this structure is dynamically efficacious: it shapes trajectories directly, through Eq.~\eqref{eq:pauli_current_general}, with no measurement involved. As we shall see in Section~\ref{sec:from_bohmian_trajectories_to_arrival_time_statistics}, DD's arrival-time distributions depend precisely on this polarization---on the direction $\hat{n}$ along which the spinor is prepared---and not on the eigenvalue any subsequent measurement would yield.} Crucially, the spin flux is divergenceless, so \eqref{eq:convective_only} and \eqref{eq:pauli_current_general} have the same divergence, and therefore, fulfill the quantum continuity equation~\eqref{eq:single_Pauli_continuity_equation}. Substituting this current into the dBB velocity formula, Eq.\ \eqref{eq:single_spin_guiding_equation}, yields a velocity field comprising a \textit{convective velocity} (arising from the first term) and a \textit{spin velocity} (arising from the second). The spin velocity induces a precessional component\footnote{This precessional motion will become explicit in our analysis of entangled states, where it gives rise to helical trajectories and backflow (see Sect.\ \ref{sec:Alice_measures_first}).} in the particle trajectories while preserving the overall equivariance of the theory.

\subsection{The \textit{de Broglie predictive method}: From Microscopic Trajectories to Empirical Predictions}\label{subsec:predictive_method}

To extract a testable numerical prediction from a quantum theory, one must connect a mathematical model built from the theory's formalism to empirical data. Ideally, one might hope to derive this connection directly, by applying the theory's own dynamical equations to the complete physical situation, including the roughly $10^{23}$ particles that make up any laboratory apparatus. This is not merely computationally impractical. The deeper obstacle is that no such calculation is even well posed: there is no canonically specified Hamiltonian that the phrase ``the apparatus'' picks out, and constructing a candidate many-particle model that could plausibly be said to represent a \textit{functioning} piece of laboratory equipment, rather than an arbitrary collection of $10^{23}$ interacting particles, would itself require substantial physical input beyond the formalism (private conversation with Will Cavendish). Consequently, every \textit{predictive method}---i.e., any recipe for extracting testable numerical predictions from the abstract formalism of a quantum theory---must instead rely on a deliberately simplified, tractable model of the situation: an \textit{idealization}, in the sense that some feature of the full physical situation is knowingly set aside for tractability (see \cite{maudlin2025actualphysicsobservationquantum} for a more detailed discussion).
\medskip

In order to predict arrival times, DD rely on a predictive method that is widely used in BM and that may be called the \textit{de Broglie predictive method}.\footnote{The idea of such a predictive method can indeed be traced back to Louis de Broglie, who, inspired by Einstein's theory of the photon, developed his theory of the double solution and the pilot wave (see, in particular, \cite[pp.\ 447-448]{de1926cr} and \cite[pp.\ 273-274]{de1927structure}). The first ``Bohmian'' formulation (i.e., pilot-wave version) of this method was proposed by Kennard in 1928 \cite{kennard1928quantum}, anticipating Bohm's results by 24 years. For a broader historical analysis related to its origin, see \cite{drezet2024did}.} It applies to experiments in which a particle propagates toward a \textit{waiting screen}, and delivers both arrival-position and arrival-time distributions (private conversation with Siddhant Das and Will Cavendish). It may be stated as follows:

\begin{quote}
\textit{For a given initial wave function $\Psi_0$ and a hypothetical initial configuration $\mathbf{X}_0$, the time and location at which a real detector fires when placed at a surface $\mathcal{D}$ is well approximated by the time and location at which the Bohmian trajectory $\mathbf{X}(t)$---computed from the same $\Psi_0$ and $\mathbf{X}_0$ but \textbf{in the absence of the detector}---first crosses the mathematical surface $\mathcal{D}$.}
\end{quote}

One feature of this statement should be emphasized, since it is what the objections discussed in Section~\ref{sec:conclusion} target. The de Broglie predictive method carries a \textit{passive-screen commitment}, expressed by the phrase ``in the absence of the detector'': the waiting screen is represented by a geometrical surface $\mathcal{D}$ which the particle may cross but which exerts no influence on its motion. Equivalently, the wave function evolves under a Hamiltonian whose external potential models the experimental setup \textit{without} the screen. Other elements of the experimental setup may of course enter that potential---the waveguide confining the particle in DD's setup (see Section~\ref{A_Brief_Outline_of_DD_Predictions}) does so---the commitment being specifically that the surface at which detection is recorded contributes nothing to it. The Bohmian formalism is thus applied to the target system alone (in DD's experiment, the spin-$\frac{1}{2}$ particle): no wave function is ascribed to any part of the measuring apparatus, which enters the description only through the surface $\mathcal{D}$.

\medskip

Concretely, for a hypothetical scenario in which the precise initial configuration $\mathbf{X}(0)=\mathbf{X}_0$ of a spin-$\frac{1}{2}$ particle were known along with its initial wave function $\Psi_0$, the method proceeds as follows. One first computes the trajectory $\mathbf{X}(t)$ by solving the Schrödinger–Pauli equation~\eqref{eq:single_particle_pauli_eq}---with the screen-free potential---together with the guiding equation~\eqref{eq:single_spin_guiding_equation}, for the initial data $(\Psi_0,\mathbf{X}_0)$. From this trajectory one then obtains the \textit{first arrival time} $\tau(\mathbf{X}_0)$ at $\mathcal{D}$ as the first crossing time:
\begin{equation}\label{eq:general_first_arrival_time}
    \tau(\mathbf{X}_0) = \min\{t \mid \mathbf{X}(t) \in \mathcal{D} \text{ and } \mathbf{X}(0) = \mathbf{X}_0 \in \operatorname{supp}[\Psi_0]\},
\end{equation}
where $\operatorname{supp}[\Psi_0]$ denotes the support of the initial wave function.

The overall structure of the de Broglie predictive method is therefore the following. Its first component is a mathematical model of the physical process unfolding in each run of the experiment, which yields an idealized trajectory $\mathbf{X}(t)$ for the target particle. Its second component connects a theoretical quantity computed from that trajectory to what the experimenter records: the first crossing time~\eqref{eq:general_first_arrival_time} is identified with the time at which the detector clicks.

It is worth noting what kind of quantity $\tau(\mathbf{X}_0)$ is. A crossing consists in the trajectory lying on one side of $\mathcal{D}$ at one time and on the other at a later time, and its being the \textit{first} crossing is a fact about the whole of the trajectory's prior history. The quantity $\tau(\mathbf{X}_0)$ is therefore \textit{diachronic}: it is a function of the trajectory $\mathbf{X}:\mathbb{R}\to\mathbb{R}^3$ as a whole, and not of the configuration at any designated instant. This is what makes the de Broglie predictive method well suited to waiting-screen experiments: each run is assigned a definite arrival time, computed from that run's trajectory, with no need to designate in advance a time \(t_f\) at which the experiment is to be interrogated. It is also what distinguishes the method from the framework underlying the POVM theorem, in which the empirical prediction is a function of the \textit{instantaneous} configuration of a macroscopic pointer at a single final time $t_f$---a \textit{synchronic} quantity (private conversation with Will Cavendish).

\medskip

Because the precise initial configuration $\mathbf{X}_0$ is not known in a single experimental run, Eq.~\eqref{eq:general_first_arrival_time} does not yield a deterministic prediction for an individual event. Instead, it defines a random variable $\tau$ whose probability distribution is obtained by averaging over the distribution of $\mathbf{X}_0$ in the ensemble of identically prepared systems.

At this point an additional, logically independent hypothesis enters: the \textit{quantum equilibrium hypothesis}. It states that for an ensemble of particles all prepared with the same wave function $\Psi_0$, the initial positions $\mathbf{X}_0$ are distributed according to $|\Psi_0|^2$. Combining this hypothesis with Eq.~\eqref{eq:general_first_arrival_time} yields the arrival‑time distribution
\begin{equation}\label{eq:arrival-time_distribution}
    \Pi_{\text{dBB}}^{\Psi_0}(\tau) = \int_{\operatorname{supp}[\Psi_0]} \delta\!\left(\tau(\mathbf{X}_0) - \tau\right) |\Psi_0(\mathbf{X}_0)|^2\ d^3\mathbf{X}_0,
\end{equation}
where \(\tau(\mathbf{X}_0)\) satisfies \eqref{eq:general_first_arrival_time}. The distribution $\Pi_{\text{dBB}}^{\Psi_0}(\tau)$ is often called an \textit{ideal} or \textit{intrinsic} arrival‑time distribution, because the influence of the detector on the particle’s motion is deliberately ignored. For a detailed exposition see \cite[Ch.~16]{durr2009bohmian}.

\medskip

It is important to note that the scope of the de Broglie predictive method is wider than arrival-time experiments. Since every trajectory crosses $\mathcal{D}$ at a definite place and a definite time, and since every trajectory inherits a probability weight from the $|\Psi_0|^2$-distribution, the primary object the method delivers is a \textit{joint} distribution of crossing positions and crossing times. Particular experiments call for particular marginals of it: the position marginal in scattering and double-slit experiments, the time marginal~\eqref{eq:arrival-time_distribution} in the arrival-time experiments of interest here, or coarser functions still, such as the classification of Stern-Gerlach detection events as ``up'' or ``down'' (private conversation with Will Cavendish).

This breadth of application matters here, because in the first two cases the predictions have been confronted with data---and they hold. Despite its passive-screen commitment, the method reproduces the measured distributions of arrival positions and the joint arrival-position-time statistics of scattering experiments (the so-called ``screen problem''~\cite{mielnik1994screen}) with high accuracy~\cite{daumer1994scattering,daumer1996scattering}, as well as the interference pattern of the double-slit experiment~\cite{das2025double}, the loss of coherence in macromolecule interferometry~\cite{viale2003analysis}, and the diffraction of atoms by surfaces~\cite{sanz2000causal}. Such empirical success lends strong support to the method as a pragmatic tool for extracting predictions from BM---though it is certainly not the only predictive method logically compatible with the theory. Like any predictive method, it rests on an idealization of the experimental situation, and its domain of empirical validity is therefore presumably bounded: there will be physical regimes in which the passive-screen commitment fails. What those regimes are, however, has not been established, and no criterion has been proposed for identifying them in advance (see Section~\ref{sec:conclusion}). In this context it seems legitimate to explore the empirical consequences of the method in a regime where it has not yet been tested.\footnote{For a potential objection to this exploration invoking the POVM theorem, we refer the reader to Section~\ref{sec:conclusion}.}

\medskip

In the remainder of this paper we adopt the de Broglie predictive method and follow out its consequences. Crucially, in the experiment proposed by DD it comes into conflict with a claim central to orthodox QM: that the statistics of any quantum experiment must be governed by a POVM. The root of the conflict is that Bohmian trajectories depend non-linearly on the initial wave function $\Psi_0$ through the guiding equation, so the distributions $\Pi_{\text{dBB}}^{\Psi_0}(\tau)$ need not be compatible with any POVM acting on $\Psi_0$. Established in detail for DD's setup by GTZ~\cite{Goldstein_2024,Goldstein_2024_SR}, the possibility of such a conflict had already been demonstrated in 2005 by Ruggenthaler \textit{et al.}~\cite{Ruggenthaler_2005}, though not in the language of POVMs: for a one-dimensional setup, they showed that the Bohmian arrival-time distribution is not a quadratic form of the initial wave function, and since any POVM induces such a form, no POVM description is available in that case.

This result did not, however, become a salient feature of the Bohmian literature prior to the work of Das and Dürr. The spin-polarization-dependent distributions they obtain are of a different character: rather than shifting the arrival-time distribution, they cause it to vanish identically beyond a finite $\tau_{\max}$---a difference in the \textit{support} of the distribution (Sec.~\ref{subsubsec:Backflow}). It is in this sense that their predictions provide the first qualitatively distinct departure from POVM statistics, opening the door to the effects analyzed in this paper.

\subsection{A Brief Outline of Das and D\"urr's Predictions}\label{A_Brief_Outline_of_DD_Predictions}

Against this background, we now turn to the specifics of the arrival-time experiment analyzed by Das, N\"oth, and D\"urr (DND) \cite{Das_2019}, which is an analytically tractable variant of the Das-D\"urr-setup \cite{das2019arrival}. This setup consists of a semi-infinite cylindrical waveguide oriented along the $\hat{z}$-axis. The waveguide is modeled by the axisymmetric potential $V_{\bot}(x,y)=\frac{m}{2}\omega^2(x^2+y^2)$ and an infinite potential barrier at $z=0$. For times $t<0$, a spin-$\frac{1}{2}$ particle is confined near $z=0$ by a harmonic potential barrier $V_{\parallel}(z)=\frac{m}{2}\omega_z^2\, z^2$.\footnote{In their initial proposal \cite{das2019arrival}, DD considered an infinite potential well between $z=0$ and $z=d$ to trap the particle. In their subsequent work with N\"oth \cite{Das_2019}, the infinite barrier at \(d\) was replaced by the smooth harmonic potential $V_{\parallel}(z)$, which merely modifies the profile of the initial spatial wave function.} The particle is taken to be initially prepared in the ground state of the potential well defined by the waveguide, the hard wall, and the harmonic barrier. The ground-state wave functions take the following general form:
\begin{equation}\label{eq:DND_initial_WF}
    \Psi_0(\mathbf{x})=\frac{\sqrt{4\,\omega}}{\pi^{3/4}}\theta(z)e^{-\frac{z^2}{2}-\frac{\omega}{2}(x^2+y^2)}\; \chi_{\hat{n}},\qquad \hat{n}\; \in S^2 
\end{equation}
in units where $\smash{\hbar=m=\omega_z=1}$. Here, $\theta(z)$ denotes the Heaviside step function, ensuring the wave function vanishes for $z<0$. Furthermore, $\chi_{s\hat{n}}\in \mathbb{C}^2$ denotes a spinor quantized along the $\hat{n}$-axis, where $s = \pm 1$ corresponds to spin-up ($\chi_{+\hat{n}}$) and spin-down ($\chi_{-\hat{n}}$) states. DND consider two possible directions for the spin: $\hat{n}=\hat{z}$ and $\hat{n}=\hat{x}$. Since the cylindrical waveguide is oriented along the $\hat{z}$ axis, $\hat{n}=\hat{z}$ is referred to as the \textit{longitudinal direction}, while $\hat{n}=\hat{x}$ is the \textit{transverse direction}.

At $t=0$, the longitudinal potential $V_{||}(z)$ is switched off, allowing the wave function to propagate ``freely'' within the waveguide toward a detection screen positioned at $z=L$. This screen is intended to record the first arrival time $\tau$ of the particle at $L$. Repeating the experiment many times ($\sim 10^5$ runs), with the same preparation, yields the arrival-time statistics. The goal is to predict this arrival-time distribution from dBB's first principles.

\medskip

Relying on the de Broglie predictive method and solving the relevant equations analytically, DND obtain two distinct spin-polarization-dependent arrival-time distributions that are central to our analysis:

\begin{enumerate}[label=(\roman*)]
    \item \textit{Heavy tailed distribution} $\Pi_{\hat{z}}(\tau)$: When the particle's spin is oriented along the waveguide axis ($\hat{n}=\hat{z}$), the arrival time distribution decays slowly, exhibiting a heavy tail.
    \item \textit{Exotic distribution} $\Pi_{\hat{x}}(\tau)$: When the spin is oriented transversely to the waveguide axis ($\hat{n}=\hat{x}$), the distribution becomes ``exotic''---it features a sharp cutoff at a maximum arrival time $\tau_{\text{max}}$ beyond which the probability density vanishes.
\end{enumerate}

\begin{figure}[htbp]
    \centering
        \includegraphics[width=13cm]{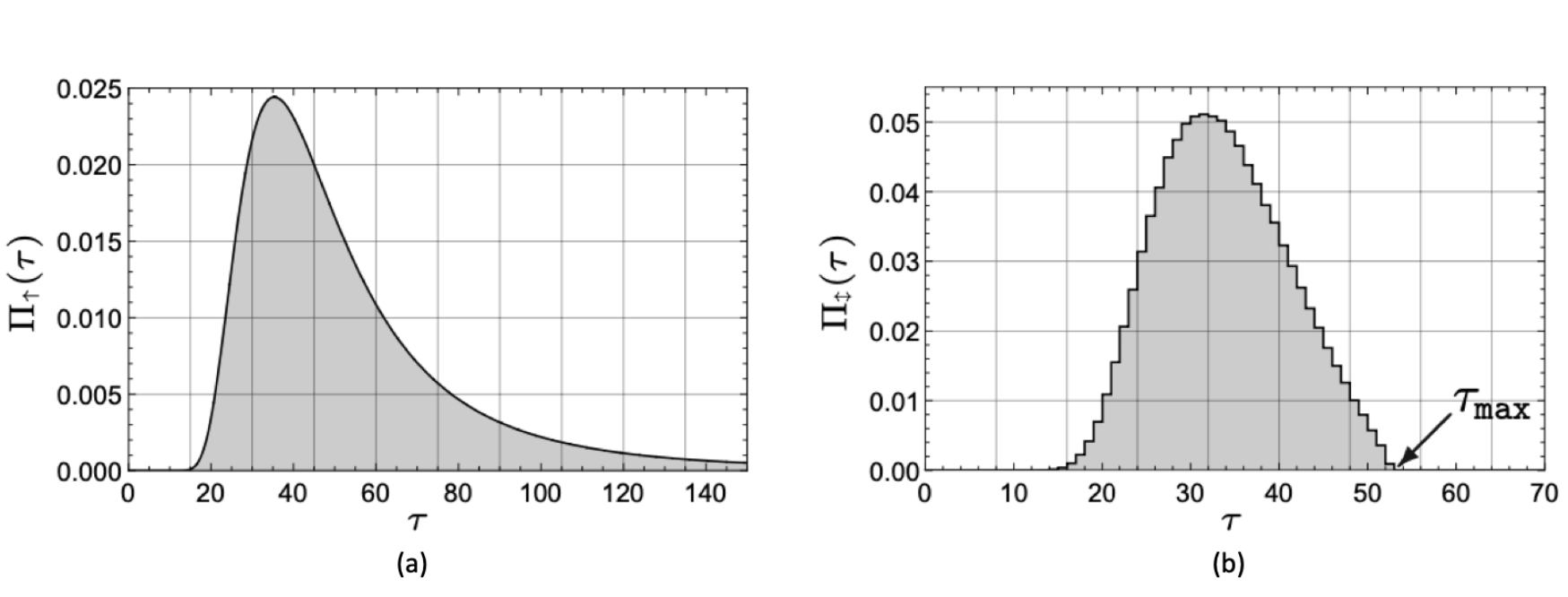}
        \caption{Spin-polarization-dependent arrival time  probability distributions predicted by DND \cite{Das_2019}. The probability distributions are obtained by assuming the \textit{de Broglie predictive method} (see text) and therefore neglect the influence of the detection screen. The choice of spin orientation relative to the waveguide axis qualitatively changes the statistical distribution of arrival times. These predictions form the empirical foundation for our proposal for foliation detection and superluminal signaling. (Reproduced from \cite{Das_2019}). (a) Longitudinal spin configuration ($\hat{n}=\hat{z}$): \textit{heavy tailed distribution} $\Pi_{\hat{z}}(\tau)$ (denoted \(\Pi_{\uparrow}(\tau)\) on the graph).  (b) Transverse spin configuration ($\hat{n}=\hat{x}$): \textit{exotic distribution} $\Pi_{\hat{x}}(\tau)$ (denoted \(\Pi_{\updownarrow}(\tau)\) on the graph) with maximum arrival time $\tau_{\text{max}}$.}
        \label{fig:spin_dependent_arrival_times_distributions}
\end{figure}

These contrasting predictions are illustrated in Fig.~\ref{fig:spin_dependent_arrival_times_distributions}. The striking difference between the two distributions, particularly the existence of a maximum arrival-time in the transverse case, provides the essential empirical signature that enables our proposed detection of the preferred foliation. We emphasize that the arrival-time statistics depend only on the direction $\hat{n}$ of the spin polarization (the axis along which the spin is prepared), not on the specific eigenvalue ($\pm 1$) obtained in a subsequent measurement. This distinction motivates the terminology ``spin-polarization-dependent'' rather than simply ``spin-dependent''.

This paper builds directly on these arrival-time predictions, that rely on the de Broglie predictive method. Crucially, as DD note, their predicted distributions are ``so extremely well articulated'' \cite{Das_2019}---the heavy-tailed and exotic distributions are so dramatically different in shape that they can be clearly distinguished even in the presence of experimental uncertainties and moderate deviations from the idealized predictions. Our proposed protocols therefore do not require exact empirical confirmation of every detail of these distributions; it is sufficient that the spin-polarization-dependent arrival-time statistics remain qualitatively distinct and approximately correct for our foliation detection and superluminal signaling protocols to work.

\subsection{GTZ's Objection}\label{subsec:GTZ_objection}

Before proceeding, we state the principal objection that has been raised against
DD's predictions, since it will occupy us in Section~\ref{sec:conclusion}. The
objection rests on the POVM theorem (see \cite{beck2025povm} and
\cite[Sec.~5.1.2]{tumulka2022foundations} for rigorous presentations), whose
conclusion is that ``the statistics of the result of every quantum experiment is
governed by a POVM'' \cite{Goldstein_2024}. Combined with a proof that DD's
arrival-time distributions are incompatible with POVM statistics, this
conclusion leads GTZ to hold that those distributions ``are actually not what
Bohmian mechanics predicts'' \cite{Goldstein_2024}. DD's distributions stand in conflict with the theorem; the contribution of GTZ \cite{Goldstein_2024,Goldstein_2024_SR} is to give that conflict a precise form, which we now recall.

Suppose the arrival-time statistics of a spin-$\frac{1}{2}$ particle prepared in
the state $\chi_{s\hat{n}}$ were governed by a POVM. Then there exists an
operator-valued measure $\mathcal{O}(d\tau)$ on the spin space $\mathbb{C}^2$
such that
\begin{equation}\label{eq:GTZ_POVM}
    \Pi_{s\hat{n}}(\tau)\,d\tau
    = \chi_{s\hat{n}}^{\dagger}\,\mathcal{O}(d\tau)\,\chi_{s\hat{n}},
    \qquad s=\pm 1 .
\end{equation}
Since $\{\chi_{+\hat{n}},\chi_{-\hat{n}}\}$ is an orthonormal basis of
$\mathbb{C}^2$ for any direction $\hat{n}$, summing \eqref{eq:GTZ_POVM} over
$s=\pm 1$ gives
\begin{equation}\label{eq:GTZ_trace}
    \Pi_{+\hat{n}}(\tau)\,d\tau + \Pi_{-\hat{n}}(\tau)\,d\tau
    = \operatorname{Tr}\left[\mathcal{O}(d\tau)\right],
\end{equation}
whose right-hand side does not depend on $\hat{n}$. Now, as emphasized above and
established in Section~\ref{subsubsec:Backflow}, DD's distributions depend only
on the polarization axis and not on the eigenvalue, so that
$\Pi_{+\hat{n}}=\Pi_{-\hat{n}}\equiv\Pi_{\hat{n}}$ for each direction, and the
left-hand side of \eqref{eq:GTZ_trace} equals $2\,\Pi_{\hat{n}}(\tau)\,d\tau$.
Applied to the two configurations DND consider, this yields
\begin{equation}\label{eq:GTZ_contradiction}
    \Pi_{\hat{x}}(\tau) = \Pi_{\hat{z}}(\tau),
\end{equation}
in flat contradiction with the distributions of
Fig.~\ref{fig:spin_dependent_arrival_times_distributions}. This contradiction
lies at the heart of the controversy between DD and DND on the one hand and GTZ
on the other. GTZ further observe that non-POVM statistics of this kind would
conflict with the no-signaling theorem \cite{Goldstein_2024_SR,drezet2024arrivaltimebohmianmechanics}.

We accept the derivation just given: DD's distributions cannot be described by a
POVM, and we take this to be established. What we do not accept is the further
step, from the non-POVM character of these distributions to the verdict that
they are not what BM predicts. That step grants the POVM theorem the highest
authority over the empirical content of BM, and our reasons for withholding it
are developed in Section~\ref{sec:conclusion}. Until then we simply adopt the de
Broglie predictive method and follow out its consequences.

\section{The Preferred Foliation: A Necessary Structure in Relativistic Bohmian Mechanics}\label{sec:the_preferred_foliation}

In this section we explain why the \textit{preferred foliation}---a space-time structure postulated by every relativistic extension of de dBB theory---emerges naturally, not only from the non-relativistic version of the theory, but even from the mathematical framework of \textit{standard} QM. Understanding this connection is essential for appreciating why the foliation, though often regarded as an extra-theoretical postulate, is in fact deeply rooted in the conceptual foundations of standard QM.

\subsection{Non-Relativistic Bohmian Mechanics and Newtonian Simultaneity}

So far our introduction of BM has been restricted to single-particle systems, yet the non-locality of the dynamics appears in multi-particle systems. In non-relativistic BM, the complete state of an $N$-particle system at time $t$ is specified by two elements: the wave function $\Psi_t$ and the actual configuration $\mathbf{X}(t) := (\mathbf{X}_1(t), \dots, \mathbf{X}_N(t))$ of this system. The dynamics of the system is governed by two deterministic equations. First, the wave function always evolves according to an $N-$particle wave equation (Schr\"odinger or Pauli), which always implies a continuity equation of the form:
\begin{equation}\label{eq:spinless_continuity_equation}
    \partial_t\rho(\mathbf{x}_1,\cdots,\mathbf{x}_N, t) + \sum_{k=1}^N \nabla_k\cdot\mathbf{J}^{(k)}(\mathbf{x}_1,\cdots,\mathbf{x}_N, t) = 0,
\end{equation}
where $\nabla_k = (\partial_{x_k}, \partial_{y_k}, \partial_{z_k})$ denotes the gradient with respect to the coordinates of the $k$-th particle, $\rho(\mathbf{x}_1,\cdots,\mathbf{x}_N, t) = |\Psi_t|^2$ is the probability density and $\mathbf{J}^{(k)}(\mathbf{x}_1,\cdots,\mathbf{x}_N, t)$ is the probability current for the $k$-th particle. Second, each particle moves along a trajectory \(\mathbf{X}_k:\mathbb{R}\to\mathbb{R}^3\) (for \(k=1,\cdots,N\)) that satisfies a \textit{guiding equation} of the generic form:
\begin{equation}\label{eq:guiding_equation}
    \dot{\mathbf{X}}_k(t) = \boldsymbol{v}_{\text{dBB}}^{(k)}\big(t,\mathbf{X}_1(t),\dots,\mathbf{X}_N(t)\big) =\frac{\mathbf{J}^{(k)}}{\rho}\big(t,\mathbf{X}_1(t),\dots,\mathbf{X}_N(t)\big).
\end{equation}
Each particle is thus guided by a velocity field $\boldsymbol{v}_{\text{dBB}}^{(k)}(\mathbf{x}_1,\cdots,\mathbf{x}_N, t) := \mathbf{J}^{(k)}/\rho$ that depends, crucially, on the instantaneous positions of \textit{all} $N$ particles.\footnote{The structure of this guiding field---i.e. \(\mathbf{v}=\mathbf{j}/\rho\) with \(\mathbf{j}\) and \(\rho\) satisfying a continuity equation \eqref{eq:spinless_continuity_equation}---ensures \textit{equivariance}: if the initial configuration $\mathbf{X}(0)$ is distributed according to $|\Psi_0|^2$, then at any later time $t$ the distribution of $\mathbf{X}(t)$ is given by $|\Psi_t|^2$. See \cite[Sec.\ 5.C]{bricmont2016making} for a rigorous proof.} This non-local coupling---the dependence of each particle's motion on the simultaneous positions of all others---embodies the essential non-locality of quantum phenomena within the Bohmian framework.
\medskip

The structure of this non-local dynamics \eqref{eq:guiding_equation} is intimately tied to the Newtonian concept of \textit{absolute simultaneity}. This connection becomes apparent when we examine the mathematical framework of standard QM, which is built upon the concept of \textit{configuration space} $\Gamma \cong \mathbb{R}^{3N}$---a point $(\mathbf{x}_1,\dots,\mathbf{x}_N) \in \Gamma$ represents a possible (ordered) spatial configuration of the $N$ particles composing the system. Here a subtle but crucial tension arises. Standard QM explicitly employs configuration space as the domain of its central mathematical object, the wave function, yet it declines to populate this space with actual configurations $(\mathbf{X}_1(t),\dots,\mathbf{X}_N(t))$ of particles. As de Broglie pointed out, ``[i]t seems a little paradoxical to construct a configuration space with the coordinates of points which do not exist'' \cite[p.~379]{bacciagaluppi2009quantumtheorycrossroadsreconsidering}. BM resolves this tension by taking configuration space literally.

In dBB theory, at any time $t$, the $N$-particle system occupies a well-defined configuration $(\mathbf{X}_1(t),\dots,\mathbf{X}_N(t))\in\mathbb{R}^{3N}$ that corresponds to a specific point in the abstract configuration space $\Gamma$. By definition, such a configuration represents the positions of all $N$ particles \textit{at a single instant}. As Maudlin emphasizes, ``[c]onfigurations are \textit{configurations at a time}; they specify where all of the particles in a system are \textit{at a given moment}'' \cite[p.~198]{maudlin2011quantum}. Consequently, by committing to the literal existence of configurations, the dBB theory commits itself to a notion of absolute simultaneity.

This point deserves emphasis because it is often overlooked in critiques of dBB theory. The formalism of orthodox QM is itself built upon configuration space and therefore \textit{already implicitly presupposes} a notion of absolute simultaneity---the same Newtonian simultaneity that underlies the time derivative in the Schr\"odinger equation. However, standard QM avoids confronting this commitment by treating configuration space as a mere mathematical convenience rather than an abstract space representing a feature of physical reality. BM, by taking the formalism seriously and populating configuration space with actual configurations of particles, makes this implicit commitment explicit.

In the non-relativistic context, this commitment is unproblematic: both Newtonian and Galilean space-time come equipped with a preferred foliation into simultaneity hyperplanes \cite[Chap. 3]{maudlin2012philosophy}. This intrinsic structure gives physical meaning to the notion of ``at a given moment'' in those classical space-times. In the relativistic context, however, the situation differs fundamentally. Because Minkowski space-time lacks any notion of absolute simultaneity, a Bohmian theory formulated in this setting must introduce a \textit{preferred foliation} of Minkowski space-time into spacelike hypersurfaces---an additional space-time structure that provides the notion of ``at a given moment'' needed to define simultaneous configurations. Thus, while many physicists criticize the preferred foliation as an unwelcome addition to relativity, it is in fact the natural extension of a commitment already present---though unacknowledged---in the standard quantum formalism.

\subsection{Hypersurface Bohm-Dirac Models: Making the Foliation Explicit}

The primary relativistic extension of dBB theory is the class of models known as Hypersurface Bohm-Dirac (HBD) models \cite{duerr_hypersurface_1999}. These models preserve the theory's commitment to deterministic particle trajectories while generalizing the dynamics to Dirac particles and achieving a form of formal Lorentz covariance---meaning the laws can be written in a coordinate-free geometric language. However, this covariance is purchased at the price of introducing an additional space-time structure: a \textit{preferred foliation} of Minkowski space-time into spacelike hypersurfaces.

\medskip

The simplest and historically first member of this class is the Bohm-Hiley-Dirac model\footnote{Often called the \textit{Bohm-Dirac model} \cite{duerr_hypersurface_1999}; we use the fuller terminology to acknowledge Basil Hiley's contribution.} \cite[Chap.~12]{bohm2006undivided}. In this model, one begins by singling out a preferred Lorentz frame \(F\), with coordinates \((t,\mathbf{x})\). For \(N\) Dirac particles, the wave function \(\Psi(t,\mathbf{x}_{(1)},\dots,\mathbf{x}_{(N)})\) satisfies the many-particle Dirac equation written in the coordinates of this frame:
\begin{equation}\label{eq:N_particles_Dirac_equation}
    i\hbar \frac{\partial \Psi}{\partial t}=\sum_{k=1}^N\big(c\boldsymbol{\alpha}_k \cdot \hat{\mathbf{p}}_k+ mc^2\beta_k\big)\Psi,
\end{equation}
where $\boldsymbol{\alpha}_k:=\mathbb{I}\otimes\cdots\otimes\mathbb{I}\otimes\boldsymbol{\alpha}\otimes\mathbb{I}\otimes\cdots\otimes\mathbb{I}$ denotes the Dirac matrices acting on the $k$-th particle, with $\boldsymbol{\alpha}=(\alpha^1,\alpha^2,\alpha^3)$ placed at the $k$-th position, and \(\beta_k\) defined in a similar manner. The particle velocities (again expressed in \(F\)) are given by the guiding equation:
\begin{equation}\label{eq:bohm_hiley_dirac_guiding}
\dot{\mathbf{X}}_{(k)}(t) = \frac{\Psi^\dagger \boldsymbol{\alpha}_k \Psi}{\Psi^\dagger \Psi}\big(\mathbf{X}_{(1)}(t),\dots,\mathbf{X}_{(N)}(t)\big).
\end{equation}
Thus, at the common time \(t\) in frame \(F\), the positions of all particles are simultaneously defined, and each particle's velocity depends on the simultaneous positions of all others. The frame \(F\) therefore provides the notion of ``at a given moment'' that is required to give meaning to a configuration at an instant.

\medskip

Each trajectory $\mathbf{X}_{(k)}:\mathbb{R}\to \mathbb{R}^3$ solving Eq.~\eqref{eq:bohm_hiley_dirac_guiding} corresponds to a worldline $X_{(k)}:\mathbb{R}\to M$ in Minkowski space-time, parametrized by the preferred frame's time coordinate: $X_{(k)}^\mu(t) := \big(t,\mathbf{X}_{(k)}(t)\big)$, with $\mu=0,1,2,3$ and $k=1,\cdots,N$. Together, these $N$ worldlines $X_{(k)}$---often called an \textit{N-path} $X:\mathbb{R}\to M\times\cdots \times M \equiv M^N$---describe the complete history of the system.

Crucially, if one were to write the Bohm-Hiley-Dirac dynamics (Eqs.~\eqref{eq:N_particles_Dirac_equation} and \eqref{eq:bohm_hiley_dirac_guiding}) in a different Lorentz frame, the resulting worldlines would differ, thus changing the predicted history of the system. Qualitatively, this can be seen from the fact that formulating these equations in a different frame would implement non-locality as an instantaneous influence in that frame---i.e., a synchronization of particle worldlines along a different family of hyperplanes \cite[Chap.\ 12]{bohm2006undivided}. More formally, in the preferred frame $F$, the guiding field in Eq.~\eqref{eq:bohm_hiley_dirac_guiding} has components:
\begin{equation}\label{eq:bohm_hiley_dirac_guiding_field}
    v^\mu_{(k)}\left(x_{(1)},\cdots,x_{(N)}\right)=\left(1, \frac{\Psi^\dagger\boldsymbol{\alpha}_{(k)}\Psi}{\Psi^\dagger\Psi}\right)=\frac{\bar{\Psi}\gamma_{(1)}^0\cdots\gamma_{(k-1)}^0\gamma_{(k)}^\mu\gamma_{(k+1)}^0\cdots\gamma_{(N)}^0\Psi}{\bar{\Psi}\gamma_{(1)}^0\cdots\gamma_{(k-1)}^0\gamma_{(k)}^0\gamma_{(k+1)}^0\cdots\gamma_{(N)}^0\Psi},
\end{equation}
with \(\bar{\Psi}:=\Psi^\dagger\gamma^0\otimes\cdots\otimes\gamma^0\) and where \(\gamma_{(k)}^\mu:=\mathbb{I}\otimes\cdots\otimes\mathbb{I}\otimes\gamma^\mu\otimes\mathbb{I}\otimes\cdots\otimes\mathbb{I}\) acts on the $k$-th particle's spinor index. This guiding field is \textit{not} a Lorentz-covariant 4-vector field. Under a Lorentz boost, its components would mix in a way that does not preserve the form of the equation: the temporal component would no longer equal $1$, and the spatial part would transform non-covariantly. Hence the functional form of the guiding equation changes when written in a different frame---meaning the Bohm-Hiley-Dirac model is not Lorentz invariant.

Consequently, the choice of $F$ to write down the equations of motion is not merely a mathematical convenience, but a genuine physical commitment to a specific reference frame. This apparent conflict with special relativity can be made less conspicuous by reformulating the dynamics in a coordinate-free geometric language, at the cost of introducing an additional space-time structure.

\medskip

The first step toward this coordinate-free reformulation is to re-express the preferred frame as a preferred \textit{foliation} of space-time. The non-local coupling in Eq.~\eqref{eq:bohm_hiley_dirac_guiding} links space-time points \(\big(t,\mathbf{X}_{(1)}(t)\big),\dots,\big(t,\mathbf{X}_{(N)}(t)\big)\) lying on the same hyperplane $\Sigma_t$ of constant time $t$ in $F$. Mathematically, the family of constant-time hyperplanes $\mathcal{F} = \{\Sigma_t\}_{t\in\mathbb{R}}$ defines a \textit{foliation} of Minkowski space-time:
\begin{equation}\label{eq:preferred_foliation}
    M = \bigcup_{t\in\mathbb{R}} \Sigma_t,\quad \Sigma_t := \{(t, \mathbf{x}) \in M \mid \mathbf{x} \in \mathbb{R}^3\}.
\end{equation}
Physically, each leaf $\Sigma_t\in\mathcal{F}$ supplies a notion of absolute simultaneity, allowing quantum non-locality to be implemented as a synchronization along these planes. At time $t$, the space-time configuration $X\left(\Sigma_t\right) := (X_{(1)}(t), \dots, X_{(N)}(t))$ lies on the leaf $\Sigma_t$, and the guiding equation~\eqref{eq:bohm_hiley_dirac_guiding} evolves configurations from $\Sigma_t$ to the next leaf $\Sigma_{t+\delta t}$.

From a mathematical standpoint, the time coordinate $t$ of the preferred frame $F$ is a scalar field $t: M \to \mathbb{R}$ on Minkowski space-time, and the constant-time hyperplanes of \(F\) correspond to the level sets of this \textit{time function} $t(x)$. Formally, for any $s\in\mathbb{R}$, the hyperplane of constant time $t=s$ in $F$ is:
\begin{equation}\label{eq:level_sets}
    \Sigma_{s} := t^{-1}(s) \equiv \{x \in M \mid t(x) = s\}.
\end{equation}
The leaves of the preferred foliation $\mathcal{F}$ are precisely these hyperplanes $\Sigma_s$. Furthermore, any hypersurface---and therefore any foliation---can be characterized by its future-pointing unit normal vector field $n(x)$. Since moving along $\Sigma_s$ does not change the value of $t(x)$, the gradient $\partial t(x)$ is everywhere orthogonal to $\Sigma_s$. This yields the relation between the future-pointing unit normal vector field and the gradient of the time function $t(x)$:
\begin{equation}\label{eq:future_pointing_unit_normal_vector_field}
    n(x) = \frac{\partial t(x)}{\sqrt{\partial t(x) \cdot \partial t(x)}},
\end{equation}
where the dot denotes the Minkowski scalar product: $\partial t(x) \cdot \partial t(x)=\partial_\mu t(x) \partial^\mu t(x)=|\partial t(x)|^2$ (using the Einstein summation convention). Because the foliation of the Bohm-Hiley-Dirac model consists of parallel hyperplanes, its normal vector field is constant throughout space-time: $n(x)=n$, with coordinates $n^\mu=(1,0,0,0)$ in the preferred frame $F$---i.e., \(n\) corresponds to the time axis unit vector of \(F\). This timelike 4-vector field is now treated as a genuine part of the space-time structure, thus enabling coordinate-free notations.

Using the unit normal vector field $n(x)$ to $\mathcal{F}$, the Bohm-Hiley-Dirac guiding field \eqref{eq:bohm_hiley_dirac_guiding_field} can be re-expressed in a coordinate-free way as:
\begin{equation}\label{eq:HBD_guiding_field}
    v_{(k)}\left(x_{(1)}, \cdots, x_{(N)}\right)=\frac{\bar{\Psi}\left(\gamma_{(1)}\!\cdot\! n\right)\cdots\left(\gamma_{(k-1)}\!\cdot\! n\right)\gamma_{(k)}\left(\gamma_{(k+1)}\!\cdot\! n\right)\cdots\left(\gamma_{(N)}\!\cdot\! n\right)\Psi}{\bar{\Psi}\left(\gamma_{(1)}\!\cdot\! n\right)\cdots\left(\gamma_{(k-1)}\!\cdot\! n\right)\left(\gamma_{(k)}\!\cdot\! n\right)\left(\gamma_{(k+1)}\!\cdot\! n\right)\cdots\left(\gamma_{(N)}\!\cdot\! n\right)\Psi},
\end{equation}
and the worldlines $X_{(k)}:\mathbb{R}\to M$ (for $k=1,\cdots,N$) satisfy the guiding equation:
\begin{equation}\label{eq:HBD_guiding_equation}
    \dot{X}_{(k)}\left(\Sigma_s\right)=v_{(k)}\left(X_{(1)}\left(\Sigma_s\right), \cdots, X_{(N)}\left(\Sigma_s\right)\right),
\end{equation}
where $\gamma\cdot n=\gamma^\mu n_\mu$ reduces to $\gamma^0$ in $F$ coordinates. The worldlines are now parametrized by the same parameter $s\in\mathbb{R}$ that labels the preferred hypersurfaces \eqref{eq:level_sets}, i.e., $X_{(k)}\left(\Sigma_s\right)\equiv X_{(k)}(s)\in\Sigma_s$, so that $\dot{X}_{(k)}\left(\Sigma_s\right)=\frac{d X_{(k)}}{ds}\left(\Sigma_s\right)$.

In essence, while we have achieved a coordinate-free notation, this dynamics remains physically equivalent to the original Bohm-Hiley-Dirac dynamics (Eqs.~\eqref{eq:N_particles_Dirac_equation} and \eqref{eq:bohm_hiley_dirac_guiding}) when written in the coordinates of the preferred frame $F$. Just as the Bohm-Hiley-Dirac model was tied to the preferred frame $F$, The HBD dynamics (Eqs.~\eqref{eq:HBD_guiding_field} and \eqref{eq:HBD_guiding_equation}) is fundamentally tied to the preferred foliation $\mathcal{F}$, changing it would alter the predicted history of the system (i.e. the N-path \(X:\mathbb{R}\to M^N\)).\footnote{However, a significant difference between the Bohm-Hiley-Dirac model and HBD models is that the latter allows for potentially curved foliations, i.e., families of curved preferred hypersurfaces, unlike the original model which was restricted to flat hyperplanes.} The preferred foliation is therefore an ontological postulate, not a mere mathematical convenience. Since the guiding law depends on the preferred frame $F$ only through its timelike
direction $n$---and not on the origin or spatial axes that a frame additionally carries---both formulations commit to the same additional geometrical structure $\mathcal{F}$ \eqref{eq:preferred_foliation}, a structure with no counterpart in the standard special-relativistic description of space-time.

Crucially for our experimental proposal, while special relativity assigns no objective temporal order to spacelike-separated events \(A\) and \(B\), the preferred foliation does. It supplies an unambiguous order determined by the sign of \(t(A)-t(B)\), where \(t:M\to\mathbb{R}\) is the time function whose level sets define the leaves of the foliation \eqref{eq:level_sets}.
\medskip

In a nutshell, the complete specification of an HBD model requires three elements: the (multi‑time) wave function \(\Psi\), the particle worldlines \(X\), and the foliation \(\mathcal{F}\). This triple \((\Psi, X, \mathcal{F})\) constitutes the ontology of the theory \cite{Tumulka_2007}. Although special relativity treats all foliations as mathematically equivalent, an HBD model elevates a particular one to the status of a physically real structure. As a geometrical structure, the preferred foliation $\mathcal{F}$ is not part of the special-relativistic description of space-time, which only postulates a light-cone structure attached to every event \cite{maudlin1996space,maudlin2011quantum,maudlin2012philosophy}.

Despite its centrality to the HBD ontology, the preferred foliation is commonly asserted to be empirically undetectable \cite[Ch. 12]{bohm2006undivided,duerr_hypersurface_1999,tumulka2018bohmian,tumulka2022foundations}. This paper challenges that assertion by presenting an experimental proposal designed to detect this supposedly invisible structure $\mathcal{F}$. We note that our approach resonates with a challenge once articulated by one of the developers of the HBD model. Roderich Tumulka, a central architect of the modern HBD framework who has since become a leading proponent of the view that $\mathcal{F}$ is empirically undetectable, initially expressed a more open stance. In the concluding passage of his influential 2007 paper \textit{The "Unromantic Pictures" of Quantum Theory}, he posed the following question:

\begin{quote}
    ``Can one observe the time foliation? That is, can one determine experimentally which
    $3-$surfaces the time leaves are? [...] It seems that a serious observation of $\mathcal{F}$ should constitute a violation of relativistic covariance. This suggests that $\mathcal{F}$ be unobservable. But presumably the model I have presented entails that there are quantum experiments observing the time foliation, as I see no reason in the model why it should be unobservable. It would be interesting to think up an experiment for which the model predicts that its result reveals the time foliation.'' \cite{Tumulka_2007}
\end{quote}

This paper takes up that very challenge. We propose a concrete experimental protocol whose outcome, if the underlying assumptions hold, would indeed reveal the preferred foliation $\mathcal{F}$---what Tumulka calls the ``time foliation.''

\medskip

Note that for the purposes of the present work, we do not need the full machinery of HBD models. Our calculations will be performed in the Bohm-Pauli theory for non-relativistic spin-\(\frac12\) particles, which is the non-relativistic limit of the Bohm-Hiley-Dirac model \cite[Sec.\ 10.4]{bohm2006undivided}. In this limit, the preferred foliation reduces to the absolute time of the frame in which the Pauli equation is written.

Given this simplification, a potential concern is that DD's predictions---and consequently our predictions---are derived from the non‑relativistic Pauli equation and might therefore be artifacts of a non‑relativistic treatment. However, Siddhant Das has subsequently shown that the key qualitative features---including spin‑dependent trajectories and the characteristic backflow---persist in a fully relativistic treatment using the Dirac equation \cite{das2021relativisticelectronwavepackets}. Thus, while we present our explicit calculations in the non‑relativistic framework for clarity, the underlying mechanism and the resulting foliation dependence are expected to survive in a relativistic extension, as required for consistency with the Hypersurface Bohm–Dirac model. We return to this point in Appendix~\ref{app:relativistic}, where we outline how the non-relativistic analysis presented here could in principle be extended to a fully relativistic treatment based on the HBD model. 

\section{The Experimental Setup}\label{sec:Setup}

\begin{figure}[h]
    \centering
    \includegraphics[width=0.5\textwidth]{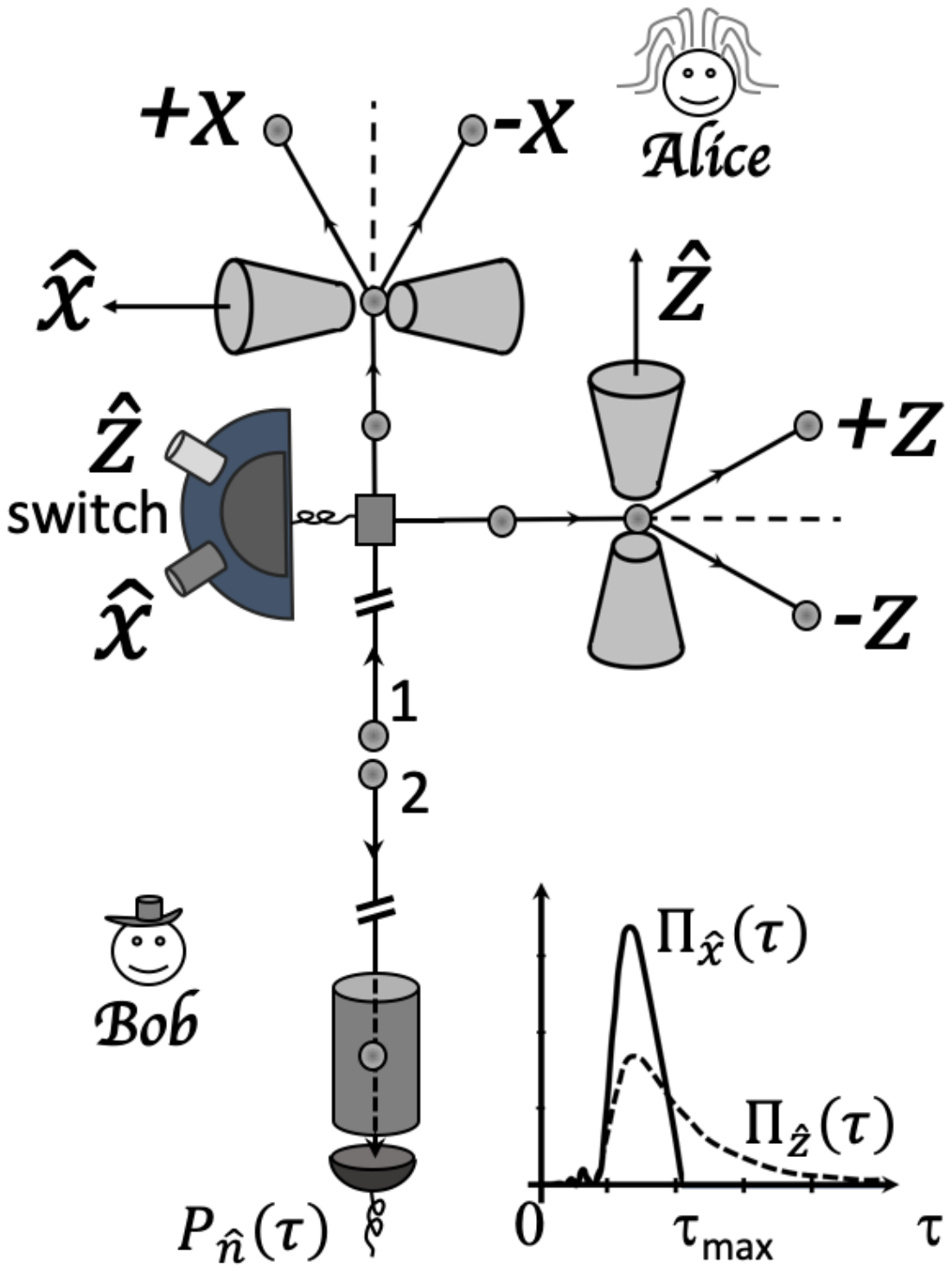}
    \caption{Schematic of the EPRB experiment with arrival-time measurement. The source produces entangled singlet-state pairs. In each run, a large ensemble ($\sim 10^5$ pairs) is emitted: for each pair, particle~1 goes to Alice, particle~2 to Bob. Alice uses a control panel (buttons 0 and 1) to select the orientation ($\hat{z}$ or $\hat{x}$) of her Stern-Gerlach magnet and perform a sequence of \(\sim 10^5\) spin measurements along that direction. Bob performs a sequence of \(\sim 10^5\) arrival-time measurements using a cylindrical waveguide along $\hat{z}$. The entanglement implies that Alice's choice projects the spin of Bob's particles, altering their Bohmian velocity field and potentially the resulting arrival-time statistics. The curves at the bottom illustrate the two distinct arrival times distributions Bob could observe. In particular beyond the time $\tau_{max}$ the `exotic' distribution $\Pi_{\hat{x}}(\tau)$ vanishes (compare with Fig.~\ref{fig:spin_dependent_arrival_times_distributions}). The entire arrangement ensures spacelike separation between Alice's and Bob's measurement events.}
    \label{fig:experimental_setup}
\end{figure}

The core of our analysis is an EPRB-type experiment augmented with arrival time measurements on one side of the setup. The same experimental arrangement, illustrated in Figure \ref{fig:experimental_setup}, enables both the superluminal signaling protocol and the foliation detection procedure we will describe. Moreover, as will become clear in Section~\ref{sec:Foliation_Detection}, the entire apparatus can be rotated and tilted about the source to obtain different orientations, a feature essential for mapping the preferred foliation.

\medskip

In this setup (see Fig.~\ref{fig:experimental_setup}) a source emits, at each experimental run, a large ensemble of spin-$\frac{1}{2}$ particle pairs: $N\approx10^5$ pairs prepared in the entangled singlet state given by Eq.~\eqref{eq:wave_function_factorization}. For each pair, particle~1, with spatial coordinate $\mathbf{x}_1$, is directed toward observer Alice, while particle~2, with coordinate $\mathbf{x}_2$, is directed toward observer Bob. Crucially, Alice's and Bob's laboratories lie sufficiently far from each others in order to ensure that, within any single run, the events corresponding to Alice's measurement on her ensemble and Bob's measurement on his ensemble are spacelike separated, i.e. cannot be connected by a light signal.

\medskip

At her station, Alice can perform a projective spin ``measurement''\footnote{We use scare quotes because, in BM, a spin ``measurement'' is a physical interaction (typically with a Stern-Gerlach magnet) that does not reveal a preexisting spin value. Spin is a contextual variable, and the particle's deflection is determined by its initial position within the wave packet. See \cite{norsen2014pilot} for a detailed explanation.} along one of two possible directions: $\hat{n}=\hat{x}$ or $\hat{n}=\hat{z}$. In each run of the experiment, before the $\sim 10^5$ particles arrive to her lab, Alice presses one of two buttons on a control panel:
\begin{itemize}
    \item \textbf{Button 0:} A beam splitter directs all incoming particles toward a $\hat{z}$-oriented Stern-Gerlach magnet.
    \item \textbf{Button 1:} The beam splitter directs all incoming particles toward a $\hat{x}$-oriented Stern-Gerlach magnet.
\end{itemize}

Thus, by pressing a button, Alice determines which Stern-Gerlach orientation (either $\hat{z}$ or $\hat{x}$) will be used for the entire ensemble in that run, i.e. performing a rapid series of $\sim 10^5$ identical spin measurements during that run. This choice effectively projects the spin of each particle onto the corresponding axis.\footnote{Since the two Stern-Gerlach magnets occupy different spatial locations (see Fig.~\ref{fig:experimental_setup}), the wave function $f_{s\hat{n}}(\mathbf{x}_1,t)$ implicitly depends on Alice's choice. A more rigorous notation $f_{s\hat{n}}^{(0,1)}(\mathbf{x}_1,t)$ would distinguish between $f_{s\hat{n}}^{(0)}(\mathbf{x}_1,t)$ (for particles directed to the $\hat{z}$-oriented magnet) and $f_{s\hat{n}}^{(1)}(\mathbf{x}_1,t)$ (for particles directed to the $\hat{x}$-oriented magnet). For readability, and because it doesn't affect the validity of our reasoning, we suppress this superscript notation in the main text.} This choice, due to the singlet entanglement \eqref{eq:wave_function_factorization}, also projects the spin state of Bob's distant particles onto the same axis.

On the other side of the setup, Bob performs a series of $\sim 10^5$ arrival-time
measurements on his ensemble of particles using a cylindrical waveguide oriented along the $\hat{z}$-axis, following DD's proposal \cite{das2019arrival,Das_2019}.\footnote{Throughout, we write $\mathbf{x}_2=(x,y,z)$ for particle~2's configuration in the Cartesian frame adapted to the waveguide, whose axis defines $\hat{z}$ and whose end face lies at $z=0$; where the cylindrical symmetry is useful we use $(\rho,\phi,z)$ with $\rho=\sqrt{x^2+y^2}$.} The waveguide is equipped with a sensitive screen that records the first arrival-time of each particle. Crucially, by using approximately $10^5$ particle pairs per experimental run, Bob can measure the complete statistical distribution of arrival times, rather than just individual arrival events.\footnote{This ensemble size is motivated by the numerical simulations of DD, who used $8\times10^5$ Bohmian trajectories to generate their arrival time histograms (see Figure \ref{fig:spin_dependent_arrival_times_distributions}) \cite{Das_2019}. Such a large ensemble ensures that Bob can reliably distinguish between the two predicted arrival time distributions. For $N$ independent events, the statistical uncertainty typically scales as $1/\sqrt{N}$, yielding approximately $0.3\%$ relative error for $N=10^5$.} After each run, Bob's screen displays a histogram (or fitted curve) representing the measured arrival-time statistics.

\bigskip

Following the trajectory-based analysis of DND \cite{Das_2019} and extending it to a two-particle entangled setting, we will show that the arrival-time distribution Bob observes at the end of each run depends jointly on Alice's choice of spin direction and on the temporal order of the two measurements relative to the preferred foliation. As we shall see in Section~\ref{sec:summary_and_analysis_of_the_results}, this joint dependence becomes an unambiguous witness of the temporal order once Alice and Bob agree that she will use the transverse direction $\hat{n}=\hat{x}$ in every run. Before deriving these predictions, we first introduce the formalism on which our derivation rests.

\section{Bohmian Mechanics for Two Entangled Spin-1/2 Particles}\label{sec:A_General_Bohmian_Formalism}

Our experimental proposal involves a system of two spin-$\frac{1}{2}$ particles. The quantum state of such a system is described by a 4-component spinor wave function residing in the tensor product space:
\[
\Psi(\mathbf{x}_1, \mathbf{x}_2, t) \in \mathcal{H}_1 \otimes \mathcal{H}_2 \equiv \left(L^2(\mathbb{R}^3) \otimes \mathbb{C}^2\right) \otimes \left(L^2(\mathbb{R}^3) \otimes \mathbb{C}^2\right) \simeq L^2(\mathbb{R}^6) \otimes \mathbb{C}^4.
\]

To describe this state efficiently, we employ a formalism based on spin projection along an arbitrary spatial direction $\hat{n} \in \mathbb{R}^3$. Let $\chi_{s\hat{n}}$ denote a one-particle spinor quantized along the $\hat{n}$-axis, where $s = \pm 1$ corresponds to spin-up ($\chi_{+\hat{n}}$) and spin-down ($\chi_{-\hat{n}}$) states. The set $\{\chi_{+\hat{n}}, \chi_{-\hat{n}}\}$ forms an orthonormal basis of $\mathbb{C}^2$, which induces the product basis for the two-particle spin space $\mathbb{C}^4$: \(\{\chi_{s_1\hat{n}} \otimes \chi_{s_2\hat{n}}\}_{s_1,s_2=\pm 1}\). In this basis, the two-particle quantum state can be expanded as:
\begin{equation}\label{eq:general_expansion}
    \Psi(\mathbf{x}_1, \mathbf{x}_2, t) = \sum_{s_1,s_2=\pm 1} \psi_{s_1\hat{n},\ s_2\hat{n}}(\mathbf{x}_1, \mathbf{x}_2, t) \left[ \chi_{s_1\hat{n}} \otimes \chi_{s_2\hat{n}} \right],
\end{equation}
where the coefficients $\psi_{s_1\hat{n}, s_2\hat{n}}(\mathbf{x}_1, \mathbf{x}_2, t)$ are spatial wave functions entangled with the corresponding spin states. According to the Born rule, $|\psi_{s_1\hat{n}, s_2\hat{n}}(\mathbf{x}_1, \mathbf{x}_2, t)|^2$ is the joint density for the configuration $(\mathbf{x}_1,\mathbf{x}_2)$ together with the outcomes $(s_1,s_2)$ of spin measurements performed along $\hat{n}$, at time $t$.\footnote{ We stress that, in BM, the indices $s_1$ and $s_2$ label components of $\Psi$ in the $\hat{n}$-basis; they do not denote spin values carried by the particles, whose only beables are their positions. Such a component acquires a connection to a measurement outcome only once a Stern-Gerlach interaction along $\hat{n}$ has spatially separated the corresponding wave packets: the outcome is then fixed by which packet contains the actual configuration, and hence---given $\Psi$---by the particle's initial position rather than by any pre-existing spin value \cite{norsen2014pilot}. This is the mechanism we appeal to in Section~\ref{sec:Alice_measures_first}. This does not make spin dynamically inert: the direction $\hat{n}$ enters the guiding field directly through the spin flux in Eq.~\eqref{eq:Jk_general}, and it is this dependence---on the polarization, not on the eigenvalue---that generates the effects studied below.}

\bigskip

This $2-$particle system obeys the Bohm-Pauli model, which is the non-relativistic limit of the Bohm-Hiley-Dirac model (Eq. \eqref{eq:N_particles_Dirac_equation} and Eq. \eqref{eq:bohm_hiley_dirac_guiding}) \cite[Sec.\ 10.4]{bohm2006undivided}. The dynamics of the two-particle wave function $\Psi(\mathbf{x}_1, \mathbf{x}_2, t)$ is governed by the Pauli equation \cite[Chap.\ 10]{holland1995quantum}:
\begin{equation}\label{eq:pauli_eq}
    i\hbar \frac{\partial \Psi}{\partial t} = \left[ \sum_{k=1}^2 \frac{1}{2m} \left( -i\hbar \nabla_k - q\mathbf{A}(\mathbf{x}_k,t) \right)^2 - \sum_{k=1}^2 \frac{q\hbar}{2m} \boldsymbol{\sigma}_k \cdot \mathbf{B}(\mathbf{x}_k,t) + V(\mathbf{x}_1, \mathbf{x}_2, t) \right] \Psi,
\end{equation}
where $\boldsymbol{\sigma}_1 := \boldsymbol{\sigma} \otimes \mathbb{I}$ and $\boldsymbol{\sigma}_2 := \mathbb{I} \otimes \boldsymbol{\sigma}$ are the spin operators for particles 1 and 2 respectively, with $\boldsymbol{\sigma} = (\sigma_x, \sigma_y, \sigma_z)$ the vector of Pauli matrices, and $\nabla_k$ denotes the gradient with respect to $\mathbf{x}_k$.

From this equation, one derives a continuity equation of the same form as in the single-particle case: \(\partial_t \rho + \nabla_1 \cdot \mathbf{J}_{\text{Pauli}}^{(1)} + \nabla_2 \cdot \mathbf{J}_{\text{Pauli}}^{(2)} = 0\), with the probability density $\rho(\mathbf{x}_1,\mathbf{x}_2,t) = \Psi^\dagger\Psi$ and the associated probability current:\footnote{We have set $\mathbf{A}=0$ in Eq.~\eqref{eq:Jk_general}; the gauge-invariant convective flux implied by Eq.~\eqref{eq:pauli_eq} is $\frac{1}{m}\operatorname{Re}
\left[\Psi^\dagger\left(-i\hbar\nabla_k-q\mathbf{A}(\mathbf{x}_k,t)\right)\Psi\right]$, which reduces to the expression above when $\mathbf{A}$ vanishes. This is legitimate throughout our calculations: every trajectory we integrate lies in a region where
$\mathbf{B}=0$---Bob's waveguide, and the field-free segments on Alice's side---so that $\mathbf{A}$ is pure gauge there and may be set to zero. The Stern-Gerlach interaction itself is never integrated explicitly; it enters our analysis only through the spatial separation of the two spin packets, Eq.~\eqref{eq:spatial_separation_WP}.}
\begin{equation}\label{eq:Jk_general}
    \mathbf{J}_{\text{Pauli}}^{(k)}(\mathbf{x}_1, \mathbf{x}_2, t) = \underbrace{\frac{\hbar}{m} \operatorname{Im}\left[ \Psi^\dagger \nabla_k \Psi \right]}_{\text{\textit{convective flux}}} + \underbrace{\frac{\hbar}{2m} \nabla_k \times \left( \Psi^\dagger \boldsymbol{\sigma}_k \Psi \right)}_{\text{\textit{spin flux}}} \quad k=1,2,
\end{equation}
which is the \(2-\)particle extension of \eqref{eq:pauli_current_general}.

The Bohmian velocity for particle $k$ is defined as the ratio of its probability current to the probability density:
\begin{equation}\label{eq:bohmian_velocity_def}
    \boldsymbol{v}_{\text{dBB}}^{(k)}(\mathbf{x}_1,\mathbf{x}_2, t) := \frac{\mathbf{J}^{(k)}_{\text{Pauli}}(\mathbf{x}_1,\mathbf{x}_2, t)}{|\Psi(\mathbf{x}_1,\mathbf{x}_2, t)|^2}, \quad k=1,2.
\end{equation}
At time $t$, the velocity of the $k-th$ particle obeys the guiding equation:
\begin{equation}\label{eq:Pauli_guiding}
    \dot{\mathbf{X}}_{k}(t)=\boldsymbol{v}_{\text{dBB}}^{(k)}\left(\mathbf{X}_1(t),\mathbf{X}_2(t), t\right), \quad k=1,2.
\end{equation}
Although the Bohm-Pauli model is formulated on Galilean space-time, the time coordinate \(t\) of the frame in which Eqs.~\eqref{eq:pauli_eq} and \eqref{eq:Pauli_guiding} are written plays exactly the same dynamical role as the preferred foliation does in a relativistic HBD model: it provides the notion of simultaneity required to define a joint configuration at an instant and to evolve it deterministically.

This is not merely an analogy. In the regime where the particle velocities \(\dot{\mathbf{X}}_{(k)}(t)\) are small compared to \(c\), the Bohm-Hiley-Dirac dynamics \eqref{eq:N_particles_Dirac_equation}--\eqref{eq:bohm_hiley_dirac_guiding}, written in the coordinates of the preferred frame \(F\), reduces to the Bohm-Pauli dynamics \eqref{eq:pauli_eq}--\eqref{eq:Pauli_guiding} written in those same coordinates \cite[Sec.\ 10.5]{bohm2006undivided}. Under this reduction the Galilean time coordinate \(t\) is not merely analogous to the foliation parameter \(s\) of Eq.~\eqref{eq:HBD_guiding_equation}: it \textit{is} that parameter, the constant-\(t\) hyperplanes are the leaves \(\Sigma_s\) of the flat foliation \(\mathcal{F}\), and each Bohm-Pauli trajectory maps to a worldline \(X_{(k)}:\mathbb{R}\to M\) with coordinates \(X^\mu_{(k)}(t):=(t,\mathbf{X}_{(k)}(t))\) in \(F\). What is approximated here is the dynamics---the Pauli guiding field is the leading term in the non-relativistic expansion of the Dirac guiding field---and not the space-time structure: a flat foliation of Minkowski space-time and the absolute time of a Galilean frame supply the same notion of simultaneity.

Consequently, the temporal ordering of spacelike-separated events relative to \(\mathcal{F}\)---an ordering that special relativity alone leaves undefined---plays a direct dynamical role in determining the trajectories and hence the arrival-time statistics. Our calculations are carried out in the Bohm-Pauli dynamics on Galilean
space-time, but by the identification just established they track precisely this ordering; it is what the protocol of Section~\ref{sec:Foliation_Detection} aims to detect.

\medskip

Equipped with this general Bohmian formalism for two spin-$\frac{1}{2}$ particles, we now specialize to the entangled \textit{singlet state} that constitutes the quantum resource for our proposed experiment.

\section{Applying the Formalism: The Entangled Singlet State}\label{sec:Bohmian_dynamics_of_entangled_singlet_state}

We begin by specifying the quantum state at the heart of our analysis. The spin part is the maximally entangled \textit{singlet state}: 
\[
\chi_{\text{singlet}} =\frac{1}{\sqrt{2}}\sum_{s=\pm1}s\left[\chi_{s\hat{n}}\otimes\chi_{-s\hat{n}}\right],
\]
expressed in the product basis \(\{\chi_{s_1\hat{n}} \otimes \chi_{s_2\hat{n}}\}_{s_1,s_2=\pm 1}\) for the $2-$particle spin space $\mathbb{C}^4$. In our setup, we consider situations where the spin and spatial degrees of freedom are also entangled. In a usual EPRB setup, the complete two-particle quantum state takes the form:
\begin{equation}\label{eq:singlet_state}
    \Psi_{\text{singlet}}(\mathbf{x}_1, \mathbf{x}_2, t) = \frac{1}{\sqrt{2}} \sum_{s=\pm 1} sf_{s\hat{n}}(\mathbf{x}_1,t) \, g_{-s\hat{n}}(\mathbf{x}_2,t) \left[ \chi_{s\hat{n}} \otimes \chi_{-s\hat{n}} \right],
\end{equation}
where $f_{s\hat{n}}(\mathbf{x}_1,t)$ and $g_{-s\hat{n}}(\mathbf{x}_2,t)$ are spatial wave functions for particles 1 and 2, respectively.\footnote{This expression is a special case of the general expansion \eqref{eq:general_expansion} with $s_1 = -s_2$ and factorized coefficients: $\psi_{s\hat{n},\,-s\hat{n}}(\mathbf{x}_1,\mathbf{x}_2,t)=sf_{s\hat{n}}(\mathbf{x}_1,t) \, g_{-s\hat{n}}(\mathbf{x}_2,t)$.} The state \eqref{eq:singlet_state} explicitly displays two distinct spin branches associated with potentially different spatial wave functions: $f_{+\hat{n}}(\mathbf{x}_1,t)g_{-\hat{n}} (\mathbf{x}_2,t)$ (for the spin branch $s=+1$) and $-f_{-\hat{n}}(\mathbf{x}_1,t)g_{+\hat{n}}(\mathbf{x}_2,t)$ (for the spin branch $s=-1$).

However, in the adapted EPRB setup considered in this paper, a crucial simplification arises from the experimental design: Bob's waveguide contains no magnetic field, so $\mathbf{B}(\mathbf{x}_2)=0$ in the Pauli equation~\eqref{eq:pauli_eq}. Consequently, particle~2's spin does not couple to its spatial wave function, and the wave packets for the two spin branches are identical: \(g_{+\hat{n}}(\mathbf{x}_2,t)=g_{-\hat{n}}(\mathbf{x}_2,t)=g_0(\mathbf{x}_2,t)\). The quantum state of the system becomes:
\begin{equation}\label{eq:wave_function_factorization}
    \Psi(\mathbf{x}_1, \mathbf{x}_2, t) = \frac{ g_{0}(\mathbf{x}_2,t)}{\sqrt{2}} \sum_{s=\pm 1} s\;f_{s\hat{n}}(\mathbf{x}_1,t) \,\left[ \chi_{s\hat{n}} \otimes \chi_{-s\hat{n}} \right],
\end{equation}
instead of \eqref{eq:singlet_state}. For later use, we express the spatial wave functions in polar form:
\begin{align}
    f_{s\hat{n}}(\mathbf{x}_1,t) &= |f_{s\hat{n}}(\mathbf{x}_1,t)| \, e^{i S^{(1)}_{s\hat{n}}(\mathbf{x}_1,t)}, \label{eq:polar_f} \\
    g_{0}(\mathbf{x}_2,t)&=|g_0(\mathbf{x}_2,t)|e^{iS_0^{(2)}(\mathbf{x}_2,t)}. \label{eq:polar_g}
\end{align}

\medskip

Our next task is to derive the Bohmian velocity fields for both particles by inserting the quantum state \eqref{eq:wave_function_factorization} into the defining equations \eqref{eq:bohmian_velocity_def} and \eqref{eq:Jk_general}. However, because the derivations of the guiding fields involve lengthy calculations, that may distract the reader from the main argument of the paper, we removed the explicit derivations in Appendix\, \ref{appendix}.

After a straightforward but lengthy calculation (see Appendix\, \ref{appendix} for details), one obtains the Bohmian velocity for particle 2:

\begin{equation}\label{eq:second_guiding_field}
    \mathbf{v}_{\text{dBB}}^{(2)}(\mathbf{x}_1, \mathbf{x}_2, t)= \displaystyle\sum_{s=\pm1} w_s(\mathbf{x}_1, t) \; \mathbf{v}^{(2)}_{-s\hat{n}}(\mathbf{x}_2, t),
\end{equation}

where two key quantities have been introduced.

First, the \textit{spin-conditional velocity field} for particle~2:

\begin{equation}\label{eq:spin_conditional_velocity}
    \mathbf{v}^{(2)}_{-s\hat{n}}(\mathbf{x}_2, t) := 
    \frac{\hbar}{m} \left( \nabla_2 S^{(2)}_{0}(\mathbf{x}_2,t) 
    - s \frac{\nabla_2 |g_{0}(\mathbf{x}_2,t)|}{|g_{0}(\mathbf{x}_2,t)|} \times \hat{n}\right), \quad s = \pm 1.
\end{equation}

This local velocity field would govern particle~2's motion if the two-particle system were in the product state $f_{s\hat{n}}(\mathbf{x}_1,t)\,g_{0}(\mathbf{x}_2,t)\left[\chi_{s\hat{n}} \otimes \chi_{-s\hat{n}} \right]$---that is, in the single branch $s$ of
the superposition \eqref{eq:wave_function_factorization}, with no entanglement between the two particles.\footnote{This field depends only on $\mathbf{x}_2$ because, in such a product state, the
two particles evolve independently under local dynamics.} There are two such conditional guiding fields: $\mathbf{v}^{(2)}_{-\hat{n}}(\mathbf{x}_2, t)$ (for the spin branch $s=+1$) and $\mathbf{v}^{(2)}_{+\hat{n}}(\mathbf{x}_2, t)$ (for the spin branch $s=-1$).

Second, the \textit{configuration-dependent weights} are defined as:\footnote{While we called them \textit{configuration-dependent weights}, due to the absence of magnetic field on Bob's side, these weights depends only on the particle 1 position at time $t$. Interestingly, in a standard EPRB setup these weights would depend on the joint configuration \((\mathbf{x}_1,\mathbf{x}_2)\) at time $t$.}

\begin{equation}\label{eq:simplified_weights}
    w_{s}(\mathbf{x}_1,t):=\frac{ |f_{s\hat{n}}(\mathbf{x}_1,t)|^2 }{ |f_{+\hat{n}}(\mathbf{x}_1,t)|^2+|f_{-\hat{n}}(\mathbf{x}_1,t)|^2 }, \quad s = \pm 1.
\end{equation}

These weights are non-negative and satisfy $w_{+1}(\mathbf{x}_1, t)+w_{-1}(\mathbf{x}_1, t)=1$ for every position $\mathbf{x}_1$ in the support of $\Psi$, forming a probability distribution over the two spin branches.

\medskip

The guiding field \eqref{eq:second_guiding_field} admits a transparent interpretation: the instantaneous velocity of particle~2 is a convex combination of the two spin-conditional velocity fields \eqref{eq:spin_conditional_velocity}. Although each spin-conditional field depends only on the local position \(\mathbf{x}_2\) of particle~2, the guiding field \eqref{eq:second_guiding_field} depends on \textit{both} \(\mathbf{x}_1\) and \(\mathbf{x}_2\) through the configuration-dependent weights \eqref{eq:simplified_weights}. The distant particle~1 thus exerts a direct, instantaneous (with respect to \(\mathcal{F}\)) influence on particle~2: its position determines how the two local velocity fields are blended to guide its entangled partner, thus implementing quantum non-locality.

\bigskip

The velocity field guiding the first particle has a somewhat similar form:
\begin{equation}\label{eq:expanded_first_current}
    \mathbf{v}_{\text{dBB}}^{(1)}(\mathbf{x}_1, t) = \displaystyle\sum_{s=\pm1} w_s(\mathbf{x}_1, t) \; \mathbf{v}^{(1)}_{s\hat{n}}(\mathbf{x}_1, t).
\end{equation}
As for particle~2, this velocity is a convex combination of two spin‑conditional velocity fields, with the same weights $w_s(\mathbf{x}_1, t)$ defined in Eq.~\eqref{eq:simplified_weights}, but with different spin-conditional velocities. The spin‑conditional velocities for particle~1 are given by:
\begin{equation}\label{eq:First_conditional_velocity_derived}
    \mathbf{v}^{(1)}_{s\hat{n}}(\mathbf{x}_1, t) = 
    \frac{\hbar}{m}\left( \nabla_1 S^{(1)}_{s\hat{n}}(\mathbf{x}_1,t) 
    + s \frac{\nabla_1 |f_{s\hat{n}}(\mathbf{x}_1,t)|}{|f_{s\hat{n}}(\mathbf{x}_1,t)|} \times \hat{n}\right), \qquad s = \pm 1.
\end{equation}
Note the sign difference in the spin term compared to the corresponding expression for particle~2, Eq.~\eqref{eq:spin_conditional_velocity}: a \textit{plus} sign appears in Eq.~\eqref{eq:First_conditional_velocity_derived} whereas a \textit{minus} sign appears in Eq.~\eqref{eq:spin_conditional_velocity}. This sign difference directly reflects the perfect anti-correlation encoded in the singlet state \eqref{eq:wave_function_factorization}: each particle's spin flux is built from its own local polarization, and in the branch $s$ these are opposite, $\langle\boldsymbol{\sigma}\rangle_{\chi_{s\hat{n}}}=s\hat{n}$ for
particle~1 and $\langle\boldsymbol{\sigma}\rangle_{\chi_{-s\hat{n}}}=-s\hat{n}$ for particle~2 (see Eq.~\eqref{eq:spin-density} in Appendix~\ref{appendix}).

More importantly---due to the absence of magnetic field on Bob's side of the setup---the two guiding fields exhibit a dynamical asymmetry. Particle~1 is guided by a local velocity field \eqref{eq:expanded_first_current}, whereas particle~2 is guided by a non-local one \eqref{eq:second_guiding_field}. This asymmetry is a feature of the present arrangement rather than of BM itself: it arises because $\mathbf{B}(\mathbf{x}_2)=0$ makes the two spatial packets on Bob's side coincide, so that the weights lose their $\mathbf{x}_2$-dependence. It is nonetheless essential to what follows, since it is the reason the influence we exploit runs from Alice to Bob and not in the reverse direction.

\section{Definition of the Temporal Order}\label{sec:definition_of_the_temporal_order}

To rigorously analyze how Bob's arrival-time distribution depends on the relative timing of Alice and Bob measurements, we must precisely define the notion of temporal order that appears in our subsequent derivations.

Although the arrival-time distribution observed by Bob results from an experimental run that involves an ensemble of \(\sim 10^5\) pairs (as described in Section~\ref{sec:Setup}), the dynamics of each pair is independent and governed by the same laws. By analyzing a single representative pair, we can determine how the trajectory of particle 2---and consequently its contribution to the statistical distribution---depends on the temporal order of the two measurements performed on that pair. The observed arrival-time distribution for the full ensemble then emerges from averaging over the initial conditions of many such pairs, as per Eq.~\eqref{eq:arrival-time_distribution}.

To describe the relative timing of the two measurements, we must first identify specific events (i.e., space-time points) that define the ``measurement'' for each particle within a single entangled pair. We therefore introduce:

\begin{itemize}
    \item \textbf{Event \(A\)}: Particle~1's exit from the inhomogeneous magnetic field inside Alice's Stern-Gerlach magnet. This event marks the completion of the spin-measurement interaction; after this moment, the two spin components of particle~1's wave function have been spatially separated (Eq.~\eqref{eq:spatial_separation_WP}), and the non-local influence on particle~2 has been fully determined.
    \item \textbf{Event \(B\)}: The release of particle~2 from its initial confinement near \(z=0\) inside Bob's waveguide. This event initiates the arrival-time measurement; it corresponds to the moment \(t=0\) in the analysis of Das and D\"urr \cite{Das_2019} (see Section~\ref{sec:from_bohmian_trajectories_to_arrival_time_statistics}). Note that Bob's measurement is not itself instantaneous: particle~2 propagates through the waveguide for a finite duration before reaching the detection surface, and its trajectory depends on the guiding field throughout that interval. Representing Bob's measurement by the single event \(B\) is therefore an idealization, whose validity we examine in Section~\ref{sec:Bob_measures_first}.
\end{itemize}

The choice of event \(A\) is motivated by the fact that the non‑local influence of Alice's spin measurement on particle~2's trajectory becomes operative only \textit{after} the two spin components of particle~1's wave function have separated spatially. A meaningful definition of temporal order in our setup must therefore distinguish whether Bob's release of particle~2 (event \(B\)) occurs \textit{before} this separation or \textit{after} it. Event \(B\) is chosen because it marks the beginning of the dynamical evolution that determines particle~2's arrival time at the detection screen.

The preferred foliation's time function \(t:M\to\mathbb{R}\) assigns coordinates \(t_A := t(A)\) and \(t_B := t(B)\) to these events. Two fundamentally distinct cases arise for each entangled pair, according to their temporal order relative to the preferred foliation:
\begin{enumerate}[label=(\roman*)]
    \item \textbf{Spin measurement before arrival-time measurement:} According to the preferred foliation, Alice's measurement event occurs before Bob's, i.e.\ \(t_A < t_B\).
    \item \textbf{Arrival-time measurement before spin measurement:} According to the preferred foliation, Bob's measurement event occurs before Alice's, i.e.\ \(t_B < t_A\).
\end{enumerate}
Strictly speaking, a third possibility exists: Alice's measurement may occur \textit{during} particle~2's transit, i.e.\ \(t_B < t_A < t_B+\tau\). As we shall see, this intermediate regime is unavoidable near the switch point of our detection protocol (Section~\ref{sec:Foliation_Detection}), and we return to it there.

A clarification is in order. As explained in Section~\ref{sec:A_General_Bohmian_Formalism}, the Galilean time coordinate of the frame in which the Bohm-Pauli dynamics is written is identified, in the non-relativistic limit, with the parameter labeling the leaves of the preferred foliation. The times \(t_A\) and \(t_B\) appearing in our derivations are accordingly the absolute time coordinates of events \(A\) and \(B\) in a Galilean frame---so that the sign of \(t_A-t_B\) is the same in every such frame---and, under that identification, they give the temporal order of \(A\) and \(B\) relative to \(\mathcal{F}\). When we construct the foliation detection protocol in Section~\ref{sec:Foliation_Detection}, we shall therefore treat \(A\) and \(B\) as points of Minkowski space-time and read the sign of \(t_A-t_B\) as the order fixed by the preferred foliation of the HBD model.

We now derive, for a single entangled pair, the explicit velocity fields governing particle~2's motion in each of the two temporal scenarios.

\section{If Alice Measures First: How Her Choice Shapes Bob's Statistics}\label{sec:Alice_measures_first}

We first analyze the scenario where, according to the preferred foliation, Alice's spin measurement at time $t_A$ precedes Bob's release of particle~2 at $t_B$, for each of the \(\sim 10^5\) pairs measured during the experimental run. The analysis proceeds in two stages. First, we determine the effective Bohmian velocity fields for particle~2 after Alice's measurement, for an arbitrary spin direction $\hat{n}\in\mathbb{R}^3$; this turns on how the configuration-dependent weights \eqref{eq:simplified_weights} evolve once Alice has measured. Second, we show that for $\hat{n}=\hat{z}$ and $\hat{n}=\hat{x}$ these fields
reproduce the two dynamical regimes analyzed by DND \cite{Das_2019}, and hence their two qualitatively different arrival-time distributions.

A caveat applies to this section and the next. The distributions in question are
\textit{ideal} ones in the sense of Section~\ref{sec:from_bohmian_trajectories_to_arrival_time_statistics}: they follow from the de Broglie predictive method, and in particular from its passive-screen commitment. We assume throughout that this recipe remains valid in the regime analyzed by DD---that what Bob's screen records approximately matches the ideal distribution. This assumption is not established; it is the antecedent of the conditional stated in Section~\ref{sec:conclusion}, and statements below to the effect that Bob ``will observe'' a given distribution are conditional upon it. Were the recorded distributions to differ, this would indicate a failure of the passive-screen commitment in this regime, not of the Bohmian dynamics.

\subsection{The Effective Guiding Field for Particle 2 After Alice's Measurement}

When particle~1 traverses Alice's Stern-Gerlach magnet, the inhomogeneous magnetic field quickly separates the essential supports of the two spin components of its wave function. After a brief decoherence time $t_A^+ > t_A$, the spatial wave packets $f_{+\hat{n}}$ and $f_{-\hat{n}}$ become essentially disjoint:
\begin{equation}\label{eq:spatial_separation_WP}
    \operatorname{supp}\left[f_{+\hat{n}}(\mathbf{x}_1, t_A^+)\right] \bigcap \operatorname{supp}\left[f_{-\hat{n}}(\mathbf{x}_1, t_A^+)\right] = \emptyset.
\end{equation}
Consequently, after the magnet, the actual particle position $\mathbf{X}_1(t_A^+)$ must lie within exactly one of these essential supports. In BM, which wave packet \(f_{s\hat{n}}\) contains the particle is determined by its initial position \(\mathbf{X}_1(0)\) within the incoming wave packet; this same initial condition fixes the eventual detection point on the screen, thereby determining the registered spin outcome (see \cite{norsen2014pilot} for details). Although we cannot predict Alice's specific outcome without knowledge of the initial position \(\mathbf{X}_1(0)\), the spin "measurement" outcome, once obtained, reliably indicates which wave packet contains the particle and which wave packet becomes an ``empty wave''.\footnote{In BM, the spin "measurement" outcome is determined by the fact that the particle lies in a specific wave packet after the spatial separation \eqref{eq:spatial_separation_WP} induced by the inhomogeneous magnetic field. Our reasoning here uses Alice's observed outcome to infer which wave packet actually guides particle~1 after the magnet. The other wave packet becomes an ``empty wave'' that no longer influences the particle's motion once decoherence has occurred.} Since \(w_s(\mathbf{x}_1, t)\propto f_{s\hat{n}}\), one of the weights \eqref{eq:simplified_weights} will become an empty weight.

Specifically, if Alice measures spin-up (conventionally associated with $s=+1$), then $\mathbf{X}_1(t_A^+) \in \operatorname{supp}\left[f_{+\hat{n}}(\mathbf{x}_1, t_A^+)\right]$ and consequently $f_{-\hat{n}}(\mathbf{X}_1(t_A^+), t_A^+)=0$. Substituting this into the expression for the weights \eqref{eq:simplified_weights} yields a Kronecker delta:
\begin{equation}\label{eq:spin_up}
    w_{s}\left(\mathbf{X}_1(t_A^+), t_A^+\right)=\delta_{s,+1}. 
\end{equation}
If Alice measures spin-down ($s=-1$), we obtain analogously:
\begin{equation}\label{eq:spin_down}
    w_{s}\left(\mathbf{X}_1(t_A^+), t_A^+\right)=\delta_{s,-1}.
\end{equation}

\medskip

When Bob releases particle~2 at $t_B > t_A^+$, Alice's outcome is already fixed and the spatial separation~\eqref{eq:spatial_separation_WP} persists. Since Bohmian trajectories cannot cross into a disjoint support, the actual configuration $\mathbf{X}_1(t)$ remains within the packet $f_{s\hat{n}}$ that contains it at $t_A^+$ for all later times, whatever its detailed motion within that packet. Substituting this configuration into the general expression~\eqref{eq:second_guiding_field}, the weights therefore retain the
Kronecker-delta form \eqref{eq:spin_up}--\eqref{eq:spin_down} throughout particle~2's transit, and not merely at the instant $t_B$, so that particle~2 is guided by a \textit{single} spin-conditional velocity field:
\begin{equation}\label{eq:Alice_first_guiding_field}
    \mathbf{v}_{\text{dBB}}^{(2)}\bigl(\mathbf{X}_1(t), \mathbf{x}_2, t\bigr)=
    \sum_{s'=\pm1}w_{s'}\bigl(\mathbf{X}_1(t), t\bigr)\, \mathbf{v}^{(2)}_{-s'\hat{n}}(\mathbf{x}_2, t)
    = \mathbf{v}^{(2)}_{-s\hat{n}}(\mathbf{x}_2, t), \qquad t \geq t_B,
\end{equation}
where \(\hat{n}\in\mathbb{R}^3\) denotes the spin direction chosen by Alice for this experimental run and \(s=\pm1\) the definite outcome she obtained---so that \(s\hat{n}\) is fixed before Bob's release and remains so throughout particle~2's transit.

\medskip

The physical interpretation is now clear. When Alice measures first along a direction \(\hat{n}\), her outcome \(s\) selects which of the two spin‑conditional velocity fields---associated with direction \(\hat{n}\)---guides particle~2. This selection is non‑local---it occurs instantaneously along the preferred foliation---and directly affects the trajectory that particle~2 follows inside Bob's waveguide. However, for our purposes the more important point is that Alice's choice of the measurement direction \(\hat{n}\) determines the \textit{type} of guiding field, independently of the specific outcome \(s\). The fields \(\mathbf{v}^{(2)}_{+\hat{n}}\) and \(\mathbf{v}^{(2)}_{-\hat{n}}\) differ only by the sign of the spin‑dependent term in Eq.~\eqref{eq:spin_conditional_velocity}; for a fixed \(\hat{n}\) they produce the same qualitative motion. But as \(\hat{n}\) varies, the structure of the guiding field changes dramatically. As we shall now see, the two directions \(\hat{n}=\hat{z}\) and \(\hat{n}=\hat{x}\) yield qualitatively different motions, leading to the distinct arrival‑time distributions predicted by DD.

\subsection{The Crucial Role of the Spin Direction}

As explained in Section\, \ref{sec:Setup}, Bob’s arrival‑time measurement is performed using a cylindrical waveguide oriented along the $\hat{z}$‑axis. For simplicity, we approximate the convective velocity by a plane‑wave motion along the waveguide axis. That is, we take $\nabla_2 S^{(2)}_{0}(\mathbf{x}_2,t) \approx k^{(2)}\hat{z}$, where $k^{(2)}$ is the constant wave number associated with the mean momentum of particle 2 along the guide. Plugging this simplification into \eqref{eq:spin_conditional_velocity} yields the tractable form for particle 2 spin-conditional velocity fields:

\begin{equation}\label{eq:Alice_First_conditional_velocity_2}
    \mathbf{v}^{(2)}_{-s\hat{n}}(\mathbf{x}_2, t) = \frac{\hbar}{m}\left( k^{(2)}\hat{z} - s\frac{\nabla_2 |g_{0}(\mathbf{x}_2,t)|}{|g_{0}(\mathbf{x}_2,t)|} \times \hat{n}\right), \quad s=\pm1.
\end{equation}

The form of this guiding field depends crucially on the cross product term $\nabla_2|g_0|\times\hat{n}$, which couples the spatial gradient of the wave packet amplitude to the spin measurement direction. The cylindrical waveguide geometry imposes a specific potential on $g_0(\mathbf{x}_2,t)$ in the Pauli equation~\eqref{eq:pauli_eq}, which determines the spatial symmetries of the amplitude $|g_0(\mathbf{x}_2,t)|$ and consequently the orientation of its gradient $\nabla_2|g_0|$. Since the spin-dependent term in the velocity is proportional to $\nabla_2|g_0|\times\hat{n}$, the relative orientation between Alice's Stern–Gerlach magnet ($\hat{n}$) and the waveguide axis becomes critical.

Specifically, the radial potential $V_{\bot}(x,y)=\frac{m}{2}\omega^2(x^2+y^2)$ \cite{Das_2019} determines an azimuthal symmetry of the amplitude $|g_0(\mathbf{x}_2,t)|$. Adopting cylindrical coordinates $(\rho,\phi,z)$ with the $\hat{z}$-axis along the waveguide, we express the wave packet as: \(g_0(\mathbf{x}_2,t) = G(\rho,z,t)\), which is independent of $\phi$ due to azimuthal symmetry. In these coordinates
\(\nabla_2|G(\rho,z,t)| = \partial_\rho |G|\,\hat{\rho} + \partial_z |G|\,\hat{z}\), therefore the spin velocity is proportional to
\begin{equation}\label{eq:cross_product_gradient_g}
\nabla_2|g_0(\mathbf{x}_2,t)|\times\hat{n} = \frac{\partial |G|}{\partial \rho}\,\hat{\rho}\times\hat{n} + \frac{\partial |G|}{\partial z}\,\hat{z}\times\hat{n}.
\end{equation}
Two spin configurations are of particular interest: the \textit{longitudinal configuration} $\hat{n} = \hat{z}$, where Alice measures the spin along the same axis as the waveguide orientation; the \textit{transverse configuration} $\hat{n} = \hat{x}$, where Alice measures the spin perpendicular to the waveguide axis. These two configurations give rise to qualitatively different motions of particle 2 inside the waveguide, as we now examine in detail.

\subsection{Longitudinal Configuration $\hat{n}=\hat{z}$: Helical Motion and Heavy-Tailed Distribution}

For the longitudinal configuration $\hat{n}=\hat{z}$, using $\hat{\rho}\times\hat{z} = -\hat{\phi}$ and $\hat{z}\times\hat{z} = 0$, Eq.~\eqref{eq:cross_product_gradient_g} simplifies to: \( \nabla_2|g_0(\mathbf{x}_2,t)|\times\hat{z} = -\partial_\rho |G|\,\hat{\phi}\). Substituting this result into Eq.~\eqref{eq:cross_product_gradient_g} gives the guiding field:

\begin{equation}\label{eq:longitudinal_velocity_final}
    \mathbf{v}^{(2)}_{-s\hat{z}}(\mathbf{x}_2, t) = \frac{\hbar }{m}\, \left(k^{(2)}\hat{z}\; +\; s\frac{1}{|G|} \frac{\partial |G|}{\partial \rho}\,\hat{\phi}\right).
\end{equation}
The resulting velocity field consists of two independent components: (i) a uniform translation along the waveguide axis with velocity $\frac{\hbar k^{(2)}}{m}$, (ii) a circulation in the azimuthal direction ($\hat{\phi}$) whose sense depends on the sign of $s$ (i.e., on Alice's spin outcome) and whose magnitude depends on the radial derivative of the wave packet's amplitude.\footnote{In their paper \cite{Das_2019}, DND calculate that this term corresponds to a circular velocity in the (transverse) $xy$-plane with angular speed $\omega$. In their paper the particle orbits the $z$-axis with constant angular speed $\omega$ in the anticlockwise direction, because they only consider the $z$-spin-up case. Since we consider $z$-spin-up and $z$-spin-down cases, the spin value $s$ shows up in our formula Eq.~\eqref{eq:longitudinal_velocity_final}, showing that the orientation of the angular motion depends on the spin eigenvalue.}
Particle~2 thus follows a helical trajectory along the $\hat{z}$-axis (see the sixth line of Table 1 in \cite{Das_2019}). The pitch of the helix is determined by the convective velocity, while its angular speed and sense of winding are governed by the spin-dependent term and Alice's measurement outcome; its radius is fixed by particle~2's initial transverse position, which is a constant of the motion \cite[Eq.~(25)]{Das_2019}.

\medskip

Equation~\eqref{eq:longitudinal_velocity_final} provides an approximate expression for the guiding field, based on the simplifying assumption of a constant convective velocity $\frac{\hbar k^{(2)}}{m}\hat{z}$. In their more accurate treatment, DND \cite{Das_2019} solve the full Pauli equation for a particle in a cylindrical waveguide and obtain a time-dependent axial velocity $\frac{tz}{1+t^2}\hat{z}$, generating the accelerated motion $Z(t)=Z_0\sqrt{1+t^2}$.\footnote{Expressions quoted from \cite{Das_2019} are written in DND's units $\hbar=m=\omega_z=1$, introduced in Eq.~\eqref{eq:DND_initial_WF}; our own expressions retain $\hbar$ and $m$ explicitly.} For an ensemble of $\sim 10^5$ particles prepared identically \eqref{eq:DND_initial_WF} and distributed according to $|\Psi_0|^2$, they simulate the corresponding Bohmian trajectories numerically and accumulate their arrival times (Eq.~\eqref{eq:arrival-time_distribution}), obtaining the \textit{heavy-tailed distribution} $\Pi_{\hat{z}}(\tau)$ of Fig.~\ref{fig:spin_dependent_arrival_times_distributions}.

What matters for our purposes is that in this configuration the spin flux leaves no trace in the statistics. The spin term is purely azimuthal, so the axial equation of motion is closed in $Z$ and the arrival time depends on $Z_0$ alone \cite[Eq.~(45)]{Das_2019}; since $Z_0$ is distributed independently of the transverse coordinates, the heavy-tailed distribution is fixed entirely by the convective flux and the $|\Psi_0|^2$ marginal in $Z_0$. The helical winding, conspicuous though it is in the trajectories, is statistically invisible---which is also why $\Pi_{\hat{z}}(\tau)$ is independent of the trapping frequency $\omega$ \cite[Sec.~V\,A]{Das_2019}, and why, as we shall see in Section~\ref{sec:Bob_measures_first}, a qualitatively different trajectory can yield the very same distribution. Our constant-velocity approximation, though it misrepresents the detailed axial motion, preserves exactly this structure.

Thus, when Alice measures first and selects the longitudinal configuration $\hat{n}=\hat{z}$, Bob will observe---after accumulating arrival times from the $\sim10^5$ pairs in a run---a histogram approximating the heavy-tailed distribution $\Pi_{\hat{z}}(\tau)$.

\subsection{Transverse Configuration $\hat{n}=\hat{x}$: Backflow and Maximum Arrival Time}\label{subsubsec:Backflow}

When Alice measures spin perpendicular to the waveguide axis ($\hat{n}=\hat{x}$),
Eq.~\eqref{eq:cross_product_gradient_g} together with $\hat{z}\times\hat{x} = \hat{y}$ and $\hat{\rho} = \cos\phi\,\hat{x} + \sin\phi\,\hat{y}$ yields $\hat{\rho}\times\hat{x} = -\sin\phi\,\hat{z}$. Hence
\begin{equation*}
    \nabla_2|g_0(\mathbf{x}_2,t)|\times\hat{x} = -\frac{\partial |G|}{\partial \rho}\sin\phi\,\hat{z} + \frac{\partial |G|}{\partial z}\,\hat{y}.
\end{equation*}
Inserting this into the spin-conditional velocity \eqref{eq:Alice_First_conditional_velocity_2} gives
\begin{equation}\label{eq:backflow_velocity}
    \mathbf{v}^{(2)}_{-s\hat{x}}(\mathbf{x}_2, t) =
    \underbrace{\frac{\hbar k^{(2)}}{m}\left(1 + s\frac{1}{k^{(2)}|G|}\frac{\partial |G|}{\partial \rho}\sin\phi\right)\hat{z}}_{\text{axial component}}
    \;-\;
    \underbrace{s\frac{\hbar}{m|G|}\frac{\partial |G|}{\partial z}\,\hat{y}}_{\text{transverse component}}.
\end{equation}
The motion therefore comprises: (i) an axial component along $\hat{z}$, whose magnitude is modulated by the spin-dependent term (through $\sin\phi$ and the radial gradient of $|G|$); (ii) a transverse component along $\hat{y}$, driven by the axial gradient of $|G|$. Unlike the longitudinal case, there is no azimuthal circulation: the field has no $\hat{x}$ component, so the trajectory remains in the plane of constant $x$ in which it started.

The contrast with the longitudinal configuration is instructive. There the spin term was purely azimuthal and left the axial motion untouched, so that the arrival time depended on $Z_0$ alone; here the spin term enters the axial component directly, while the transverse component depends in turn on $z$. The two motions are coupled, and it is this coupling that---as we now recall---generates a maximum arrival time.

The spin-dependent modulation can even reverse the axial velocity when
\begin{equation}\label{eq:back_flow}
    s\frac{1}{k^{(2)}|G|}\frac{\partial |G|}{\partial \rho}\sin\phi < -1,
\end{equation}
a phenomenon known as \textit{quantum backflow}: the particle temporarily moves opposite to its mean momentum direction. This condition can be made explicit using DND's solution \cite[Eq.~(16)]{Das_2019}. Their transverse profile is Gaussian, $|G|\propto \exp\!\left(-m\omega\rho^{2}/2\hbar\right)$, so that $\partial_\rho|G|/|G| = -m\omega\rho/\hbar$; inserting this together with $\sin\phi = y/\rho$ into \eqref{eq:back_flow} yields the more tractable condition
\begin{equation}\label{eq:tractable_back_flow}
    s\,\omega y > \frac{\hbar k^{(2)}}{m},
\end{equation}
that is, backflow occurs when the axial component of the spin velocity exceeds the convective velocity $\hbar k^{(2)}/m$ of our approximation. A trajectory exhibiting this behavior is displayed in Figure~6 of \cite{Das_2019}.

With $\omega$ fixed by the waveguide, whether a given trajectory undergoes backflow depends on Alice's outcome $s$ and on the particle's $y$-coordinate, itself driven by the transverse component of \eqref{eq:backflow_velocity}. The two outcomes are not, however, statistically distinguishable. The trajectory remains in the plane $x=X_0$, and since $|G|$ depends on $y$ only through $\rho$, the equations of motion \eqref{eq:backflow_velocity} map into one another under $s\to-s$ together with $y\to-y$: if $\bigl(y(t),z(t)\bigr)$ solves them for outcome $s$, then $\bigl(-y(t),z(t)\bigr)$ solves them for $-s$. The reflected trajectory has the same axial motion, hence the same arrival time, and $|\Psi_0|^2$ is symmetric under $y\to-y$ (see \eqref{eq:DND_initial_WF}). The two outcomes therefore yield identical arrival-time statistics---which is what allows Alice's choice of $\hat{n}$, and not her uncontrollable outcome, to determine what Bob observes.
\medskip

In their full analysis, DND \cite{Das_2019} go beyond the constant-velocity approximation. Solving the Pauli equation for an $\hat{x}$-polarized particle in the cylindrical waveguide, they obtain an axial velocity $\bigl(\omega y + \frac{tz}{1+t^2}\bigr)\hat{z}$ (Eq.~(21c) in \cite{Das_2019}), which after the substitution $Z(t)=\xi(t)\sqrt{1+t^2}$ (Eq.~(28) in \cite{Das_2019}) leads to bounded oscillations $\xi_s \le \xi(t) \le \xi_b$ (Eq.~(38) in \cite{Das_2019}): the trajectory oscillates between the two dispersing envelopes $\xi_s\sqrt{1+t^2}$ and $\xi_b\sqrt{1+t^2}$.

The existence of a maximum arrival time follows from two facts. First, whenever the trajectory touches the \textit{upper} envelope after the earliest possible crossing time, it necessarily lies beyond the detector plane $z=L$; the first crossing must therefore occur before that moment. Second---and this is what makes the bound uniform---the time required to complete a half-cycle of the oscillation is bounded independently of the initial position. Combining the two, DND obtain an explicit upper bound (Eq.~(59) in \cite{Das_2019}): every Bohmian trajectory, whatever its initial position, strikes the detector before a finite
$\tau_{\text{max}}$ fixed by the waveguide parameters alone. This is the origin of the \textit{exotic arrival-time distribution} $\Pi_{\hat{x}}(\tau)$ displayed in Fig.~\ref{fig:spin_dependent_arrival_times_distributions}, which DND obtain by simulating an ensemble of $\sim10^5$ trajectories with initial positions distributed according to $|\Psi_0|^2$.

Thus, when Alice measures first and selects the transverse configuration ($\hat{n}=\hat{x}$), we predict that Bob's accumulated arrival times will approximate the exotic distribution $\Pi_{\hat{x}}(\tau)$, with its characteristic maximum arrival time.

\section{Bob Measures First: Heavy Tailed Distribution}\label{sec:Bob_measures_first}

We now analyze the reverse temporal ordering: Bob's arrival time measurement at $t_B$ occurs \textit{before} Alice's spin measurement at $t_A$ according to the preferred foliation ($t_B < t_A$).

At the moment $t_B$ when Bob performs his measurement, neither particle has yet interacted with an inhomogeneous magnetic field. In the Pauli equation~\eqref{eq:pauli_eq}, we have $\mathbf{B}(\mathbf{x}_1) = \mathbf{B}(\mathbf{x}_2) = 0$ at time $t_B$. Just as for particle~2, the spatial wave functions for the two spin branches of particle~1 are identical at this time:

\begin{equation}\label{eq:particle_1_simplification}
    f_{s\hat{n}}(\mathbf{x}_1,t_B) = f_0(\mathbf{x}_1,t_B) = |f_0(\mathbf{x}_1,t_B)|e^{iS_0^{(1)}(\mathbf{x}_1,t_B)}, \quad s = \pm 1.
\end{equation}

\medskip

Substituting this identity into the expression for the weights (Eq.~\eqref{eq:simplified_weights}) yields:

\begin{equation*}
    w_s\left(\mathbf{x}_1,t_B\right)=\frac{ |f_{0}(\mathbf{x}_1,t_B)|^2}{ |f_{0}(\mathbf{x}_1,t_B)|^2 + |f_{0}(\mathbf{x}_1,t_B)|^2 } = \frac{1}{2}, \quad s = \pm1.
\end{equation*}

Thus, at the measurement time $t_B$, the weights lose their spatial dependence and become uniform: $w_{+1}(\mathbf{x}_1,t_B) = w_{-1}(\mathbf{x}_1,t_B) = 1/2$. Consequently, the velocity field~\eqref{eq:second_guiding_field} for particle~2 simplifies to a simple average of the two spin-conditional velocity fields:

\begin{equation}\label{eq:Bob_before_Alice_second_velocity_field}
    \mathbf{v}_{\text{dBB}}^{(2)}(\mathbf{x}_2, t) = \frac{1}{2}\left[\mathbf{v}^{(2)}_{-\hat{n}}(\mathbf{x}_2,t_B) + \mathbf{v}^{(2)}_{+\hat{n}}(\mathbf{x}_2,t_B)\right]=\frac{\hbar}{m}\nabla_2 S^{(2)}_0(\mathbf{x}_2,t)\approx \frac{\hbar k^{(2)}}{m}\hat{z}.
\end{equation}

The spin-dependent terms, being linear in $s$ with opposite signs, cancel exactly. Consequently, when Bob measures first, particle~2 is guided \textit{solely} by the convective velocity derived from the spatial phase $S^{(2)}_0$. Physically, this means particle~2 behaves as if it were a non-entangled spinless particle with wave function $g_0(\mathbf{x}_2,t)$ \eqref{eq:polar_g}.

\medskip

This purely axial velocity field should be contrasted with the helical motion that arises in the Alice-first, $\hat{n}=\hat{z}$ case (Eq.~\eqref{eq:longitudinal_velocity_final}). While the trajectories differ qualitatively (straight versus helical), the arrival-time at a detector placed perpendicular to the waveguide axis depends only on the axial component of motion. Consequently, both cases yield the same arrival-time distribution: the \textit{heavy tailed distribution} $\Pi_{\hat{z}}(\tau)$ shown in Fig.~\ref{fig:spin_dependent_arrival_times_distributions}.

Therefore, when Bob measures first for each of the \(\sim 10^5\) measurements performed during the run, we predict that he will always observe this heavy tailed distribution $\Pi_{\hat{z}}(\tau)$.\footnote{Regarding the subsequent evolution of particle~1, the simplification in Eq.~\eqref{eq:particle_1_simplification} holds only until particle~1 enters Alice's Stern-Gerlach magnet. At a time $t_A^- < t_A$ just before entering the magnet, the magnetic simplification \eqref{eq:particle_1_simplification} no longer holds, particle~1's wave function recover its spin-dependent form \eqref{eq:polar_f} and its velocity is given by Eq.~\eqref{eq:expanded_first_current}. The subsequent passage through the magnet will separate the spin components, leading to a definite outcome. Given the quantum equilibrium hypothesis, the Born rule probabilities apply: Alice will measure spin-up along $\hat{n}$ with probability $1/2$ and spin-down with probability $1/2$.}

\section{Toward Foliation Detection: The Empirical Signature of Temporal Order}\label{sec:summary_and_analysis_of_the_results}

Let us begin this section by summarizing, in Table~\ref{tab:temporal_order}, the full range of possible arrival-time distributions for each ensemble of $\sim 10^5$ particles, assuming that the same temporal order---either $t_A < t_B$ or $t_B < t_A$---holds for each of the \(\sim 10^5\) pairs of measurements performed during the experimental run.

\begin{table}[h]
    \centering
    \begin{tabular}{lcc}
        \hline
        \textbf{Temporal order} & \textbf{Alice's choice (button)} & \textbf{Bob's distribution} \\ 
        \hline
        Alice first ($t_A < t_B$) & 0 ($\hat{z}$) & Heavy Tailed \\ 
        Alice first ($t_A < t_B$) & 1 ($\hat{x}$) & \textbf{Exotic Distribution} \\ 
        \hline
        Bob first ($t_B < t_A$) & 0 or 1 (any) & Heavy Tailed \\ 
        \hline
    \end{tabular}
    \caption{Arrival-time distributions observed by Bob for different temporal orders and Alice's measurement choices. The predictions summarized here rest on the validity of the de Broglie predictive method (Section~\ref{subsec:predictive_method}).}
    \label{tab:temporal_order}
\end{table}

To construct a foliation detection protocol, we need to extract an empirical signature of the preferred foliation from the predictions summarized in Table~\ref{tab:temporal_order}. Specifically, we are looking for an unambiguous witness of the temporal order defined by \(\mathcal{F}\).

First, from the outcome \(\tau\) of a single arrival-time measurement, Bob cannot unambiguously determine the temporal order of the two measurement events \(A\) and \(B\). To overcome this limitation, our experimental protocol requires measuring the arrival times of an entire ensemble of \(\sim 10^5\) entangled pairs per run. This accumulates sufficient statistics for Bob to reliably distinguish between the heavy-tailed and exotic distributions. We require that these two distributions be perfectly correlated with the temporal order that held during the run. This perfect correlation would allow Bob to deduce the temporal order from the distribution displayed on his screen.

From Table~\ref{tab:temporal_order}, if Bob observes the exotic distribution \(\Pi_{\hat{x}}(\tau)\) (Fig.~\ref{fig:spin_dependent_arrival_times_distributions}), he can immediately conclude that during that run Alice's measurements occurred before his according to the preferred foliation (\(t_A < t_B\)). However, if he observes the heavy-tailed distribution \(\Pi_{\hat{z}}(\tau)\) (Fig.~\ref{fig:spin_dependent_arrival_times_distributions}), the interpretation is ambiguous: it could indicate either ``Bob first'' (\(t_B < t_A\)) or ``Alice first with \(\hat{n} = \hat{z}\)'' during that run.

\begin{figure}[hbtp]
    \centering
    \includegraphics[width=0.6\textwidth]{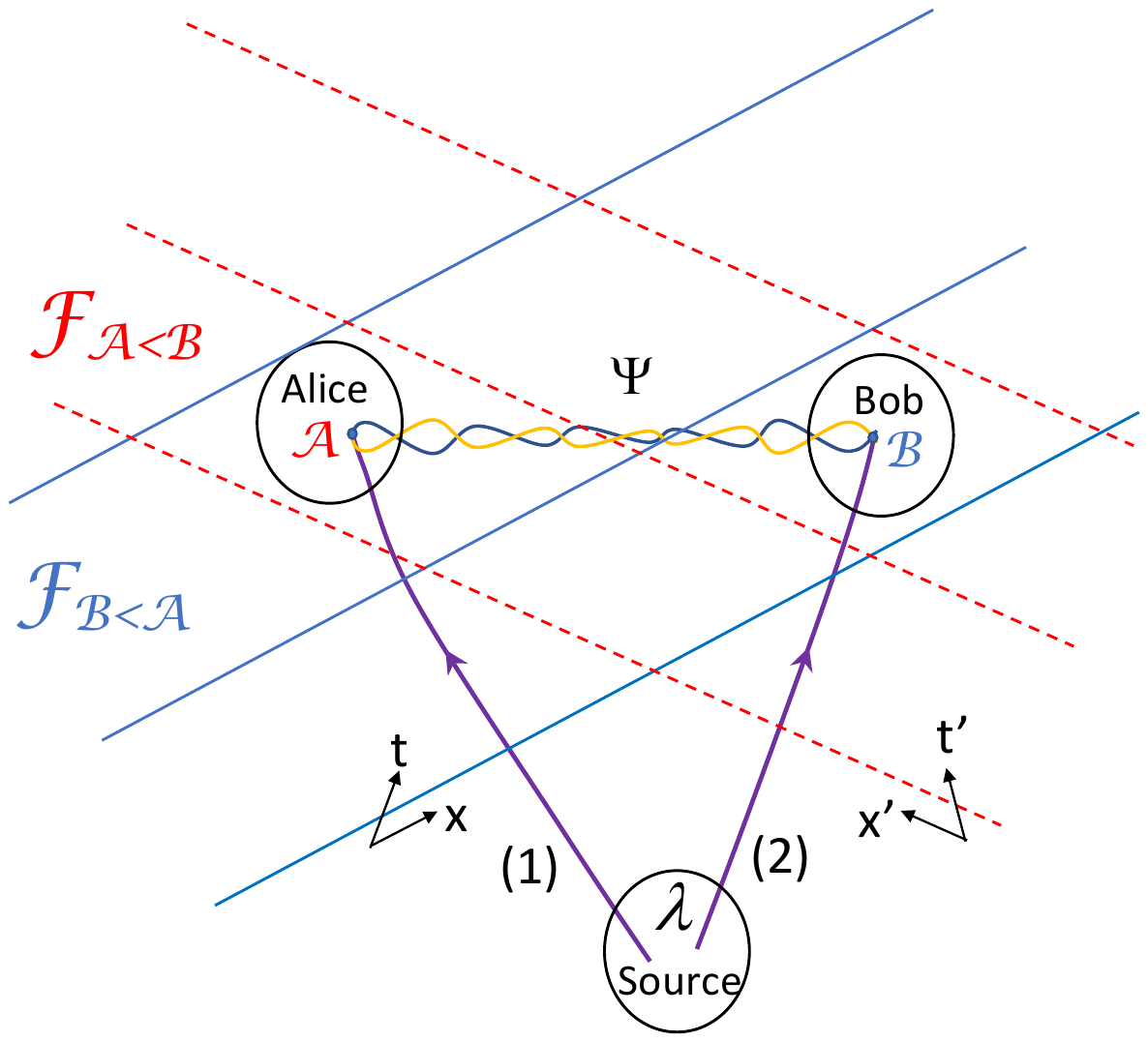}
    \caption{Event \(\mathcal{A}\): Alice's \(\sim 10^5\) \(\hat{x}\)-spin measurements. Event \(\mathcal{B}\): Bob's \(\sim 10^5\) arrival-time measurements. Their space-time positions are fixed. Special relativity provides no objective temporal order for such spacelike-separated events, yet as Table~\ref{tab:temporal_order} shows, the predicted arrival-time distribution depends crucially on whether \(t_A < t_B\) or \(t_B < t_A\) according to the foliation. The red foliation (\(t_A < t_B\)) predicts the exotic distribution for Bob; the blue foliation (\(t_B < t_A\)) predicts the heavy-tailed distribution. These incompatible predictions demonstrate that the red and blue foliations cannot both be mere conventions---one of them corresponds to the actual structure of space-time. Crucially, Bob's observed arrival-time distribution reveals which: if he sees the heavy-tailed distribution, the existing foliation cannot be the red one; if he sees the exotic distribution, it cannot be the blue one. This observed distribution thus provides the first experimental datum for reconstructing the preferred foliation \(\mathcal{F}\). (For related discussions on the necessity of a preferred foliation, see \cite{Drezet_2019,Hardy,Gisin_2011}.)}
    \label{fig:Different_Foliations}
\end{figure}

To resolve this ambiguity, Alice must fix her measurement direction throughout the experiment. By pre‑agreeing that Alice will consistently use the transverse configuration (\(\hat{n} = \hat{x}\)) in every experimental run, the observed distributions become an unambiguous binary indicator:
\begin{itemize}
    \item Exotic distribution \(\Pi_{\hat{x}}(\tau)\) \(\Rightarrow\) Alice measured first (\(t_A < t_B\));
    \item Heavy-tailed distribution \(\Pi_{\hat{z}}(\tau)\) \(\Rightarrow\) Bob measured first (\(t_B < t_A\)).
\end{itemize}
Thus, with Alice's spin direction fixed to \(\hat{n} = \hat{x}\), the curve observed by Bob serves as an unambiguous witness to the temporal order that held during that run. This is not yet a full detection of the foliation, but it tells us that the existing foliation is one that yields the temporal order implied by Bob's arrival-time distribution. This is illustrated in Fig.~\ref{fig:Different_Foliations}, which shows how Bob's observed distribution selects between two candidate foliations that assign different temporal orders to the same events. This provides the seed of the foliation detection protocol that we develop in the next section.

Before turning to the detection protocol, one might object that the analysis leading to Table~\ref{tab:temporal_order} is non-relativistic, whereas the proposal itself involves spacelike-separated observers and a preferred foliation that may move at an arbitrary subluminal speed relative to the common frame $\Lambda$. Appendix~\ref{app:relativistic} outlines the procedure that would be followed in a fully relativistic treatment based on the HBD model, and explains why the logical inferences of Table~\ref{tab:temporal_order} are expected to survive it.

\section{Mapping a Flat Preferred Foliation: An Experimental Protocol}\label{sec:Foliation_Detection}

The predictions summarized in Table~\ref{tab:temporal_order} can be systematically exploited to map the geometry of the preferred hypersurfaces. While HBD models generally allow curved foliations, our detection protocol is specifically designed for a flat foliation---i.e., a family of parallel hyperplanes.

As established in Section~\ref{sec:the_preferred_foliation}, any foliation of Minkowski space \(M\) is completely characterized by its future-pointing unit normal vector field \(n(x)\) (see Eq.~\eqref{eq:future_pointing_unit_normal_vector_field}). Measuring \(n(x)\) is therefore equivalent to detecting the foliation, which is precisely what our protocol aims to achieve. Before describing the protocol itself, we first establish the mathematical framework underlying it.

For the flat foliation \(\mathcal{F}\) considered here, the normal vector field is constant throughout space-time, i.e., \(n(x) = n\). In the preferred frame \(F\), its components are simply \((1,0,0,0)\). During the first stage of the protocol (introduced below), Alice and Bob will establish a common Lorentz frame \(\Lambda\) using standard synchronization procedures. In this frame \(\Lambda\), the future-pointing unit normal vector of the preferred foliation has unknown coordinates
\[
(n^0, n_x, n_y, n_z) \equiv (n^0, \mathbf{n}),
\]
where \(\mathbf{n} = (n_x, n_y, n_z)\) denotes the spatial part. The objective of our detection protocol is to determine these four components---a task that requires four independent constraints.

The first and most obvious constraint is the normalization condition imposed by the Minkowski metric (with signature \((+,-,-,-)\)):
\[
|n|^2 = (n^0)^2 - (n_x^2 + n_y^2 + n_z^2) = 1.
\]
Since the foliation is future-oriented, \(n^0 > 0\), which fixes the sign when solving for \(n^0\):
\begin{equation}\label{eq:first_condition}
    n^0 = \sqrt{1 + n_x^2 + n_y^2 + n_z^2}.
\end{equation}
Thus, once the three spatial components \(\mathbf{n}\) are determined, the time component \(n^0\) follows directly from Eq.~\eqref{eq:first_condition}. The problem therefore reduces to finding three independent equations for \(\mathbf{n}\).

The key physical constraint is the simultaneity condition itself. Two events \(P_A\) and \(P_B\) that are simultaneous with respect to the preferred foliation lie on a common preferred hyperplane \(\Sigma \in \mathcal{F}\). Their separation vector \(P_A - P_B\) is therefore tangent to \(\Sigma\) and, by definition, orthogonal to \(n\) in Minkowski space. For a pair of simultaneous events \((P_A, P_B)\) with coordinates \((T_A, \mathbf{X}_A)\) and \((T_B, \mathbf{X}_B)\) in the common Lorentz frame \(\Lambda\), this orthogonality condition reads:
\begin{equation}\label{eq:second_condition}
    (P_A - P_B) \cdot n = (T_A - T_B)\, n^0 - (\mathbf{X}_A - \mathbf{X}_B) \cdot \mathbf{n} = 0,
\end{equation}
where \((\mathbf{X}_A - \mathbf{X}_B) \cdot \mathbf{n} = (X_A - X_B)n_x + (Y_A - Y_B)n_y + (Z_A - Z_B)n_z\).

Given two additional pairs of simultaneous space-time points, \((P'_A, P'_B)\) and \((P''_A, P''_B)\), the simultaneity condition provides two further equations analogous to~\eqref{eq:second_condition}:
\begin{align}
        (P'_A - P'_B) \cdot n &= (T'_A - T'_B)\, n^0 - (\mathbf{X}'_A - \mathbf{X}'_B)\cdot\mathbf{n} = 0, \label{eq:third_condition} \\
        (P''_A - P''_B) \cdot n &= (T''_A - T''_B)\, n^0 - (\mathbf{X}''_A - \mathbf{X}''_B)\cdot\mathbf{n} = 0. \label{eq:fourth_condition}
\end{align}
Inserting \(n^0=\sqrt{1+\mathbf{n}^2}\) from Eq.~\eqref{eq:first_condition} into Eqs.~\eqref{eq:second_condition}–\eqref{eq:fourth_condition} yields three equations for the three unknowns \(n_x,n_y,n_z\).

These equations have a unique solution if and only if the three separation vectors \(P_A - P_B\), \(P'_A - P'_B\), \(P''_A - P''_B\) are linearly independent. In other words, they must not all lie in a common 2‑dimensional subspace of Minkowski space-time; equivalently, they must span the full 3‑dimensional tangent space of the hyperplane \(\Sigma\).\footnote{Analogously, to determine the unit normal vector of a 2‑plane in \(\mathbb{R}^3\) one needs two non‑collinear tangent vectors. For a 3‑hyperplane in Minkowski space, three linearly independent tangent vectors are required.}

Thus, given three pairs of simultaneous space-time points whose separation vectors are linearly independent, the 4-vector \(n\) is mathematically determined by the system of four equations---Eqs.~\eqref{eq:first_condition}–\eqref{eq:fourth_condition}. Solving this system yields the orientation of the preferred foliation \(\mathcal{F}\) relative to the common Lorentz frame \(\Lambda\).

We now turn to the experimental protocol that will allow Alice and Bob to detect three pairs of simultaneous space-time points. This protocol proceeds in five stages.

\subsection*{Stage 1: Establishing a Common Reference Frame}

Alice and Bob first establish a common Lorentz frame $\Lambda$ by exchanging repeated light signals and synchronizing their clocks according to the standard Einstein synchronization procedure \cite[pp.~88--91]{maudlin2012philosophy}. This allows them to assign consistent coordinates $(t', \mathbf{x})$ to all relevant events that will occur during the experimental protocol.\footnote{This time coordinate $t'$ in $\Lambda$ should not be confused with the time coordinate $t$ of the preferred foliation, which we denote by $t$ without a prime.}

For each entangled pair prepared in state \eqref{eq:wave_function_factorization} and shared in the setup of Fig.~\ref{fig:experimental_setup}, the two measurements are represented by space-time points $A$ and $B$ as defined in Section~\ref{sec:definition_of_the_temporal_order}. In the common frame $\Lambda$, Alice records the coordinates $(t'_A, \mathbf{X}_A)$ of event $A$ (her spin measurement), and Bob records the coordinates $(t'_B, \mathbf{X}_B)$ of event $B$ (his release of particle~2). These coordinates are recorded for each of the $\sim 10^5$ pairs in a run. Crucially, to avoid confusion, the times $t'_A$ and $t'_B$ in $\Lambda$ must be distinguished from the preferred foliation times $t_A$ and $t_B$ used in Sections~\ref{sec:Alice_measures_first} and~\ref{sec:Bob_measures_first}.

\subsection*{Stage 2: From Measurement Sequences to Space-Time Points}

We now explain how Alice and Bob extract an ordered pair of space-time points from each experimental run.

During a run, Alice performs a sequence of $\sim 10^5$ spin measurements while Bob simultaneously performs a sequence of $\sim 10^5$ arrival-time measurements. Consequently, they record $\sim 10^5$ individual pairs of space-time points $(A,B)$, where $A=(t'_A, \mathbf{X}_A)$ and $B=(t'_B, \mathbf{X}_B)$. If a single pair of measurements takes a duration $\delta t$ in the common Lorentz frame $\Lambda$, the entire sequence extends over a time interval $\Delta t \sim \delta t \cdot 10^5$. Each sequence thus occupies a finite space-time region: denote by $\mathcal{R}_A$ the region containing Alice's measurements and by $\mathcal{R}_B$ the region containing Bob's. The spatial separation between Alice's and Bob's laboratories is chosen to be sufficiently large that both the spatial extent of each laboratory and the duration $\Delta t$ of each measurement sequence are negligible compared to the spacelike interval separating $\mathcal{R}_A$ and $\mathcal{R}_B$. Under this condition, we approximate Alice's entire sequence of measurements during a given run as a single space-time point $\mathcal{A}$, and Bob's sequence as a point $\mathcal{B}$. This idealization is justified because any two points within $\mathcal{R}_A$ are spacelike-separated from any two points within $\mathcal{R}_B$ by approximately the same large interval.

Thus, each experimental run is represented by two space-time points $\mathcal{A}$ and $\mathcal{B}$. While these idealized points are not directly measurable, they can be naturally defined from the recorded collections. For each run, we define $\mathcal{A}$ as the space-time point with coordinates $(T_A, \mathbf{X}_A)$ in $\Lambda$, where $T_A$ is the mean of the $\sim 10^5$ individual measurement times $t'_A$ recorded by Alice (i.e., the average time at which her particles exit the Stern-Gerlach magnet during this run). Similarly, $\mathcal{B}$ is defined as the space-time point with coordinates $(T_B, \mathbf{X}_B)$ in $\Lambda$, where $T_B$ is the mean of the $\sim 10^5$ individual release times $t'_B$ recorded by Bob (i.e., the average time at which his particles are released from initial confinement). This convention captures the central tendency of each sequence and is robust against small fluctuations in individual measurement times.

\medskip

As explained in the previous section, with Alice's measurement direction fixed to $\hat{n} = \hat{x}$, the observed arrival-time distribution allows Bob to infer the temporal order that held during the run, and therefore the temporal order of $\mathcal{A}$ and $\mathcal{B}$ according to the preferred foliation. Specifically, if Bob observes the exotic distribution $\Pi_{\hat{x}}(\tau)$, he infers that $\mathcal{A}$ precedes $\mathcal{B}$ according to the foliation: $t_\mathcal{A}<t_\mathcal{B}$; if he observes the heavy-tailed distribution $\Pi_{\hat{z}}(\tau)$, he infers $\mathcal{B}$ precedes $\mathcal{A}$: $t_\mathcal{B}<t_\mathcal{A}$. Here $t_\mathcal{A}:=t(\mathcal{A})$ and $t_\mathcal{B}:=t(\mathcal{B})$, with $t:M\to\mathbb{R}$ the time function whose level sets \eqref{eq:level_sets} define the leaves of $\mathcal{F}$.\footnote{Crucially, this temporal order is defined by the sign of $t(\mathcal{A}) - t(\mathcal{B}) \equiv t_\mathcal{A} - t_\mathcal{B}$. This sign need not coincide with the sign of $T_A - T_B$, because $\mathcal{A}$ and $\mathcal{B}$ are spacelike-separated and the foliation $\mathcal{F}$ and the common frame $\Lambda$ are generally different Lorentz frames. The observed distribution thus reveals the order relative to $\mathcal{F}$, not relative to $\Lambda$.}

In this way, every experimental run yields a pair of space-time points $(\mathcal{A}, \mathcal{B})$ whose objective temporal order is revealed to Bob through the arrival-time distribution displayed on his screen. This enables the next stage: the iterative detection of simultaneity points.

\subsection*{Stage 3: Locating a Pair of Simultaneous Events}

\begin{figure}[hbpt]
    \centering
    \includegraphics[width=0.6\textwidth]{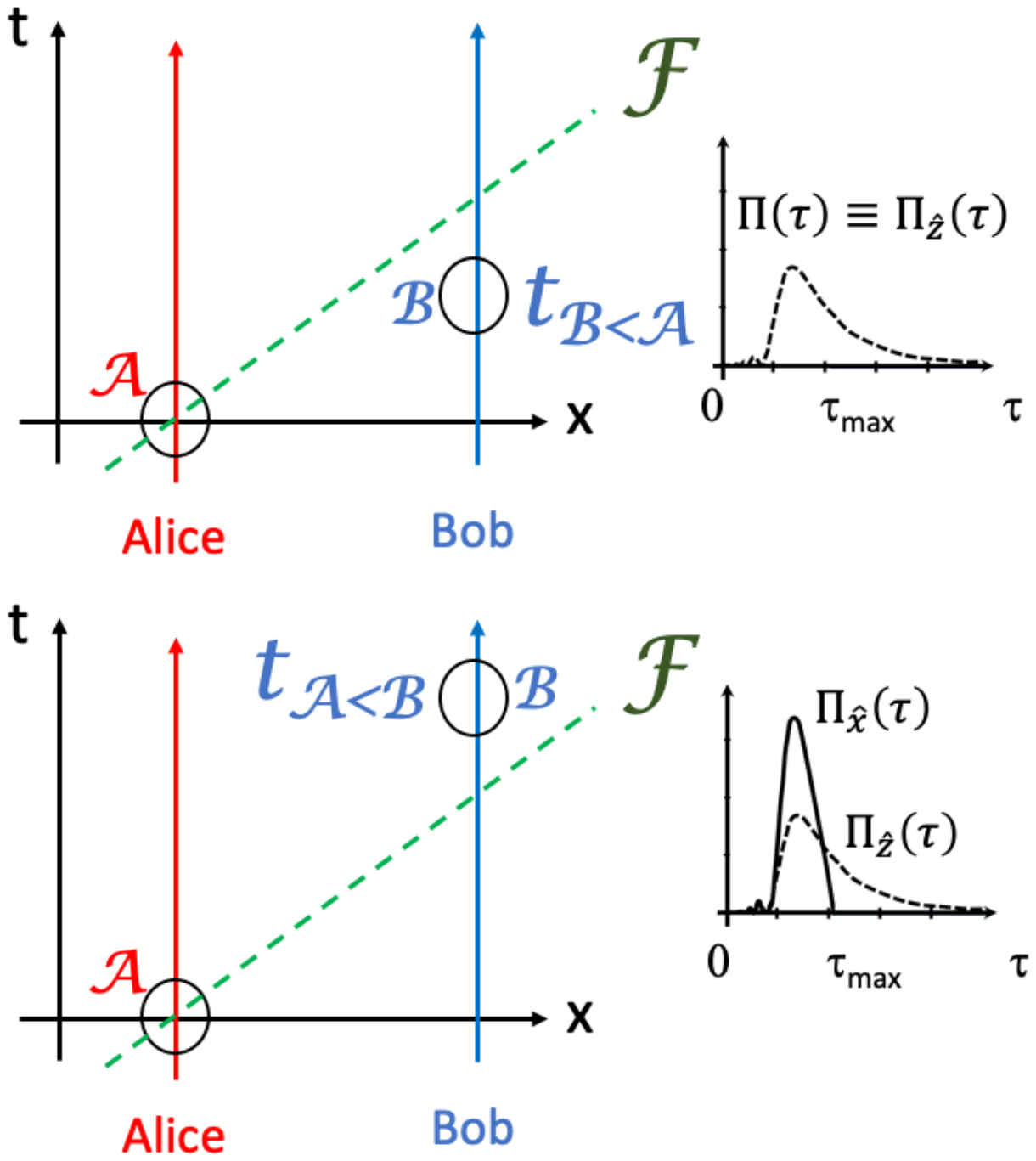}
    \caption{Space-time diagrams illustrating the detection of the switch point. The red vertical line is the worldline of Alice's magnet; the blue vertical line is the worldline of Bob's waveguide. Event $\mathcal{A}$ denotes Alice's series of spin measurements, event $\mathcal{B}$ Bob's series of arrival-time measurements. The green dashed line represents the preferred hypersurface containing $\mathcal{A}$. (a) When $\mathcal{B}$ lies below this leaf ($t_\mathcal{B} < t_\mathcal{A}$), Bob observes the heavy-tailed distribution $\Pi_{\hat{z}}(\tau)$. (b) When $\mathcal{B}$ lies above the leaf ($t_\mathcal{A} < t_\mathcal{B}$), Bob observes the exotic distribution $\Pi_{\hat{x}}(\tau)$. The switch point (not shown) occurs when $\mathcal{B}$ lies exactly on the green dashed line. In the iterative search, Bob adjusts his waveguide position to transition between cases (a) and (b), thereby roughly locating the switch point, then refines his position to improve accuracy. For clarity, the diagram simplifies the protocol in two ways: first, it keeps the spatial separation (in $\Lambda$) between the two worldlines fixed, whereas in reality Bob moves his waveguide, so this horizontal distance varies across runs; second, it treats $\mathcal{A}$ as a fixed point, though different runs correspond to different $\mathcal{A}$ points (since time elapses between the completion of different runs) even with Alice's magnet fixed at $\mathbf{X}_A$. The real protocol therefore does not simply move $\mathcal{B}$ along a fixed worldline to intersect a fixed $\mathcal{A}$, but the essential logic is captured by this idealized representation.}
    \label{fig:Switch_point}
\end{figure}

The goal of this stage is to identify a pair of space-time points $(P_A, P_B)$ that are (approximately) simultaneous according to the preferred foliation. The core idea is to perform several experimental runs, adjusting Bob's waveguide position $\mathbf{X}_B$ relative to the source after each run, until the observed arrival-time distribution switches from one type to the other. As illustrated in Fig.~\ref{fig:Switch_point}, this switch occurs precisely when Bob's measurement event $\mathcal{B}$ lies on the same preferred hypersurface as Alice's measurement event $\mathcal{A}$---i.e., when the two events are simultaneous according to $\mathcal{F}$. For a fixed space-time point $\mathcal{A}$ corresponding to Alice's sequence of measurements in a given run, the \textit{switch point} $\mathcal{B}_{\mathcal{A}}^*$ is the unique point on Bob's worldline that satisfies $t(\mathcal{A}) - t(\mathcal{B}_{\mathcal{A}}^*) = 0$, where $t:M\to\mathbb{R}$ is the time function whose level sets define the preferred foliation via \eqref{eq:level_sets}. In the common Lorentz frame $\Lambda$, this switch point has coordinates $(T_B^*, \mathbf{X}_B^*)$. The experimental task is to determine these coordinates by iteratively adjusting Bob's waveguide position until the distribution flips, thereby locating $\mathcal{B}_{\mathcal{A}}^*$ on his worldline.\footnote{For simplicity, we assume that Alice's sequence of measurements $\mathcal{A}$ remains effectively fixed across runs, while only Bob's sequence $\mathcal{B}$ moves along his worldline as he adjusts his waveguide position. In practice, different runs yield slightly different $\mathcal{A}$ points, but because Alice's magnet is held fixed relative to the source and the measurement procedure is identical, these variations are negligible. Treating $\mathcal{A}$ as fixed simplifies the exposition without affecting the essential logic.} This is accomplished in two steps.

\begin{enumerate}[label=\textbf{Step \arabic*:}, leftmargin=*, align=left]
    
    \item \textbf{Iterative search for the switch point.} Alice and Bob perform a series of $N$ experimental runs (the number $N$ is determined adaptively). Alice's magnet remains fixed at position $\mathbf{X}_A$ throughout. Between runs, Bob adjusts his waveguide position according to the following rule:
    
    \begin{itemize}
        \item If Bob observes the \textit{exotic distribution} $\Pi_{\hat{x}}(\tau)$, he infers $t_\mathcal{A} < t_\mathcal{B}$. To make his measurement event earlier in the next run, he moves his waveguide \textit{closer} to the source along the line connecting the source to his laboratory, thereby reducing the travel time of particles~2.\footnote{We assume the particles travel at approximately the same speed from run to run. Moving the waveguide closer (farther) causes Bob to receive his $\sim10^5$ particles earlier (later) in the next run, while Alice receives her particles at roughly the same time.}
        
        \item If Bob observes the \textit{heavy-tailed distribution} $\Pi_{\hat{z}}(\tau)$, he infers $t_\mathcal{B} < t_\mathcal{A}$. To make his measurement event later in the next run, he moves his waveguide \textit{farther} from the source, increasing the travel time of particles~2.
    \end{itemize}
    
    Bob begins with coarse adjustments to locate the approximate region where the distribution flips, then progressively refines the step size to pinpoint the transition with increasing precision. Once he deems his sequence of measurements $\mathcal{B}$ to be sufficiently close to simultaneity with $\mathcal{A}$, he sends a light signal to Alice indicating completion of this step.
    
    \item \textbf{Recording the simultaneous pair.} Using the definitions from Stage~2, Alice and Bob compute the coordinates of the events $\mathcal{A}$ and $\mathcal{B}$ corresponding to the final experimental run before Bob's light signal. Denote these space-time points as $P_A := (T_A, \mathbf{X}_A)$ for Alice's last sequence of spin measurements, and $P_B := (T_B, \mathbf{X}_B)\approx (T_B^*, \mathbf{X}_B^*)$ for Bob's last sequence of arrival-time measurements approximating the switch point $\mathcal{B}_{\mathcal{A}}^*$. By construction, $P_A$ and $P_B$ are nearly simultaneous according to the preferred foliation $\mathcal{F}$.
\end{enumerate}

\subsection*{Stage 4: Obtaining Two Further Simultaneous Pairs}

Alice and Bob then rotate and tilt the entire setup about the source. From this new configuration, they repeat Stage 3 to detect a second pair of simultaneous space-time points \((P'_A,P'_B)\), with \(P'_A:=(T'_A,\mathbf{X}'_A)\) and \(P'_B:=(T'_B,\mathbf{X}'_B)\). Rotating and tilting the setup a second time, they repeat Stage 3 once more to detect a third pair \((P''_A,P''_B)\), where \(P''_A:=(T''_A,\mathbf{X}''_A)\) and \(P''_B:=(T''_B,\mathbf{X}''_B)\). Each rotation and tilt is chosen to ensure that the separation vectors \((P_A-P_B)\), \((P'_A-P'_B)\), and \((P''_A-P''_B)\) are non‑coplanar, i.e., they cannot all lie in a single two‑dimensional plane in \(M\).

\subsection*{Stage 5: Solving for the Normal Vector}

Having obtained three pairs of simultaneous space-time points, Alice and Bob substitute the measured coordinates
\[
(T_A,\mathbf{X}_A),\;(T_B,\mathbf{X}_B),\;(T'_A,\mathbf{X}'_A),\;(T'_B,\mathbf{X}'_B),\;(T''_A,\mathbf{X}''_A),\;(T''_B,\mathbf{X}''_B)
\]
into the system of equations Eqs.~\eqref{eq:first_condition}–\eqref{eq:fourth_condition} and solve for \((n^0,n_x,n_y,n_z)\), the components of the normal vector \(n\) in the common Lorentz frame \(\Lambda\). If our predictions (see Table~\ref{tab:temporal_order}) are correct, this procedure yields the orientation of the preferred foliation \(\mathcal{F}\) relative to \(\Lambda\), thereby providing an empirical detection of the foliation.

\medskip

The foliation detection protocol described above assumes a flat foliation, characterized by a constant normal vector field \(n(x)=n\). Extending this method to a curved foliation would be considerably more complex, as the normal vector field \(n(x)\) would then vary from point to point. In that case, determining the foliation would require multiple point-by-point measurements of \(n(x)\) across space-time, rather than the global determination of a single constant vector.

\subsection*{Temporal Resolution and Practical Limitations}

A few remarks are in order concerning the implementation of this protocol and
its ability to locate the switch point. The setups used by Alice and Bob have
their own temporal limitations, and these constrain the spatiotemporal
resolution of the experiment.

Consider first Alice's side. Event $A$ was defined in
Section~\ref{sec:definition_of_the_temporal_order} as particle~1's exit from the
inhomogeneous field, at which point the two spin components of its wave function
have been spatially separated, Eq.~\eqref{eq:spatial_separation_WP}. Writing
$t_A^-$ for the moment at which particle~1 enters the magnet, the duration of
the Stern-Gerlach interaction is $\tau_{\text{SG}} := t_A - t_A^-$. This
quantity matters because the experimenter controls $t_A^-$ rather than $t_A$:
the separation must be complete before Bob releases his particle, so the
Alice-first configuration requires $t_B > t_A^- + \tau_{\text{SG}}$. Were the
two components still overlapping at $t_B$, the weights
\eqref{eq:simplified_weights} would not yet have collapsed to the Kronecker-delta
form \eqref{eq:spin_up}--\eqref{eq:spin_down}, and the inference of
Table~\ref{tab:temporal_order} would not go through. There is, however, no
fundamental limit on $\tau_{\text{SG}}$, which can in principle be made
arbitrarily small.

\medskip

What cannot be guaranteed is that every trajectory completes its transit in
time. Because $\Pi_{\hat{z}}$ is heavy-tailed, some fraction $\varepsilon$ of
the $\sim 10^{5}$ pairs in a run will still be propagating when particle~1
enters the magnet, and this fraction cannot be made to vanish for any finite
delay $\Delta := t_A^- - t_B$. The relevant time scale against which $\Delta$
must be judged is $\tau_{\max}$, the cutoff of the exotic distribution, which as noted in Section~\ref{subsubsec:Backflow} is independent of the initial position and fixed by the setup alone: taking $\Delta \gg \tau_{\max}$ makes
$\varepsilon$ small, though never zero.

It need not be zero. What Bob reads off his screen is not the detailed shape of
the histogram but the presence or absence of counts beyond $\tau_{\max}$: the
exotic distribution vanishes identically there, whereas the heavy-tailed one
does not. Any run in which counts are registered past the cutoff is therefore
decoded as heavy-tailed, and the verdict is insensitive to the precise value of
$\varepsilon$.

\medskip

This also settles the intermediate regime anticipated in
Section~\ref{sec:definition_of_the_temporal_order}, and turns it to advantage.
Near the switch point $\mathcal{B}^{*}_{\mathcal{A}}$, different pairs within a
single run realize different foliation-relative orders, so the accumulated
histogram is a mixture $(1-\varepsilon)\,\Pi_{\hat{x}} +
\varepsilon\,\Pi_{\hat{z}}$ rather than either distribution alone, and Bob
observes a gradual crossover rather than a sharp flip as the waveguide is
displaced. The weight of counts beyond $\tau_{\max}$ is then proportional to
$\varepsilon$, and is directly measurable: it varies from zero, deep in the
Alice-first regime, to its full value once every pair is Bob-first. The switch
point can accordingly be located as the waveguide position at which this excess
reaches half its asymptotic value, and the width of the crossover sets the
precision with which $\mathcal{B}^{*}_{\mathcal{A}}$, and hence the leaf of
$\mathcal{F}$ through $\mathcal{A}$, can be determined.

\medskip

Finally, a remark on the duration of a run. In treating Alice's and Bob's
measurement sequences as single space-time points (Stage~2) we assumed the total
duration $\Delta t \sim N\delta t$ to be negligible compared with the spacelike
interval separating the two laboratories. Since $\delta t$ is bounded below by
$\tau_{\text{SG}}$ and by the transit time in the waveguide, and since
$N\sim 10^5$, this assumption is not innocuous. Performing the measurements
sequentially is not the only option: they could equally be carried out in
parallel, at the cost of a correspondingly large number of experimental setups.
Doing so would reduce $\Delta t$, shrink the space-time regions
$\mathcal{R}_A$ and $\mathcal{R}_B$, and thereby improve the resolution with
which $\mathcal{A}$ and $\mathcal{B}$ are defined, in principle down to the
limit set by $\tau_{\text{SG}}$ and the transit time alone. These remarks are
intended only as guidelines for a possible implementation; experimenters would
doubtless find improvements and simplifications.

\section{A Superluminal Signaling Protocol}\label{sec:superluminal_signaling}

\subsection{The Protocol}

Superluminal signaling becomes possible when Alice's measurement precedes Bob's measurement according to the preferred foliation. In this temporal ordering, if our predictions (Table~\ref{tab:temporal_order}) are correct, the setup of Fig.~\ref{fig:experimental_setup} allows Alice to transmit information to Bob faster than light. The protocol proceeds as follows.

\medskip

\noindent\textbf{Preliminary calibration.} Alice and Bob first ensure that in their configuration Alice's measurements occur before Bob's according to the preferred foliation. To this end, Alice fixes her choice to button~1 (transverse configuration $\hat{n}=\hat{x}$) and they perform trial runs. Bob adjusts the distance of his waveguide from the source until he consistently observes the exotic distribution $\Pi_{\hat{x}}(\tau)$. This confirms that $t_A < t_B$ holds for this configuration of the experimental setup.

\medskip

\noindent\textbf{Code agreement.} Once calibrated, Alice and Bob agree on a binary code:
\begin{itemize}
    \item \textsc{Code ``0''}: Bob observes the heavy-tailed distribution $\Pi_{\hat{z}}(\tau)$ (Fig.~\ref{fig:spin_dependent_arrival_times_distributions}).
    \item \textsc{Code ``1''}: Bob observes the exotic distribution $\Pi_{\hat{x}}(\tau)$ (Fig.~\ref{fig:spin_dependent_arrival_times_distributions}).
\end{itemize}

\medskip

\noindent\textbf{Signaling attempt.} For each attempt, a full experimental run is performed: $\sim10^5$ particle pairs are prepared in the singlet state \eqref{eq:wave_function_factorization} and distributed to both stations. Just before the particles arrive, Alice presses either button~0 or button~1, activating the $\hat{z}$-oriented or $\hat{x}$-oriented Stern–Gerlach magnet, respectively. Her choice projects the spin of Bob's distant particles onto the corresponding axis.

Bob collects the arrival times of his $\sim10^5$ particles and constructs the statistical distribution. He then decodes:
\begin{itemize}
    \item Heavy-tailed distribution $\Pi_{\hat{z}}(\tau)$ $\rightarrow$ ``0''
    \item Exotic distribution $\Pi_{\hat{x}}(\tau)$ $\rightarrow$ ``1''
\end{itemize}

Because the temporal order is fixed ($t_A < t_B$) by the preliminary calibration, Bob's observed distribution directly reflects Alice's button choice, enabling superluminal information transfer.

\subsection{A Historical Comparison with the FLASH Proposal}

\begin{figure}[h]
    \centering
    \includegraphics[width=0.4\textwidth]{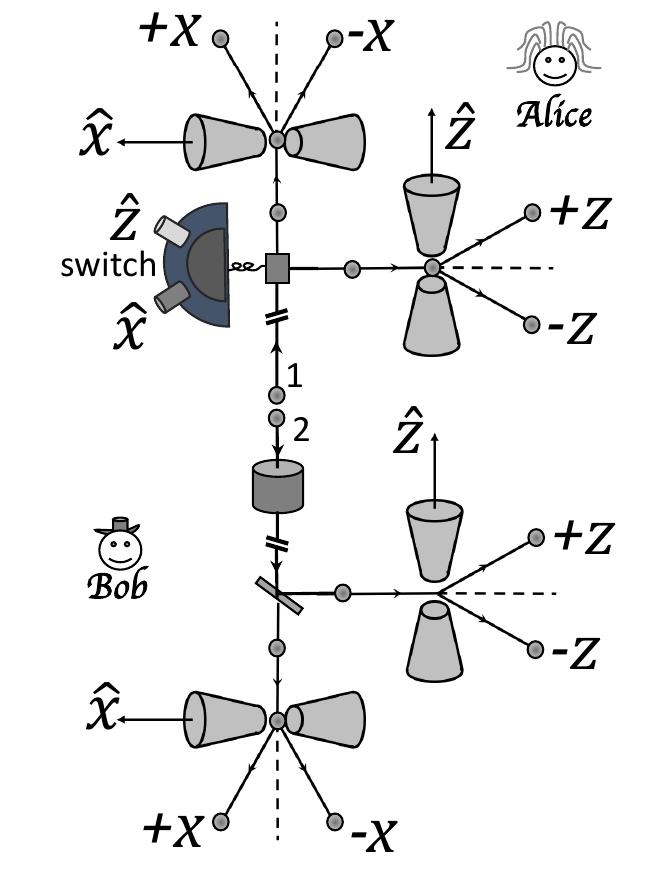}
    \caption{Schematic of the FLASH experiment \cite{herbert1982flash} (compare with Fig.~\ref{fig:experimental_setup}). Alice and Bob share an entangled pair prepared in the singlet state \eqref{eq:singlet_state}. By pressing a button, Alice selects either the $\hat{x}$ or $\hat{z}$ direction for her spin measurement. On Bob's side, the particle passes through a hypothetical ``gain tube'' that multiplies it into a large ensemble of $N \gg 1$ identical copies. A beam splitter then directs half of these copies toward a Stern--Gerlach magnet oriented along $\hat{x}$, and the other half toward a magnet oriented along $\hat{z}$. If the gain tube functioned as assumed and Alice measured first, the resulting statistics would in principle enable faster-than-light signaling.}
    \label{fig:Flash_setup}
\end{figure}

In order to critically assess both the reach and the limitations of our proposed protocol, it is instructive to compare it with a historically influential---and ultimately unsuccessful---proposal for superluminal signaling. In 1982, Nick Herbert put forward a scheme for faster-than-light communication known as FLASH (an acronym for First Laser-Amplified Superluminal Hookup) \cite{herbert1982flash}. The setup, illustrated in Fig.~\ref{fig:Flash_setup}, shares a striking resemblance to ours: Alice and Bob are each sent one particle of an entangled pair prepared in the singlet state \eqref{eq:singlet_state}, and Alice attempts to transmit information by choosing the orientation of her Stern--Gerlach magnet. Moreover, as in our own signaling protocol, the FLASH proposal relies on the assumption that Alice's measurement occurs \textit{before} Bob's.

In the FLASH setup, just as in ours, Alice operates a control panel that allows her to choose between two spin measurement directions (say, $\hat{x}$ or $\hat{z}$) by pressing the corresponding button. The crucial difference lies in Bob's apparatus: instead of measuring arrival times, he attempts to \textit{clone} his particle using a hypothetical ``gain tube'' that produces a macroscopic number $N$ of identical copies (e.g., $N \approx 10^5$). These $N$ particles are then directed toward a beam splitter, which sends half of them toward a Stern--Gerlach magnet measuring spin along $\hat{x}$, and the other half toward a magnet measuring spin along $\hat{z}$.

If such a scheme were feasible, it would enable Alice to transmit a superluminal signal to Bob. By partitioning the $N$ particles into four channels labeled \(\alpha \in \{+\hat{x}, -\hat{x}, +\hat{z}, -\hat{z}\}\) and recording the count $N_\alpha$ in each at the end of a run, Bob can construct the quantity
\begin{equation*}
    \eta = \frac{1}{2}\left(1 + \frac{|N_{+\hat{x}} - N_{-\hat{x}}|}{N_{+\hat{x}} + N_{-\hat{x}}} - \frac{|N_{+\hat{z}} - N_{-\hat{z}}|}{N_{+\hat{z}} + N_{-\hat{z}}}\right),
\end{equation*}
which takes the value $0$ if Alice presses button $0$ (measuring along $\hat{z}$) and $1$ if she presses button $1$ (measuring along $\hat{x}$).\footnote{Statistical fluctuations due to the beam splitter introduce relative errors of order $1/\sqrt{N}$ on the signal.} For example, when Alice measures along $\hat{z}$, the $N/2$ particles sent to the $\hat{z}$-oriented magnet all emerge in a single channel (either $+\hat{z}$ or $-\hat{z}$), while the $N/2$ particles directed to the $\hat{x}$-oriented magnet are distributed equally between $+\hat{x}$ and $-\hat{x}$ (approximately $N/4$ each), giving \(\eta = 0\). Conversely, if Alice measures along $\hat{x}$, the roles are interchanged and Bob obtains \(\eta = 1\). By computing $\eta$ after each run, Bob could therefore decode Alice's bit, enabling faster-than-light communication.

\medskip

Historically, it is worth noting that Herbert based his proposal on the contemporary understanding of laser physics and ``gain tubes'' \cite{herbert1982flash}. More importantly, his line of reasoning directly stimulated the formulation of the celebrated no-cloning theorem by Wootters and Zurek \cite{wootters1982single}, Dieks \cite{dieks1982communication}, and Ghirardi \cite{peres2003no}. This theorem forbids the cloning of Bob's particles regardless of the measurement basis chosen by Alice, thereby invalidating Herbert's argument and ruling out the FLASH proposal as a means of superluminal communication. Furthermore, an experimental test carried out in 2007 with entangled photons \cite{de2007experimental} conclusively demonstrated that FLASH does not work.

Given this historical record, one might infer that our own proposal for superluminal signaling and foliation detection is doomed to failure. Yet there is a crucial difference. The FLASH proposal was introduced by Herbert from the perspective of a reference frame in which Alice measures first. What happens if we instead adopt a frame in which Bob measures first?
\medskip

Assuming a sufficient delay between their respective measurement sequences, Bob would clone his particle into $N$ copies \textit{before} Alice's particle even enters her apparatus. The difficulty arises from the rotational invariance of the singlet state \eqref{eq:singlet_state}, which takes the same form in both the \(\{\chi_{s_1\hat{x}} \otimes \chi_{s_2\hat{x}}\}_{s_1,s_2=\pm 1}\) and \(\{\chi_{s_1\hat{z}} \otimes \chi_{s_2\hat{z}}\}_{s_1,s_2=\pm 1}\) bases. Consequently, if Bob clones in the $\hat{z}$ basis, he will obtain $N/2$ particles in either $+\hat{z}$ or $-\hat{z}$, and $N/4$ particles in each of $+\hat{x}$ and $-\hat{x}$. If he clones in the $\hat{x}$ basis, the statistics are simply interchanged.

When Bob clones first, Herbert's device is not merely difficult to build: its specification is incomplete. A gain tube is defined as amplifying a given input spin state, but in this ordering Bob's particle has no spin state of its own: it is half of an entangled pair, and its reduced state is maximally mixed. The device is therefore silent as to which of the two distributions it should produce, and FLASH yields no prediction at all. Ultimately, the no-cloning theorem explains why the gain tube cannot operate as required, so the question of what it would do never arises.

Our proposal differs from FLASH precisely here. The preferred foliation postulated by a HBD model fixes a unique temporal order for any pair of spacelike-separated events, and we have derived well-defined predictions (see Table \ref{tab:temporal_order}) for both temporal orders. Where FLASH yields no prediction at all in the Bob-first ordering, ours yields the heavy-tailed distribution. In other words, contrary to FLASH, our experimental proposal can be consistently and meaningfully analyzed in a Bohmian framework. The remaining uncertainty concerns neither the apparatus nor the dynamics, but the predictive method of Section~\ref{subsec:predictive_method} used to pass from Bohmian trajectories to detection records.

\subsection{Apparent Retrocausality Without Paradox}

A superluminal signal from Alice to Bob would appear \textit{retrocausal} to a third observer. Consider Charlie, in a Lorentz frame for which the temporal order of the two experiments is reversed relative to $\Lambda$: in his description, Bob's screen displays the distribution encoding Alice's choice before Alice has made it, so that Bob appears to act in anticipation of a message not yet sent. This looks like an action directed into the past, and therefore like something that ought to be forbidden.

No logical paradox arises, however. Within an HBD model, the preferred foliation fixes the true causal order of events, and it is with respect to $\mathcal{F}$, not to Charlie's frame, that control is exercised: as shown in Section~\ref{sec:Alice_measures_first}, the influence runs from Alice to Bob because her measurement precedes his according to $\mathcal{F}$, and the dynamical asymmetry of the guiding fields \eqref{eq:second_guiding_field}--\eqref{eq:expanded_first_current} ensures that it cannot run the other way. Relative to $\mathcal{F}$, the action propagates from past to future. What Charlie describes is a genuine physical process represented in a frame whose notion of simultaneity is not the physically distinguished one; the retrocausality is an artifact of that choice. There is accordingly no means of altering the past in this theory, and none of the contradictions usually associated with closed timelike curves. For closed timelike curves in a Bohmian context, see \cite{Drezet_2019,bohm2006undivided}.

\section{Conclusion}\label{sec:conclusion}

To summarize, we have shown that extending DD's arrival-time predictions to two entangled spin-$1/2$ particles in Minkowski space-time yields a protocol for experimentally detecting the preferred foliation postulated by the Bohm-Hiley-Dirac model, or equivalently by a HBD model equipped with a flat foliation. The same extension also provides a protocol for superluminal signaling. In essence, our predictions, summarized in Table~\ref{tab:temporal_order}, establish the following logical implication:

\begin{quote}
\textit{If the idealized arrival-time distributions of Das and D\"urr are approximately correct, then both the detection of the preferred foliation and superluminal signaling become possible.}
\end{quote}

As with any logical implication involving empirically uncertain propositions, one can adopt a \textit{modus ponens} stance, accepting the antecedent and thus the consequent, or a \textit{modus tollens} stance, rejecting the consequent and thus the antecedent. The maxim that ``one person's \textit{modus ponens} is another's \textit{modus tollens}'' applies with particular force here. Our conditional has the potential to partition the Bohmian community into two camps:

\begin{itemize}
    \item \textbf{The \textit{modus tollens} camp} holds that superluminal signaling and the empirical detectability of a preferred foliation are so deeply in conflict with the known data supporting special relativity that they are physically impossible. For this camp, our result constitutes a \textit{reductio ad absurdum} of DD's arrival-time distributions.
    \item \textbf{The \textit{modus ponens} camp} accepts the antecedent: the de Broglie predictive method, which has proven remarkably reliable across a wide range of scattering and interference phenomena, remains a valid idealization even in the presence of robust quantum backflow. For this camp, our two-particle extension of DD's predictions reveals a striking consequence: the preferred foliation of every relativistic extension of BM could, in principle, be detected.
\end{itemize}

A clarification is needed about what \textit{modus tollens} would reject here. The antecedent of our conditional is not the Bohmian dynamics alone, but a conjunction: DD's distributions follow from the Pauli equation and the guiding equation, together with the quantum equilibrium hypothesis and the de Broglie predictive method. A Bohmian who denies the consequent retains the first two (nothing in the proposed experiment bears on the existence of trajectories or on the $|\Psi_0|^2$ distribution of initial configurations) and is therefore committed to locating the failure in the third, and more precisely in the \textit{passive-screen commitment} of the de Broglie predictive method (Section~\ref{subsec:predictive_method}). In other words, the first crossing of the geometrical surface $\mathcal{D}$ by a screen-free trajectory would not, in this physical regime, approximate the moment at which a real detector fires. That is a substantive claim about the scope of a predictive method, not a refutation of BM. We return to this claim below, once the more radical objection has been addressed.

\medskip

Our aim here is not to advocate for either camp, but to argue for the value of an empirical test of both DD's predictions and our own. We do, however, wish to argue against a third response, which stands in neither camp and rejects the antecedent of our conditional on independent grounds.\footnote{To be more rigorous, the expression ``on independent grounds'' might be problematic, since postulating no-signaling might logically imply the universality of POVM statistics. An attempt to establish this is currently being undertaken by Will Cavendish \cite{cavendish2026}, but has not yet been published.} Although GTZ \cite{Goldstein_2024} do invoke the \textit{modus tollens} version of our implication, their principal objection to DD's predictions is not framed in terms of signaling at all: it rests on the POVM theorem, as recalled in Section~\ref{subsec:GTZ_objection}. On this view our implication is sound but idle: its antecedent is already known to be false, and no experiment need be performed to establish this. The same verdict would then apply to the predictions of Table~\ref{tab:temporal_order}, ruling out our proposal from the outset as unworthy of empirical scrutiny. We announced in Section~\ref{subsec:GTZ_objection} that we would not grant the theorem this authority, and we now give our reasons. It is worth noting that Detlef Dürr was among the architects of the Bohmian account of operators as observables underlying the POVM theorem \cite{durr2004quantum}, and was aware of its potential conflict with the de Broglie predictive method \cite[Remark~7.5]{durr2020understanding}, yet nonetheless derived and published the predictions at issue.

\medskip

Following Maudlin's recent analysis of Einstein's views on scientific methodology \cite{maudlin2025actualphysicsobservationquantum}, the methodological tension at the heart of this debate can be brought out through two of Einstein's remarks to Heisenberg. On one side stands the famous admonition that ``it is the theory which alone decides what is measurable'' \cite{heisenberg1971physics}. With this remark, Einstein was objecting to Heisenberg's claim that a quantum theory should be built solely upon observable quantities. A stronger reading of the claim, however, is this: one must construct a precise theoretical framework in order to determine which experiments are feasible within it. This is exactly what the Bohmian version of the POVM theorem sets out to do for BM. Its reasoning is as follows. For any quantum experiment, if one applies the Bohmian dynamics to ``all the particles involved in the experiment'' \cite[p.~137]{durr2020understanding}, that is, to the \textit{composite system} comprising both the target system and the $\sim10^{23}$ particles composing the apparatus, the theory can \textit{in principle} predict the distribution of the final configuration occupied by all those particles. By equivariance, at quantum equilibrium that configuration is distributed according to the Born rule applied to the wave function of the composite system. If experimental outcomes are then identified, via a calibration function,\footnote{The calibration function is expressed mathematically as \(F:\Gamma\to\mathcal{Z}\), where \(\Gamma\) is the configuration space of the composite system and \(\mathcal{Z}\) the outcome space.} with the final configuration of the apparatus or its pointer, one can prove that BM yields only POVM statistics. We shall refer to this recipe (ascribe a wave function to the composite system of target and apparatus, evolve it to the final time, and read the outcome statistics off the Born measure of the corresponding region of its configuration space) as the \textit{composite-system analysis}.

Crucially, however, the composite-system analysis rests on a premise that has never been subjected to direct empirical test: that the empirical predictions of a quantum experiment are fully captured by applying the Born rule to the wave function of the \textit{composite system}, on the relevant subset of that system's configuration space (see, e.g., Eq.~(5.37b) in \cite{tumulka2022foundations}). While this reasoning is faithful to BM's first principles, it does not reflect how the theory is actually applied. In practice, no one ascribes an initial wave function to a macroscopic measurement device, evolves it through the measurement process, and applies the Born rule to its pointer coordinates. Moreover, for the reasons given in Section~\ref{subsec:predictive_method}, this is not merely a matter of computational cost: there is no canonically specified Hamiltonian that the phrase ``the apparatus'' picks out, so it is not even settled \textit{which} dynamics one would be solving. The empirical success of QM, and of BM, rests instead on applying some version of the Born rule to systems of very limited size (individual particles, atoms, small molecules) and on connecting the resulting distributions to data by means of practical rules. Existing data therefore support the application of BM and of orthodox QM to systems of very limited size; they do not support the application of the Born rule at the scale the POVM theorem takes for granted, namely to ``all the particles involved in the experiment''. This does not disprove the claim that a real detector would strongly influence the outcome of DD's experiment, nor does it refute the assertion that the true arrival-time statistics will ultimately be described by a POVM. It does, however, call into question the theorem's authority as a decisive argument against DD's predictions and, by extension, against our own.

That the composite-system analysis cannot be applied numerically to a real physical experiment is not an abstract complaint; it has a visible symptom in the debate itself. Invoking the theorem, GTZ conclude that DD's distributions are not what BM predicts, but the theorem does not put them in a position to say what BM predicts instead. The constraint \eqref{eq:GTZ_contradiction} tells us that the two distributions must coincide; it does not tell us what either of them is, and no POVM for DD's setup is derived from applying the composite-system analysis to that setup. A concrete, computed distribution is thus excluded on the authority of a result that supplies no concrete distribution of its own. Proponents of the POVM theorem are accordingly left with the task of finding a detector model (for recent efforts in this direction, based on idealized models, see \cite{drezet2024arrivaltimebohmianmechanics,Joz2026}), applicable to DD's waveguide setup, from which POVM-compatible arrival-time statistics could actually be computed and compared with DD's predictions.

\medskip

On the other side of the methodological tension lies Einstein's complementary insight, offered immediately after the remark quoted above: ``[One] must appreciate that observation is a very complicated process'' \cite{heisenberg1971physics}. As Maudlin emphasizes, Einstein later stressed the necessity of idealizations for making numerical predictions in physics \cite{einstein1957albert,maudlin2025actualphysicsobservationquantum}. Because measurement cannot be analyzed with full rigor within any existing theoretical framework, one is always compelled to introduce well-informed idealizations and phenomenological rules that the theory is expected to justify at a later stage. The quantum treatment of a measurement interaction must therefore stop somewhere, and every quantum prediction rests on some idealization. This point was already made in Section~\ref{subsec:predictive_method}, where it motivated the notion of a predictive method in the first place. Within BM, the standard tool for waiting-screen experiments is the de Broglie predictive method, and its passive-screen commitment is precisely such an idealization. It is nonetheless remarkably effective, reproducing the statistics of scattering~\cite{daumer1994scattering,daumer1996scattering}, double-slit interference~\cite{das2025double,viale2003analysis}, and atom diffraction~\cite{sanz2000causal}. Moreover, beyond its direct empirical support, the de Broglie predictive method can be used to \textit{derive} the exact POVMs known to describe many standard experiments, including those of scattering theory and momentum measurements (see \cite[Ch.~16]{durr2009bohmian} and \cite{daumer1994scattering}), a task the composite-system analysis is unable to perform.

Section~\ref{subsec:predictive_method} noted that the domain of empirical validity of the de Broglie predictive method is presumably bounded, and that no criterion has been proposed for identifying in advance the regimes in which its passive-screen commitment fails. We can now say more precisely what is at issue in DD's setup. The experiments against which the method has been confronted with data share a common structure: the detection surface lies in the far field relative to the support of the initial wave function, and no long-range potential acts on the particle in the region where detection occurs.\footnote{Siddhant Das has pointed out to us (private communication) that this condition is too strong as stated: Demir \cite{Demir:2022jvc} applies the de Broglie predictive method to particles falling in a gravitational potential, a
long-range potential that does not vanish at the detection surface.} Under these conditions the wave function approaches a plane wave near the screen and the idealized Bohmian trajectories approximate straight lines of constant velocity; they are effectively outgoing, and recrossings of the detection surface are negligible. So far as we can tell, it is this quasi-classical behavior near the screen that has been taken to license neglecting the detector's back-action in such experiments. The reasoning is nowhere made explicit, however, and it is heuristic: it rests on a classical analogy rather than on a derivation, and no rigorous account has been given of \textit{why} the passive-screen commitment succeeds where it does (private conversation with Will Cavendish).

DD's setup satisfies the first of these conditions but not the second. The screen lies in the far field, but the waveguide imposes a confining potential that does not vanish in its vicinity, so the trajectories remain strongly shaped by the spin-dependent term of the guiding field right up to the moment of detection. In the transverse configuration this is the backflow regime of Section~\ref{subsubsec:Backflow}, in which the axial velocity can reverse outright, condition~\eqref{eq:tractable_back_flow}: the trajectories are not simply outgoing, and the quasi-classical picture that underwrites the passive-screen commitment elsewhere does not apply. One may therefore legitimately ask whether that commitment remains empirically adequate here. What one cannot presently do is answer the question. Since the reasons for the method's success in the familiar regime have never been made precise, there is no criterion by which its failure in this one could be established in advance. A \textit{modus tollens} response that supplies no concrete model of a real detector, showing that the screen's back-action would erase the spin dependence in DD's setup, asserts that a boundary exists without locating it.

Read in this light, the experiment DD propose can be viewed as a test of the
\textit{scope} of the de Broglie predictive method.\footnote{We owe this way of framing the proposal to an anonymous referee.} If the exotic distribution is recorded, the passive-screen commitment survives into a regime in which a persistent long-range potential induces robust quantum backflow and highly non-classical trajectories. The consequences traced in this paper would then follow. If it is not, the method has been shown to have a boundary, and the experiment will have gone some way toward locating it. Either outcome is informative about a question that has so far been settled only by analogy, and neither can be reached by argument alone.

\medskip

We are therefore left with two competing extrapolations, neither of which can claim priority over the other on grounds available to theoretical argument alone. On one side, every quantum experiment performed to date has yielded statistics compatible with a POVM. This uniform record supports an induction toward the universality of POVM statistics, and hence the conclusion that DD's predictions are very probably wrong. We wish to emphasize that this constitutes a reasonable argument for the universality of POVM statistics, but that it rests on an induction, not on the demonstration the POVM theorem claims to provide. On the other side, the de Broglie predictive method has an empirical record of its own, earned in the far-field, potential-free regime, and extending it to the backflow regime is likewise an induction. The two inductions diverge exactly where DD's experiment lies.

The current debate surrounding DD's arrival-time predictions reflects a broader division within the Bohmian community about the theory's future. A majority is inclined to confine BM to reproducing the results of orthodox QM, an approach strongly supported by the predictive power of that formalism.\footnote{It is worth noting that attempts to refute BM by exhibiting a departure from orthodox QM have so far failed; see the recent claim in \cite{NatureKlaers} and the response in \cite{NatureDrezet}.} Others are willing to relax certain assumptions of the theory, and Valentini's proposal is the clearest case in point. On the standard view, quantum equilibrium suffices to ensure no-signaling and to preclude any detection of the preferred foliation; Valentini obtains both by relaxing it \cite{valentini2001hiddenvariablesstatisticalmechanics,ValentiniAFLDB,valentini2010quantum,Valentinibook} (see also
\cite{DrezetEntropy} for a review of possible modifications to BM). Our proposal does not proceed in this way. Even with quantum equilibrium preserved throughout, the de Broglie predictive method, taken to be valid in this regime, implies that superluminal signaling is possible in principle and, more importantly for present purposes, that the preferred foliation could be detected. The two routes therefore locate the required departure at different points: in the distribution of initial configurations for Valentini, and in the passage from screen-free trajectories to detector records for us.

\medskip

To conclude, despite decades of peaceful coexistence between the de Broglie predictive method and the Bohmian version of the POVM theorem, DD's arrival-time predictions have brought to the surface a latent incompatibility whose possibility was already demonstrated in 2005 (Section~\ref{subsec:predictive_method}). We have argued that this incompatibility cannot be resolved in advance by appeal to the POVM theorem: the theorem does not rule out DD's predictions, since its authority rests on a premise, the application of the Born rule at the scale of the composite system, that has never been tested. Only the experiment proposed by Das and D\"urr can settle the debate over the correctness of their arrival-time predictions. As long as those predictions remain unrefuted, and remain firmly anchored in the well-established, empirically successful applications of BM, it is legitimate to explore their revolutionary consequences and to re-examine critically the authority of theorems that would rule them out. Such an attitude is all the more justified given the novelty of the physical regime investigated by DD (robust, controllable backflow) and the absence of any well-established theoretical framework for arrival-time experiments.

In any case, the logical implication established in this paper confirms the boldness of DD's prediction, a boldness that is no defect:

\begin{quote}
   ``Bold ideas, unjustified anticipations, and speculative thought, are our only means for interpreting nature: our only organon, our only instrument, for grasping her. And we must hazard them to win our prize. Those among us who are unwilling to expose their ideas to the hazard of refutation do not take part in the scientific game.'' \cite[p.~281]{popper2005logic}
\end{quote}

\section*{Acknowledgments}

We are deeply grateful to Siddhant Das and Will Cavendish for their insightful comments on an earlier draft of this paper. Siddhant also generously clarified key aspects of DD's arrival-time predictions, and Will provided invaluable discussions on the ongoing debate about arrival times within the Bohmian community. We also thank Tim Maudlin for stimulating public discussions that motivated this investigation and for highlighting the connection between arrival times and the preferred foliation. Finally, we thank the anonymous referees for their careful reading and constructive criticism; one of them suggested framing the proposal as a test of the scope of the de Broglie predictive method, a formulation we have adopted.

\appendix
\section{Guiding fields derivation}
\label{appendix}
We begin by computing the probability density for our quantum state \eqref{eq:wave_function_factorization}, which follows directly from the orthonormality of the spin basis:
\begin{align}
    \rho(\mathbf{x}_1, \mathbf{x}_2, t) &= \Psi^\dagger(\mathbf{x}_1, \mathbf{x}_2, t) \Psi(\mathbf{x}_1, \mathbf{x}_2, t) \nonumber\\
    &=\frac{|g_{0}(\mathbf{x}_2,t)|^2}{2} \sum_{s,s'=\pm 1}  ss' f_{s\hat{n}}^*(\mathbf{x}_1,t) \; f_{s'\hat{n}}(\mathbf{x}_1,t)\; \left[\chi_{sn}\otimes\chi_{-sn}\right]^\dagger \left[\chi_{s'n}\otimes\chi_{-s'n}\right] \nonumber\\
    &= \frac{|g_{0}(\mathbf{x}_2,t)|^2}{2} \sum_{s=\pm1} |f_{s\hat{n}}(\mathbf{x}_1,t)|^2. \label{eq:probability_density}
\end{align}
In the last step we used \(\left[\chi_{sn}\otimes\chi_{-sn}\right]^\dagger \left[\chi_{s'n}\otimes\chi_{-s'n}\right]=\left( \chi_{s\hat{n}}^\dagger\chi_{s'\hat{n}} \right) \left( \chi_{-s\hat{n}}^\dagger\chi_{-s'\hat{n}} \right)\) along with the orthonormality relations
\begin{equation}\label{eq:orthonormality_conditions_spin_basis}
    \chi_{s\hat{n}}^\dagger \chi_{s'\hat{n}} = \delta_{s,s'},
\end{equation}
which eliminate all cross terms, and eventually used $s^2=+1$. The result is a simple weighted sum of the probability densities associated with each spin branch.

We now derive the velocity field for particle~2, which is more central to our analysis than that of particle~1.

\medskip

The Bohmian velocity for particle~2, given by Eq.~\eqref{eq:bohmian_velocity_def}, is determined by the probability current $\mathbf{J}^{(2)}_{\text{Pauli}}$ defined in Eq.~\eqref{eq:Jk_general}. Substituting the singlet state \eqref{eq:wave_function_factorization} into this expression yields a structured form that reveals how the spatial and spin degrees of freedom intertwine in the guiding field of the second particle. We proceed by evaluating the \textit{convective flux} and the \textit{spin flux} separately, then combine them to obtain the total current and, finally, the Bohmian velocity.

\bigskip

We first evaluate the convective current (first term in Eq.~\eqref{eq:Jk_general}) for the singlet state ~\eqref{eq:wave_function_factorization}:
    \begin{align}
            \mathbf{J}^{(2)}_{\text{conv}}(\mathbf{x}_1, \mathbf{x}_2, t)&=\frac{\hbar}{m}\operatorname{Im}\left[ \Psi^\dagger \nabla_2 \Psi \right]= \frac{ |g_{0}(\mathbf{x}_2,t)|^2}{2}\,  \sum_{s=\pm 1} |f_{s\hat{n}}(\mathbf{x}_1,t)|^2\; \frac{\hbar}{m}\; \nabla_2 S^{(2)}_{0}(\mathbf{x}_2,t) \label{eq:conv_current}
    \end{align}
where we used the spin orthogonality Eq.~\eqref{eq:orthonormality_conditions_spin_basis}, and the polar form Eq.~\eqref{eq:polar_g} that gave us:
\[
    g_{0}^*(\mathbf{x}_2,t) \, \nabla_2 g_{0}(\mathbf{x}_2,t)=|g_{0}(\mathbf{x}_2,t)|\; \nabla_2|g_{0}(\mathbf{x}_2,t)|+i\; |g_{0}(\mathbf{x}_2,t)|^2 \;\nabla_2 S^{(2)}_{0}(\mathbf{x}_2,t). 
\]

\bigskip

To compute the spin current for particle~2 (i.e. the second term in Eq.\eqref{eq:Jk_general}), we first evaluate the spin expectation density $\Psi^\dagger \boldsymbol{\sigma}_2 \Psi$, a vector-valued density giving the local expectation value of particle $2$'s spin at each configuration point $(\mathbf{x}_1,\mathbf{x}_2)$. Substituting the singlet state~\eqref{eq:wave_function_factorization} and using $\boldsymbol{\sigma}_2 := \mathbb{I} \otimes \boldsymbol{\sigma}$, we obtain:
\begin{align}
    \Psi^\dagger \boldsymbol{\sigma}_2 \Psi 
    =|g_{0}(\mathbf{x}_2,t)|^2\hat{n}\; \frac{1}{2}\sum_{s=\pm 1} -s|f_{s\hat{n}}(\mathbf{x}_1,t)|^2 \label{eq:spin-density}
\end{align}
where we used the orthonormality condition \eqref{eq:orthonormality_conditions_spin_basis}, $s^2=1$, and \(\chi_{-s\hat{n}}^\dagger \boldsymbol{\sigma} \chi_{-s\hat{n}} = \langle \boldsymbol{\sigma} \rangle_{\chi_{-s\hat{n}}} = -s\hat{n}\); because $\chi_{-s\hat{n}}$ is an eigenstate of $\hat{n}\cdot\boldsymbol{\sigma}$ with eigenvalue $-s$.

Taking the curl of Eq.\ \eqref{eq:spin-density} and multiplying by $\hbar/2m$ yields the spin current:
\begin{align}
    \mathbf{J}^{(2)}_{\text{spin}}(\mathbf{x}_1, \mathbf{x}_2, t) =\frac{\hbar}{4m}\left[\sum_{s=\pm 1} -s|f_{s\hat{n}}(\mathbf{x}_1,t)|^2\right]  \nabla_2\times\left(|g_{0}(\mathbf{x}_2,t)|^2\hat{n}\right)\label{eq:intermediary_spin_current}
\end{align}
To evaluate the curl, we apply the vector identity $\nabla \times (a \mathbf{b}) = a (\nabla \times \mathbf{b}) + (\nabla a) \times \mathbf{b}$ with $a = |g_{0}|^2$ and $\mathbf{b} = \hat{n}$. Since $\hat{n}$ is a constant vector, $\nabla_2 \times \hat{n} = 0$, leaving: \(\nabla_2 \times \left( |g_{0}|^2 \hat{n} \right) = \left( \nabla_2 \left(|g_{0}|^2\right) \right) \times \hat{n}.\) Since $\nabla_2 \left(|g_{0}|^2\right) = 2|g_{0}|\; \nabla_2 \left(|g_{0}|\right)$, we obtain: 
\[
\nabla_2 \times \left( |g_{0}(\mathbf{x}_2,t)|^2\; \hat{n} \right)=2|g_{0}|\, \left(\nabla_2 |g_{0}|\right)\times\hat{n}=2|g_{0}|^2\, \left(\frac{\nabla_2 |g_{0}|}{|g_{0}|}\times\hat{n}\right).
\]
Substituting this back into Eq.\ \eqref{eq:intermediary_spin_current}, we obtain:
\begin{align}
    \mathbf{J}^{(2)}_{\text{spin}}(\mathbf{x}_1, \mathbf{x}_2, t) =\frac{\hbar}{2m}|g_{0}(\mathbf{x}_2,t)|^2 \sum_{s=\pm 1} |f_{s\hat{n}}(\mathbf{x}_1,t)|^2  \left[-s\frac{\nabla_2|g_{0}(\mathbf{x}_2,t)|}{|g_{0}(\mathbf{x}_2,t)|}\times\hat{n}\right]\label{eq:spin_current}
\end{align}
This expression reveals that the spin current is a weighted sum of contributions from each spin branch, each proportional to the cross product of the logarithmic gradient\footnote{The expression "logarithmic gradient" refers to the fact that $\frac{\nabla|\psi|}{|\psi|}=\nabla\operatorname{ln}(|\psi|)$.} of the spatial amplitude with the quantization axis $\hat{n}$.

\bigskip

With both the convective and spin currents computed, we now assemble the total probability current for particle~2 and extract the corresponding Bohmian velocity field. Summing the convective current \eqref{eq:conv_current} and spin current \eqref{eq:spin_current} yields the complete Pauli current:
\begin{equation}\label{eq:total_current_2}
    \mathbf{J}^{(2)}_{\text{Pauli}} = \frac{ |g_{0}(\mathbf{x}_2,t)|^2}{2}\sum_{s=\pm 1} |f_{s\hat{n}}(\mathbf{x}_1,t)|^2 \,\left[ \frac{\hbar}{m} \nabla_2 S^{(2)}_{0}(\mathbf{x}_2,t) - s\frac{\hbar}{m} \frac{\nabla_2 |g_{0}(\mathbf{x}_2,t)|}{|g_{0}(\mathbf{x}_2,t)|} \times \hat{n} \right].
\end{equation}

Substituting this expression \eqref{eq:total_current_2} together with the probability density \eqref{eq:probability_density} into the guiding equation \eqref{eq:bohmian_velocity_def} (with $k=2$) yields the Bohmian velocity field for particle~2 \eqref{eq:second_guiding_field}.

\medskip

An analogous derivation for particle~1 yields its Pauli current and Bohmian velocity field. Following the same procedure as for particle~2, we obtain the Bohmian guiding field for particle~1 \eqref{eq:expanded_first_current}.

\section{Toward a Relativistic Treatment}
\label{app:relativistic}

The calculations of Sections~\ref{sec:A_General_Bohmian_Formalism}--\ref{sec:Bob_measures_first} are carried out in the Bohm-Pauli theory on Galilean space-time. Since our thought experiment requires spacelike-separated observers sharing an EPR state, and since the preferred foliation $\mathcal{F}$ may in principle move at an arbitrarily large (though subluminal) speed relative to the common Lorentz frame $\Lambda$, one may ask whether a directly relativistic analysis based on the Dirac equation would not be more appropriate. Although this is not the main focus of the present work, we outline here the procedure that could be followed in the general case, and explain why our results are expected to remain valid.

\medskip

To this end, we assume that the EPR state of two fermions initially at rest and obeying the Pauli exclusion principle is given by the singlet state:
\begin{eqnarray}
  \Psi_{\alpha,\beta}^{\textrm{in}}(x_0,y_0)=\frac{1}{\sqrt{2}}[f(x_0)g(y_0)+g(x_0)f(y_0)]\cdot\frac{1}{\sqrt{2}}[\chi^{\uparrow}_{\alpha}\chi^{\downarrow}_{\beta}-\chi^{\downarrow}_{\alpha}\chi^{\uparrow}_{\beta}]\label{spinor}
\end{eqnarray}
Here, $\alpha$ and $\beta$ are two integer indices ranging from 0 to 3, corresponding to the components of the bispinors, while $f(x), g(y) \in \mathbb{C}$ are two wave functions that are well localized in space at the initial time $t_0$. Spacetime points are denoted by $x=(t,\mathbf{x})$ in the Lorentz frame $\Lambda$, and similarly for $y$. By definition we use the two bispinors for the fermions at rest:  
\begin{eqnarray}
   \chi^{\uparrow} =\left(\begin{array}{c}1\\0\\0\\0
    \end{array}\right),  & \textrm{  }\chi^{\downarrow} =\left(\begin{array}{c}0\\1\\0\\0
    \end{array}\right).
\end{eqnarray}
Note that, for completeness, the symmetry of the spatial part of $\Psi_{\alpha,\beta}(x_0,y_0)$ is explicitly included in Eq.~\ref{spinor}.
Moreover, in the thought experiments considered here, Alice and Bob have developed two experimental setups designed to accelerate the two initially resting spins and direct them toward their respective workstations (laboratories). In the multi-time Dirac representation, at spacetime points $x$, $y$, the quantum state of the system can be unambiguously written as
\begin{eqnarray}
  \Psi_{\alpha,\beta}^{\textrm{interaction}}(x,y)=\frac{1}{2}[f^{\uparrow}_{\alpha}(x)g^{\downarrow}_{\beta}(y)-f^{\downarrow}_{\alpha}(x)g^{\uparrow}_{\beta}(y)+g^{\uparrow}_{\alpha}(x)f^{\downarrow}_{\beta}(y)-g^{\downarrow}_{\alpha}(x)f^{\uparrow}_{\beta}(y)]\label{spinor2}
\end{eqnarray}
where the bispinor $f^\uparrow_\alpha(x)$ is obtained through the unitary evolution: $f(x_0)\chi^{\uparrow}_{\alpha}\rightarrow f^\uparrow_\alpha(x)$ and similarly for the three other terms of Eq.~\ref{spinor2}.\\
\indent It is worth noting that, within the multi-time  Dirac formalism used here, the times $t_x$ and $t_y$ associated with the spacetime points $x$ and $y$ are, in general, no longer required to be equal.  Furthermore, since the two fermions, within the classical electromagnetic field approximation, do not interact with each other but only with external fields, there is no need to require that $x$ and $y$ be spacelike separated. This condition is nevertheless satisfied at later times, once the $f$ and $g$ wave packets have become well separated. The precise shape of the wave packets is not particularly relevant here. We assume that, immediately before entering the Stern–Gerlach apparatus for Alice’s particle and the time-of-arrival detector for Bob’s particle, the fermions are described by wave functions of the form corresponding to particles that have undergone a Lorentz boost:
\begin{eqnarray}
   f^{\uparrow}(x) \sim F(x)e^{-ipx}\left(\begin{array}{c}1\\0\\\frac{\boldsymbol{\sigma}\cdot \mathbf{p}}{E+m}\left(\begin{array}{c}1\\0\end{array}\right)
    \end{array}\right),  f^{\downarrow}(x) \sim F(x)e^{-ipx}\left(\begin{array}{c}0\\1\\\frac{\boldsymbol{\sigma}\cdot \mathbf{p}}{E+m}\left(\begin{array}{c}0\\1\end{array}\right)
    \end{array}\right)\nonumber\\
    g^{\uparrow}(x) \sim G(x)e^{-ip'x}\left(\begin{array}{c}1\\0\\\frac{\boldsymbol{\sigma}\cdot \mathbf{p'}}{E'+m}\left(\begin{array}{c}1\\0\end{array}\right)
    \end{array}\right),  g^{\downarrow}(x) \sim G(x)e^{-ip'x}\left(\begin{array}{c}0\\1\\\frac{\boldsymbol{\sigma}\cdot \mathbf{p'}}{E'+m}\left(\begin{array}{c}0\\1\end{array}\right)
    \end{array}\right)\label{spinor3}
\end{eqnarray} In this formula, $F(x)$ and $G(x)$ denote two wave packets that are strongly localized in Alice’s and Bob’s laboratories, respectively. They are characterized by the mean four-momenta $p = (E, \mathbf{p})$ and $p' = (E', \mathbf{p'})$, corresponding to particles with energy $E$ and momentum $\mathbf{p}$ on Alice’s side, and energy $E'$ and momentum $\mathbf{p'}$ on Bob’s side. Finally $\boldsymbol{\sigma}$ is a vector regrouping the three Pauli matrices. \\
\indent As can be seen from Eq. \ref{spinor3}, after the Lorentz boost, the lower components ($\alpha=2,3$) of the bispinor are no longer zero, making the analysis with a Stern–Gerlach apparatus (on Alice’s side) or a time-of-arrival detector (on Bob’s side) more involved than in the non-relativistic case.
There are two possible alternatives. One can either accelerate Alice’s and Bob’s laboratories as well, so that the spinors recover a non-relativistic form in their respective rest frames, or decelerate the particles just before they enter Alice’s and Bob’s laboratories. In either case, the aim is to reduce the bispinors to a simpler form, allowing the non-relativistic analysis presented in Section~\ref{sec:A_General_Bohmian_Formalism} to be carried out again. Without this modification, the presence of the lower components of the bispinors of the two particles could make the analysis considerably more challenging. This is essential for the protocols discussed in this article, which rely on the perfect anticorrelation of the EPR singlet state and in particular this is key if we want to recover the logical inferences of Table \ref{tab:temporal_order}.

\medskip

This analysis allows us to safely apply the HBD model, which is based on the application of the multi-time Dirac formalism to spacelike hypersurfaces. As discussed in Section \ref{sec:the_preferred_foliation}, a preferred foliation of spacetime is introduced in order to define these spacelike hypersurfaces. In the present case, this foliation, denoted by $\mathcal{F}$, allows us to calculate the probability current associated with the two particles and, consequently, to define the dBB velocities as four-vectors, $v_1^\mu(x,y)$ for particle 1, and analogously for particle 2.
Within the HBD formalism, $x$ and $y$ denote the spacetime coordinates of the two particles on the same leaf of the spacetime foliation. This amounts to imposing a nonlocal synchronization between the two particles and consequently induces a nonlocal dynamical coupling between them. Developing this formalism for concrete physical problems is in general demanding, and for that reason we have restricted the analysis of this paper to the non-relativistic regime. Under the conditions described above, however, the bispinors reduce to their rest-frame form, the analysis of Sections~\ref{sec:A_General_Bohmian_Formalism}--\ref{sec:Bob_measures_first} applies as written, and the qualitative distinction between $\Pi_{\hat{z}}(\tau)$ and $\Pi_{\hat{x}}(\tau)$ on which our protocols depend is preserved. This expectation is consistent with Das's demonstration that spin-dependent trajectories and persistent backflow survive a fully relativistic treatment based on the Dirac equation \cite{das2021relativisticelectronwavepackets}. More elaborate calculations could of course further substantiate our results in the relativistic regime, while taking into account the assumptions about Alice's and Bob's setups discussed above.




\end{document}